\pdfoutput=1 
\documentclass[aps,prx,superscriptaddress,twocolumn,10pt]{revtex4-2} 
\usepackage[utf8]{inputenc}
\usepackage[margin=1in]{geometry}
\usepackage{amsmath,amsfonts,amssymb}
\usepackage[T1]{fontenc}
\usepackage{titletoc}
\usepackage{dsfont}
\usepackage{xcolor}
\definecolor{forestgreen}{rgb}{0, 0.7, 0}
\definecolor{lightpurple}{rgb}{0.8, 0, 0.8}
\definecolor{pink}{rgb}{1, 0.2, 0.5}
\usepackage{physics}
\usepackage{graphicx}
\usepackage{stackengine}
\usepackage{hyperref}[hypertexnames=false]
\hypersetup{
	colorlinks   = true, 
	urlcolor     = pink,
	linkcolor    = pink, 
	citecolor   = pink 
}
\usepackage[all]{hypcap} 
\usepackage{enumitem} 
\usepackage{titlesec} 
\usepackage{booktabs,tabularx}
\usepackage{multirow}
\usepackage{verbatim} 
\usepackage{bbm}
\usepackage{bm}
\usepackage[export]{adjustbox}
\usepackage[caption=false,position=top,labelformat=empty]{subfig} 

\usepackage{tikz}
\newcommand*\detector{\tikz[]{\draw[fill=gray!30] (0,0) -- (0,-0.3) -- (0.15, -0.3) arc[start angle=-90, end angle=90, radius=0.15]  -- cycle;}}

\titleformat{\section}{\bfseries\centering\uppercase}{\thesection.}{1em}{}
\titlespacing{\section}{0pt}{1em}{1em}
\titlespacing{\subsection}{0pt}{1em}{1em}

\titlecontents{section}
  [0em]                                  
  {\addvspace{0.5\baselineskip}\bfseries} 
  {\contentslabel{2.0em}}                
  {}                                     
  {\hfill\contentspage}                  

\titlecontents{subsection}
  [2.0em]
  {\small}
  {\contentslabel{2.5em}}                
  {}
  {\hfill\contentspage}

\newcommand{\printAppendixTOC}{%
    \startcontents[appendix]
    \printcontents[appendix]{}{1}{\section*{Appendix Contents}} 
}

\newcommand{\ee}{\end{equation}}
\newcommand{\be}{\begin{equation}}
\newcommand{\p}[1]{\left( #1 \right)}
\newcommand{\br}[1]{\left[#1\right]}

\newcommand{\hc}[1]{#1^\dagger} 

\renewcommand\bra[1]{{\langle{#1}|}}
\makeatletter
\renewcommand\ket[1]{%
	\@ifnextchar\bra{\k@t{#1}\!}{\k@t{#1}}%
}
\newcommand\k@t[1]{{|{#1}\rangle}}
\makeatother
\newcommand{\kb}[2]{\ket{#1} \bra{#2}}

\newcommand{\subalign}[1]{%
	\begin{subequations}
		\begin{align}
			#1
		\end{align}
	\end{subequations}
}
\renewcommand{\eqref}[1]{Eq.~(\ref{#1})} 
\newcommand{\secref}[1]{Sec.~\ref{#1}} 
\newcommand{\figref}[1]{Fig.~\ref{#1}} 
\newcommand{\appref}[1]{App.~\ref{#1}} 
\newcommand{\refref}[1]{Ref.~\citenum{#1}}

\newcommand{\ra}{\rightarrow}
\newcommand{\om}{\omega}
\newcommand{\g}{\gamma}
\newcommand{\kint}{\kappa_\textrm{int}}

\newcommand{\overbar}[1]{\mkern 1.5mu\overline{\mkern-1.5mu#1\mkern-1.5mu}\mkern 1.5mu}

\newcommand{\cB}{\mathcal{B}}
\newcommand{\cC}{\mathcal{C}}
\newcommand{\cE}{\mathcal{E}}
\newcommand{\cF}{\mathcal{F}}
\newcommand{\cO}{\mathcal{O}}

\newcommand{\tin}{\textrm{in}}
\newcommand{\tout}{\textrm{out}}
\newcommand{\nvec}{\vec{n}}

\newcommand{\hvec}{\vec{h}}

\DeclareMathOperator*{\Motimes}{\text{\raisebox{0.25ex}{\scalebox{0.8}{$\bigotimes$}}}} 

\newcommand{\ketm}[1]{ \ket{#1}\!\rangle} 
\newcommand{\bram}[1]{ \langle\!\bra{#1}}

\newcommand{\kbm}[2]{ \ketm{#1}\!\bram{#2} }
\newcommand{\vac}{\ketm{0}} 
\newcommand{\red}[1]{\textcolor{red}{#1}}

\newcommand{\blue}[1]{\textcolor{blue}{#1}}

\definecolor{turquoise}{rgb}{0.0, 0.85, 0.82}

\newcommand{\snlNM}{Quantum Algorithms and Applications Collaboratory (QuAAC), Department of Quantum Computer Science, Sandia National Laboratories, Albuquerque, NM 87185, USA} 
\newcommand{\snlCA}{Quantum Algorithms and Applications Collaboratory, Sandia National Laboratories, Livermore, CA 94550, USA}


\begin{document}
	
\title{
    Limits of heralded photonic Bell-state generation in the presence of loss
} 
\author{Kevin J. Randles}
\email{kjrandl@sandia.gov}
\affiliation{\snlNM}
\author{Manuel H. Mu\~noz-Arias}
\affiliation{\snlCA}
\author{Mohan Sarovar}
\affiliation{\snlCA}

\begin{abstract}
    High-quality entangled states of photons underlie quantum information science (QIS) applications across communication, sensing, and computing. In many discrete-variable photonic QIS architectures, large application-ready states (e.g., cluster states, repeater graph states) are constructed via fusion measurements on small entangled seed states, of which Bell states are the fundamental example. The quality of seed-state generation therefore sets a baseline for application performance, making it crucial to understand this process under realistic error mechanisms, particularly in integrated photonics experiments. In this work, we analyze five heralded schemes for generating event-ready photonic Bell states, contrasting their heralding probabilities, fidelities, and error robustness. We develop a hierarchy of error models, progressing from an analytically tractable lumped-loss model to realistic heralded single-photon sources with multiphoton emission errors and finally to integrated frequency-bin implementations with architecture-dependent loss. Across these models, we find that schemes based on higher-order multiphoton interference provide superior fidelity robustness in low-loss implementations, while lower-photon-number schemes can be preferable when probabilistic sources or lossy beamsplitters dominate the resource cost. Our results provide design guidance for integrated discrete-variable quantum photonics, with particular relevance for frequency-bin architectures where active beamsplitter loss can determine the optimal resource-state-generation strategy.
\end{abstract}
	
\maketitle

\section{Introduction} 
Photons have long been understood as natural carriers of quantum information and entanglement, especially for quantum communication and networking applications \cite{gisin2007quantum}. Facilitated by advancements in hardware and theory, photonic approaches have further emerged as promising means of scaling quantum computers---by acting as conduits linking matter-based quantum systems for distributed quantum processing \cite{main2025distributed}---and of doing quantum computing itself \cite{maring2024versatile,psiquantum2025manufacturable}, especially with the recent advent of fusion-based quantum computing \cite{bartolucci2023fusion}. 
Photonic quantum computing (PQC) is promising for several reasons \cite{rudolph2017optimistic}:
the level of stochastic noise it exhibits can be incredibly small compared to matter-based approaches, it is naturally very fast (natively operating at gigahertz clock cycles) and modular, and many integrated photonics devices are compatible with established semiconductor fabrication (CMOS) technologies, enabling large-scale manufacturability.

Fundamental to photonic quantum information processing---including PQC as well as quantum networking applications such as all-photonic quantum repeaters \cite{azuma2015all}---is the ability to controllably generate high-quality entangled resource states of photons (typically taken to be stabilizer states, which are locally Clifford equivalent to graph states).
However, unlike the circuit model of quantum computation for matter-based qubits, there is no \emph{deterministic} linear optical entangling gate such as a CNOT gate. 
Accordingly, other methods are needed to controllably generate high-quality entangled photonic states.

One method, which is closely aligned with standard linear-optical PQC \cite{knill2001scheme,bartolucci2023fusion}, is to use measurement-induced nonlinearities in consonance with linear optical circuits to perform \emph{probabilistic} entangling operations.
Such operations enable the heralded generation of entangled resource states. Therein, entangled output states are probabilistically prepared by passing single photons through tailored photonic circuits with success heralded by the detection pattern observed on a subset of measured modes \cite{forbes2025heralded,hartnett2026automated}.
Moreover, projective probabilistic entangling measurements called \emph{fusions} can be used to construct larger entangled resource states from smaller seed states such as Bell states or 3-qubit Greenberger--Horne--Zeilinger (GHZ) states \cite{browne2005resource,bartolucci2023fusion}.

Photon loss is a pervasive error mechanism in photonic quantum information processing. Accordingly, the faithful modeling of how photon loss impacts resource-state generation and of how corresponding errors propagate is crucial for assessing the feasibility of a given implementation.
In this paper, we are concerned with the impact of realistic photon loss mechanisms on heralded resource-state generation in integrated optics platforms. We use heralded Bell-state generation (HBSG) as a benchmark for understanding how source loss, multiphoton emission, detector loss, and beamsplitter loss constrain integrated photonic resource-state generation. In particular, we analyze five heralded schemes for generating dual-rail encoded Bell states, contrasting their heralding probabilities, fidelities, and robustness across various error regimes. We perform strong linear-optical simulations of the lossy schemes using Quandela’s Perceval software \cite{heurtel2023perceval} and develop custom modeling tools to reduce the overhead of simulating loss and multiphoton sources. 

A central goal of this work is to connect component-level imperfections to resource-state-generation performance in a way that is both analytically transparent and adaptable to specific integrated-photonic architectures. We therefore present a hierarchy of error models, beginning with an encoding-agnostic lumped-loss model for dual-rail HBSG. This model captures regimes in which photon loss can be effectively summarized by source, circuit, and detector efficiencies, and it allows us to derive analytic expressions that clarify how loss affects the heralding probability and Bell-state fidelity of each scheme. We then incorporate SFWM-compatible heralded single-photon sources (HSPSs), where source loss and imperfect heralding introduce both vacuum and multiphoton contributions to the input state. Finally, we specialize the framework to integrated frequency-bin implementations, where active frequency-domain beamsplitters introduce architecture-dependent, nonuniform loss that cannot generally be reduced to the lumped model and qualitatively changes which schemes are preferred. Across these levels of modeling, we identify which HBSG schemes are favored in different error regimes, show that schemes based on higher-order multiphoton interference exhibit enhanced fidelity robustness in low-loss implementations, and quantify a tradeoff between the output fidelity and heralded generation probability.

Dual-rail encoding with frequency modes has become increasingly feasible over the past decade, with significant strides in both the theoretical framework and experimental realization of frequency-bin processing and PQC \cite{lukens2017frequency,lu2023frequency,clementi2023programmable,myilswamy2025chip,lukens2026paradigm,congia2026fully}. Frequency encodings naturally support massive multiplexing, as a single fiber or waveguide can support hundreds of naturally phase-stable frequency bins \cite{joshi2018frequency}. This advantage, leveraged classically in wavelength-division multiplexing as a cornerstone of high-speed telecommunications, is equally attractive for quantum information processing. However, frequency-bin encodings also introduce distinct implementation challenges. In particular, linear-optical primitives such as beamsplitters require active, driven devices to engineer coherent inter-frequency interference. While such devices can enable native multimode operations, they can also introduce significant loss relative to passive spatial- or polarization-mode beamsplitters. In this work, we build on recent detailed models of ring-resonator-based frequency-domain beamsplitters \cite{munoz2026modeling} to quantify how imperfect active beamsplitters affect HBSG in integrated frequency-bin platforms.
 
The paper is organized as follows. In Secs.~\ref{sec:herBSG} and \ref{sec:sources}, we review the HBSG schemes considered here and the source models used in the analysis. In \secref{sec:lumpedHBSG}, we introduce the generic lumped-loss model and derive analytic expressions for heralding probabilities and fidelities. In \secref{sec:HBSGwHSPSs}, we extend the analysis to lossy HSPSs with multiphoton errors. In \secref{sec:frequencyBinHBSG}, we specialize the framework to integrated frequency-bin implementations with lossy active beamsplitters. Finally, in \secref{sec:conclusion}, we summarize implications for integrated photonic resource-state generation and discuss extensions to larger resource states.

\section{Background: Heralded Bell-State Generation}\label{sec:herBSG} 
Bell states are useful throughout much of quantum information science, e.g., enabling quantum teleportation. 
They are the prototypical entangled resource state and a building block for more complex resource states in fusion-based approaches. Moreover, they can be used as the starting point for nearly ballistic percolation schemes for generating large cluster states \footnote{
    See Ref.~\citenum{bartolucci2021creation} for details including the introduction of boosted type-I fusions as well as a caveat regarding the need for a single step of local adaptivity, which is not needed for larger starting states such as $n\geq 3$ qubit GHZ states.
}.
Accordingly, modeling heralded Bell-state generation (HBSG) is not only well suited for theory, but is also crucial for guiding early experiments and subsequent modeling of larger resource-state generation. 
Moreover, despite its relative simplicity, the faithful modeling of lossy HBSG already exhibits a rich character and is surprisingly nontrivial as strong linear-optical simulation gets rapidly more difficult (in terms of both computation time and memory) as the number of photons and modes being simulated increases \cite{heurtel2023strong}. 
A review of heralded approaches for photonic entanglement generation is given in \refref{forbes2025heralded}.

Note that a similar analysis contrasting HBSG circuits in the presence of loss and other non-idealities was conducted in \refref{shaw2023errors}. Therein, they use a first-quantized, continuous-variable modeling framework 
to analyze the impact on HBSG schemes of errors inducing ``non-computational leakage,'' where more than one photon occupies the dual rails meant to encode a qubit, especially as induced by the photons being partially distinguishable in degrees of freedom (DoFs) other than the qubit encoding.
Meanwhile, we work directly within the second-quantized, discrete-variable formalism and are primarily concerned with the impact of photon loss and multiphoton errors from heralded single-photon sources. Our analytical treatment brings conceptual clarity to the effects of these imperfections on HBSG, and allows us to pinpoint the zero-transmission law~\cite{tichy2010zero} as the origin of the observed robustness of some of the analyzed schemes. Additionally, we analyze the impact of nonuniform beamsplitter loss as is particularly relevant in frequency-domain implementations. Thus, we find that our focus and methodology differ appreciably from that of \refref{shaw2023errors}, yet our findings are complementary, leading to a rather complete picture of which HBSG schemes are favored across different contexts and error regimes.

To keep this article self-contained we restrict our analysis to comparing HBSG schemes, though similar analyses can be performed for the generation of other seed states. For instance, \refref{wiesner2024influence} investigates the impact of imperfections (photon loss, multiphoton errors, and photon distinguishability) on the 3-qubit GHZ state generation scheme of \refref{varnava2008good}. 

\subsection{Dual-rail encoding} 
We work in the dual-rail encoding with qubit basis states, as defined by mode occupations, of the form 
\begin{subequations}
    \label{eq:dualRailQubitDef}
    \begin{align}
        \ket{0} &\equiv \ketm{1_j 0_k} = \hc{a}_j \vac, \\
        \ket{1} &\equiv \ketm{0_j 1_k} = \hc{a}_k \vac,
    \end{align}
\end{subequations}
where $\hc{a}_j$ and $\hc{a}_k$ are the creation operators for orthogonal modes $j$ and $k$, respectively. Here we use $\ketm{\cdot}$ to denote Fock basis states with well-defined mode occupation numbers, and $\ket{\cdot}$ to denote qubit states as well as superpositions of Fock states. 
We assume different modes are disjoint in the DoF(s) being used for the encoding, but are otherwise indistinguishable such that their creation and annihilation operators satisfy the standard bosonic commutation relations $[a_j, \hc{a}_k] = \delta_{jk}$ and $[a_j, a_k] = 0$. We will typically take qubits to be defined on adjacent modes (e.g., waveguides or frequency bins for spatial and frequency encodings, respectively), $k = j+1$. 

Then the canonical Bell basis for maximally entangled two-qubit states is 
\subalign{
    \ket{\Phi^\pm} &= \frac{\ket{00} \pm \ket{11}}{\sqrt{2}}
    = \frac{ \ketm{1010} \pm \ketm{0101}  }{\sqrt{2}}, \\
    \ket{\Psi^\pm} &= \frac{\ket{01} \pm \ket{10}}{\sqrt{2}}
    = \frac{ \ketm{1001} \pm \ketm{0110} }{\sqrt{2}}.
}
Along with the canonical Bell states, we will also declare success upon generating states of the form
\be
    \ket{\chi^\pm} = \frac{ \ketm{1100} \pm \ketm{0011} }{\sqrt{2}},
\ee	
which are Bell states with the modes permuted. For instance, $\ket{\chi^\pm}$ and $\ket{\Phi^\pm}$ are equivalent up to a permutation of the central two modes or equally by defining the two qubits on modes of same parity. 

\subsection{Circuit notation}
When drawing photonic circuits a horizontal line corresponds to a distinct mode. We focus on the class of two-mode beamsplitters of the form
\be\label{eq:generalBS_matrix}   
    BS(\theta, \phi) = 
    \begin{pmatrix}
        \cos{\theta} & i e^{i \phi} \sin{\theta} \\
        i e^{-i \phi} \sin{\theta} & \cos{\theta}
    \end{pmatrix},
\ee
though one can easily translate into other beamsplitter and phase-shifter implementations (see \appref{app:beamsplitter_identities}). Graphically, we represent such a beamsplitter by a vertical line adorned with a circle at each end (acting on modes $j$ and $k$):
\be
    BS_{jk}(\theta, \phi) = \raisebox{-0.4\totalheight}{\includegraphics[width=0.45\linewidth]{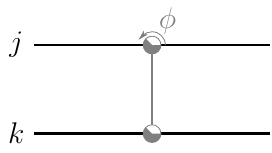} } \nonumber
\ee
with the angles $\theta$ and $\phi$ indicated via the beamsplitter color and the fill pattern of the circles marking the modes the beamsplitter is acting on, respectively [see Fig.~\ref{fig:BSG_circuits}(f)]. We present each circuit using only beamsplitters of the form of \eqref{eq:generalBS_matrix}, where we allow for beamsplitters between non-adjacent modes to avoid the need for explicit mode swaps. In the frequency domain such non-adjacent mode beamsplitters could be implemented natively, however, in other encodings, such as a spatial-mode encoding, this would require optical switching or routing \footnote{In the 4P5M circuit, non-adjacent mode beamsplitters and thus optical switching can be avoided if one is ok with the output Bell state being defined on non-adjacent modes \cite{fldzhyan2021compact}.}. 

\subsection{Ideal HBSG schemes}
\begin{figure}[htp]
    \hfill \subfloat[(a) 4P5M]{\includegraphics[height=2.25cm]{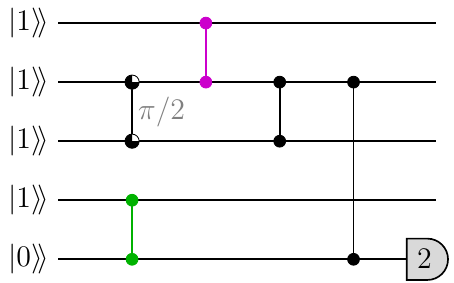}} \hfill
    \subfloat[(b) 5P5M]{\includegraphics[height=2.35cm]{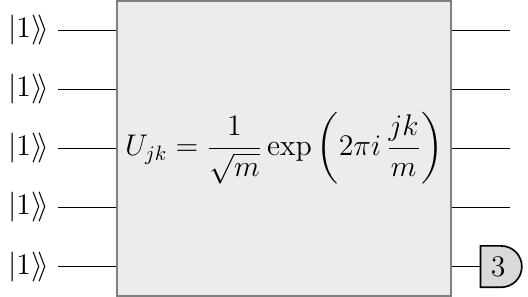}} \hfill \\ 
    \hfill \subfloat[(c) 4P6M]{\includegraphics[height=2.7cm]{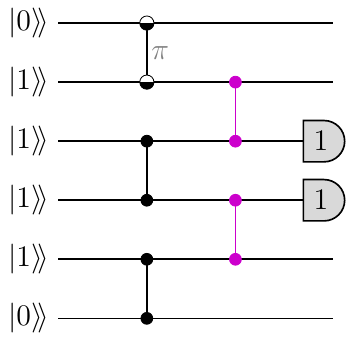}} \hfill 
    \subfloat[(d) 6P6M]{\includegraphics[height=2.7cm]{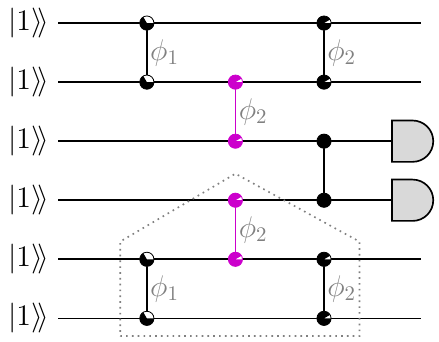}} \hfill \\ 
    \hfill \subfloat[(e) 4P8M]{\includegraphics[height=3.6cm]{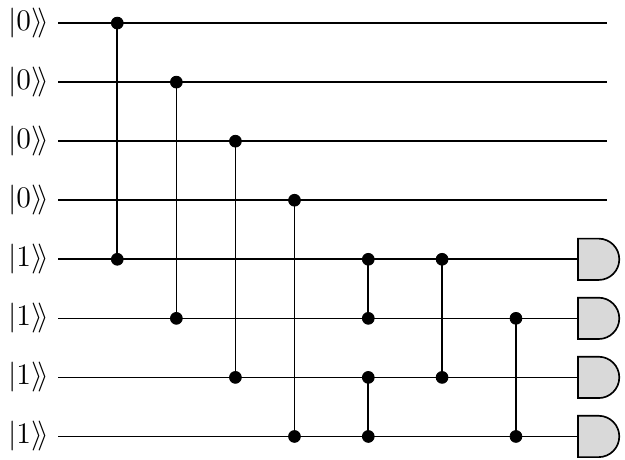}} \hfill
    \subfloat[(f) Legend]{\includegraphics[height=3.2cm]{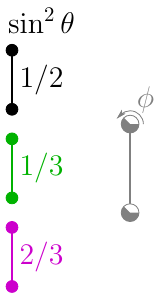}}
    \caption{
    Class of heralded Bell-state generation schemes considered in this work. 
    (a)-(f) See main text and Table \ref{tab:BSG_circuit_overview}.
    (a) HBSG scheme using minimal number of photons and modes (one for the heralding measurement and four for the output Bell state) \cite{stanisic2017generating}.
    (b) This circuit is an $m \times m$ discrete Fourier transform (DFT) with $m=5$. We simply specify the matrix elements on the graphic as there are a variety of ways to implement such an interferometer (see \appref{app:circuitVariants}).
    (d) This circuit uses two $m=3$ DFT circuits (the bottom of which is highlighted in the dashed region) followed by a single 50:50 beamsplitter and detection on the corresponding modes. Each DFT is decomposed into three beamsplitters with phases $\phi_1 = 2\pi/3$ and $\phi_2 = \pi/6$ as shown. 
    }\label{fig:BSG_circuits}
\end{figure}

A survey of passive, dual-rail HBSG schemes found in the literature (and compiled in \refref{forbes2025heralded}) are shown in Fig.~\ref{fig:BSG_circuits}. They can be characterized via the number of single-photon inputs they use, $N$, and the number of modes they operate on, $M$, and hence we will label them using the notation $N$P$M$M. The corresponding references and parameter specifications are provided in Table~\ref{tab:BSG_circuit_overview}. The target input states are indicated to the left of each circuit. For the first three schemes only a single outcome heralds success; as such, said outcome is indicated on the detectors,  shown as \detector, in Fig.~\ref{fig:BSG_circuits}(a)-(c). Meanwhile, for the other two schemes several detection patterns herald success. 

\begin{table*}[htp]
    \begin{tabular}{cccccccc}
        Ref(s) & $N$ & $M$  & $p_\textrm{ideal}$ & $n_{\rm BS}$ & $n_\textrm{states}$ & Heralding click pattern(s): $C_h$ \\ \hline
        \citenum{fldzhyan2021compact,hartnett2026automated} & 4  & 5 & $1/9 \approx 11.11\%$ & 5 & 1 & (2) \\
        \citenum{paesani2021scheme} & 5 & 5 & $12/125 = 9.60\%$  & 10 & 1 & (3) \\ 
        \citenum{carolan2015universal,fldzhyan2021compact} & 4 & 6 & $2/27 \approx 7.41\%$ & 5 & 1 & (11) \\ 
         \citenum{bhatti2025heralding} & 6 & 6 & $4/27 \approx 14.81\%$ & 7 & 2 & (40), (31), (13), (04) \\ 
        \citenum{zhang2008demonstration,bartolucci2021creation} & 4 & 8  & $3/16 = 18.75\%$  & 8 & 3 & 
        (1100), (0011),
        (1001), (0110),
        (1010), (0101) \\ 
    \end{tabular}		
    \caption{
        Parameter specifications for the HBSG circuits of Fig.~\ref{fig:BSG_circuits}. 
        $N$ is the number of input photons across $M$ modes,
        $p_\textrm{ideal}$ is the ideal probability of heralding a Bell-like state, 
        $n_{\rm BS}$ is the number of two-mode beamsplitters needed to implement the circuit, and
        $n_\textrm{states}$ is the number of orthogonal Bell-like states that can be heralded by the circuit.
        For each circuit, successful generation of a Bell-like state is heralded by a detector click pattern in the corresponding set $C_h$ whose elements are shown.         
    }
    \label{tab:BSG_circuit_overview}
\end{table*}

Additional circuit details are given in \appref{app:HBSG_extra} including
the specific state(s) heralded by each of these schemes, 
related implementation variants,
and 
practical tradeoffs and limitations. 
Here we note that each of these schemes requires detectors with some photon-number-resolving (PNR) capabilities. Namely, even the 4P6M and 4P8M schemes whose heralding click patterns consist solely of single-photon detections should be able to discern between 0, 1, and $\geq 2$ photons (in contrast with what is indicated in Table 1 of \refref{forbes2025heralded}), so that $\geq 2$ photon outcomes do not masquerade as single-photon detection events. Thus, we will assume PNR detectors are used, though we account for them being inefficient. A brief discussion of photodetection theory using positive operator-valued measures (POVMs) is given in App.~\ref{app:detectorPOVMs}.

\subsection{Photon loss}\label{sec:photonLoss}
We faithfully model photon loss using bosonic amplitude damping (AD) channels, wherein single-mode loss is characterized by a dimensionless loss rate $\g$ or equally by an efficiency $\eta \equiv 1 - \g$.
Such single-mode AD channels, denoted $\cE_\eta(\rho)$, are equivalent to the \textit{beamsplitter loss model}, where the input state, $\rho$, is mixed with the vacuum on a beamsplitter with transmissivity $\cos^2{\theta} = \eta$, in the notation of \eqref{eq:generalBS_matrix}, and the vacuum-initialized loss mode is traced over. 
Schematically,
\begin{subequations}
		\begin{align}\label{eq:singleModeAD}
			\cE_\eta(\rho) \; &= \raisebox{-0.45\totalheight}{ \includegraphics[width=0.4\linewidth]{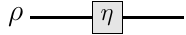} } \\
            &= \raisebox{-0.4\totalheight}{ \includegraphics[width=0.45\linewidth]{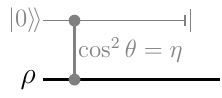} }, \label{eq:singleModeBSLossModel}
		\end{align}
\end{subequations} 
where the parallel vertical lines on the (gray) auxiliary loss mode indicate that it is being traced over.

Photon loss can be expensive to model as it is a nonunitary transformation. Accordingly, it is prudent to minimize the number of additional loss modes, which can be accomplished by leveraging two properties of photon loss channels: 
\begin{enumerate} 
    \item Single-mode loss is multiplicative. \label{item:AD_multiplicative}
    \be
			 \raisebox{-0.45\totalheight}{ \includegraphics[width=0.35\linewidth]{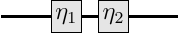} } 
            = \raisebox{-0.45\totalheight}{ \includegraphics[width=0.35\linewidth]{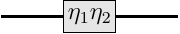} } \nonumber
    \ee
    \item Uniform loss commutes with linear optics. \label{item:uniformAD_commutes}
    \be
        \raisebox{-0.45\totalheight}{ \includegraphics[width=0.44\linewidth]{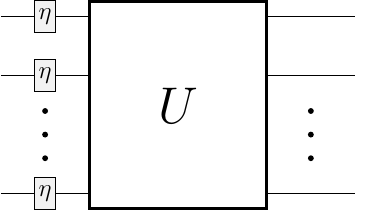} } 
        = \raisebox{-0.45\totalheight}{ \includegraphics[width=0.44\linewidth]{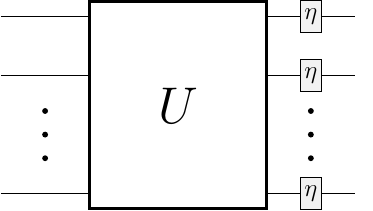} } \nonumber
    \ee
\end{enumerate}
Together these properties can be used to simplify loss modeling by effectively shifting loss to photon sources and detectors. In \appref{app:lossyCircuits} we provide a Kraus representation of such AD channels and show that they satisfy the above properties.

\subsection{Modeling framework}\label{sec:modelingFramework}
For each of the schemes shown in Fig.~\ref{fig:BSG_circuits}, we model three stages (a)-(c), that we will now discuss in turn.

\phantomsection
\subsubsection*{(a) Input-state generation}\label{stage:inputStateGen}
First, one must generate the $N \geq 4$ photons that are to be incident on the $M$-mode circuit. As detailed in \secref{sec:sources}, we consider two different single-photon source models: 
ideal generation followed by uniform photon loss, and 
heralded generation via biphoton states. The former is idealized yet allows us to gain analytical traction on the behavior of the various HBSG schemes. Meanwhile, the latter is a more refined model tailored to on-chip implementations including those within the frequency domain. 
The full state of the source, denoted $\rho_s$, is ideally the specific $N$ photon input state, $\ketm{\vec{N}}$, dictated by the circuit (as indicated on the left of each circuit in Fig.~\ref{fig:BSG_circuits}, e.g., $\ketm{\vec{N}} = \ketm{11110}$ for the 4P5M circuit). 

\phantomsection
\subsubsection*{(b) HBSG circuit}\label{stage:HBSG_circuit}
After the input state is generated, the state of the signal modes, $\rho_s$, along with $M_h = M - 4$ vacuum-initialized herald modes, $\ketm{\vec{0}_h}$, are passed into the lossy HBSG circuit, $U$. Ideally $U$ corresponds to an $M$-mode HBSG linear-optical unitary. However, when present, we account for nonuniform photon loss in the circuit by including a number, $M_l$, of additional vacuum-initialized loss modes, $\ketm{\vec{0}_l}$.
Namely, we suitably dilate the ideal unitary to purify the AD channels under the beamsplitter loss model. Accordingly, the input to the lossy circuit is $\rho_\tin = \rho_s \otimes \ketm{\vec{0}_e}\bram{\vec{0}_e}$ with $\ketm{\vec{0}_e} = \ketm{\vec{0}_h} \otimes \ketm{\vec{0}_l}$, which is then passed through the dilated unitary, $U$, yielding the output $\rho_\tout = U \rho_\tin \hc{U}$. Such evolution of a density matrix through a photonic circuit can be directly simulated in Perceval. We simply have to find the circuit corresponding to the lossy unitary.
Moreover, to reduce the dimension of $U$ and ease the numerics, we use properties of AD channels (see Sec.~\ref{sec:photonLoss}) to minimize $M_l$ by shifting the losses to the beginning and end of the circuit to the extent possible. Then, we must additionally compute the corresponding action of these AD channels on $\rho_\tin$ and $\rho_\tout$, respectively.

\phantomsection
\subsubsection*{(c) Heralding}\label{stage:heralding}
The herald-mode photons are then detected leading to a click pattern $\hvec = (h_1, \cdots, h_{M_h})$ with probability $p_{\hvec} = {\Tr}(\rho_\tout \hat{\Pi}_{\hvec})$, where $\hat{\Pi}_{\hvec}$ is the POVM element corresponding to click pattern $\hvec$. The heralded state on the remaining modes is
$ 
    \rho_{\hvec} = \Tr_e(\rho_\tout \hat{\Pi}_{\hvec} )/p_{\hvec}, 
$
where the partial trace is over the extra modes, $e$, i.e., both the measured herald modes and the inaccessible loss modes. 

If $\hvec$ is in the set of desired click patterns $C_h$ (see Table \ref{tab:BSG_circuit_overview}), then this post-herald output state should ideally be a Bell-like state, $\ket{\cB_{\hvec}} \in \{\ket{\Phi^\pm}, \ket{\Psi^\pm}, \ket{\chi^\pm} \}$, 
that can be used for subsequent operations. 
The fidelity 
\be
    \cF_{\hvec} = \bra{\cB_{\hvec}} \rho_{\hvec} \ket{\cB_{\hvec}}
\ee
serves as a measure of the quality of this output state. When $|C_h| > 1$ (for the 4P8M and 6P6M circuits), we are ultimately concerned with the overall heralding probability for a circuit
\be
    p_{\textrm{her}} = \sum_{\hvec \in C_h} p_{\hvec}
\ee
(the ideal values of which are reported in Table \ref{tab:BSG_circuit_overview}) as well as the weighted-average fidelity 
\be\label{eq:averagedBellFidelity}
    \overbar{\cF} = \sum_{\hvec \in C_h} \frac{ p_{\hvec} }{ p_{\textrm{her}} } \cF_{\hvec}.
\ee
The overall probability of success of the HBSG scheme is the product of this heralding probability and the probability that the $N$-photon input state generation was successful, $p_{\rm succ}=p_{\rm her} \times p_N$.

\section{Photonic sources}\label{sec:sources}
For the input state generation of stage \hyperref[stage:inputStateGen]{(a)}, we consider two models of single-photon production.
Before introducing them, however, it will be helpful to note two common approaches for producing single photons.
The generation of a single photon can be heralded from biphoton sources implemented using spontaneous parametric down conversion (SPDC) or spontaneous four-wave mixing (SFWM). 
Alternatively, matter-based emitters can be used, wherein the coherent production of a photon is induced via a classical drive such as a laser \footnote{
    Solid-state emitter sources are promising, especially in their potential to be near-deterministic, have suppressed multiphoton errors, and directly generate resource states by mediating effective photon-photon interactions via matter-based entanglement \protect\cite{maring2024versatile,esmann2024solid,huet2025deterministic}. 
    However, they tend to inherit stochastic errors from the emitter \protect\cite{rudolph2017optimistic} and suffer increased partial distinguishability errors, their on-chip integration is challenging \protect\cite{signorini2020chip}, and, within the frequency domain, the engineering of many emitter-based sources across disjoint frequencies would be a formidable task. 
    Thus, while promising in many facets, the proper treatment of emitter-mediated approaches would require disjoint modeling from this work and is best left to dedicated works.
}. 

We first introduce an error model that by virtue of its simplicity is applicable to both emitter-based and heralded biphoton single-photon sources. 
Then we proceed to focus on and develop a more refined error model for heralded biphoton sources, especially SFWM-based microring resonator sources, due to their on-chip compatibility and flexibility in terms of generating highly indistinguishable and pure photons across many frequencies
\cite{vernon2015spontaneous,vernon2017truly,chuprina2019generating,psiquantum2025manufacturable,lukens2026beyond,qin2026high}. 

\subsection{Amplitude-damped single photons}
The simplest model for a lossy single-photon source is one that yields the desired single photon with some probability, $\eta$, and otherwise, with probability $\g = 1 - \eta$, the photon scatters into another physically inaccessible mode and is lost, yielding the vacuum:
\be\label{eq:simple_rho1}
    \rho_1(\eta) = \eta \kbm{1}{1} + \g \kbm{0}{0}.
\ee
For an emitter-based source this loss could be a result of scattering during the photon production or absorption by the bulk when trying to couple onto a specific waveguide. This simplified lossy source model is similarly obtained for heralded biphoton single-photon sources if the heralding of the idler mode is done perfectly yet the output signal mode experiences loss. Notably this source model is equivalent to an ideal single photon state undergoing an AD channel with efficiency $\eta$:
\be
    \rho_1(\eta) = \cE_\eta(\kbm{1}{1}). 
\ee

\subsection{Lossy biphoton sources}\label{sec:lossyBiphotonSources}
Both SFWM and SPDC can be employed to create biphoton sources that, in the limit of unit purity, i.e., for a separable joint spectral amplitude, generate the two-mode-squeezed vacuum state,
\be\label{eq:TMSW_pure}
    \ket{\textrm{TMSV}(\lambda)} = \sqrt{1-|\lambda|^2} \sum_{n=0}^\infty  \lambda^n \ketm{n, n}_{s,i}, 
\ee
on a pair of modes deemed the signal, $s$ and idler, $i$. Here $\lambda \in \mathbb{C}$ is a squeezing parameter that can be tuned by varying the power of the pump field driving the nonlinear process, $0 \leq |\lambda| < 1$. Ideally, the detection of $n$ photons in the output of one arm, say the idler, heralds the presence of $n$ photons on the other arm, the signal, with probability $P_n = \p{1 - |\lambda|^2} |\lambda|^{2n}$. Accordingly, such biphoton sources can be used as heralded single-photon sources (HSPSs), where, as one can easily show, the maximum single-photon generation probability is $P_1^{(\textrm{max})} = 1/4$ for $|\lambda|^2 = 1/2$. 

Photon loss on the signal and idler arms, along with detector inefficiency, will degrade the quality of the output state relative to the target single-photon state. We model this as signal and idler mode photon loss with efficiencies $\eta_s$ and $\eta_i$, respectively, followed by idler mode measurement using a PNR detector with efficiency $\eta_d$. Here we assume negligible dark counts, which is reasonable for high-quality modern detectors (e.g., superconducting nanowire single-photon detectors \cite{natarajan2012superconducting}) that can have mHz-kHz dark count rates as compared to order nanosecond detection windows. Some discussion on the inclusion of dark counts and their impacts is given in \appref{app:detectorPOVMs}. 

Then, with the shorthand $\g_s = 1 - \eta_s$,
$\eta_{id} = \eta_i \eta_d = 1 - \g_{id}$,
and
$\xi_i = |\lambda|^2 \g_{id}$,
one finds that a 1-photon detection occurs with probability 
\be\label{eq:p1HSPS}
    p_1 = \frac{\eta_{id} (1 - |\lambda|^2) |\lambda|^2}{\p{1-\xi_i}^{2}},  
\ee
heralding the state of the signal mode to be
\be\label{eq:heraldedSPS}
    \varrho_1 = \sum_{n=0}^\infty 
    c_n(\eta_s, \xi_i)
    \ketm{n}\!\bram{n}
\ee
with 
\be\label{eq:all_HSPS_coefficients}
    c_n =  
    \frac{ \p{1 - \xi_i}^2 \p{n + \g_s \xi_i } \p{ \eta_s \xi_i}^n }{ \xi_i \p{1- \g_s \xi_i}^{n+2} }
\ee
as derived in \appref{app:sourceTheory} (along with additional discussion of biphoton sources and photodetection).

We expand these coefficients for small $\g_s$ and $\g_{id}$, assuming they are comparably small, i.e., both proportional to a parameter $\g$ that we are expanding in terms of:
\begin{subequations}
    \label{eq:HSPS_coefficients}
    \begin{align}
        c_0 &= \g_s \p{1 - 2 \xi_i } + \cO(\g^3), \\
        c_1 &= 1 - \sum_{n \neq 1} c_n, \\
        c_2 &= 2 \xi_i \br{ 1 - 2 \p{\g_s + \xi_i }  }  + \cO(\g^3), \label{eq:c2HSPS} \\
        c_{n \geq 3} &= n \xi_i^{n-1} + \cO(\g^n).
    \end{align}
\end{subequations}
Thus, to leading order signal loss leads to vacuum contributions while idler and heralding loss lead to multiphoton events.
Accordingly, multiphoton contamination, as dictated by $\xi_i = |\lambda|^2 \g_{id}$, can be suppressed 
by some combination of reducing heralding loss, $\g_{id}$, as well as the pump power and thus squeezing parameter, $|\lambda|$ (at the expense of reducing the generation rate). In the limit of negligible idler and heralding loss we reobtain the simpler source model of \eqref{eq:simple_rho1} on a successful herald, namely, $\lim_{\g_{id} \ra 0} \varrho_1 = \rho_1(\eta_s)$ with $p_1 \ra P_1(|\lambda|) \leq 1/4$.

\subsection{Lossy \textit{N}-photon state preparation}
Returning to the consideration of an $N$P$M$M scheme, we are concerned with the generation of the desired $N$-photon state, $\ketm{\vec{N}}$, which can be accomplished simply with $N$ single-photon sources using either of the above methods. We focus on two scenarios, where all the sources are either amplitude damped single photons 
\be\label{eq:NPhotonSourceA}
    \rho^{(A)}_s = \Motimes_{j=1}^N \rho_1(\eta_j)
\ee
or they are HSPSs
\be\label{eq:NPhotonSourceB}
    \rho^{(B)}_s = \Motimes_{j=1}^N \varrho_1\p{ \g_s^{(j)}, \xi_i^{(j)} }.
\ee
In the latter case, the probability of independently generating $N$ photons in a single shot is 
\be
    p_N^{(B)} =  \prod_{j=1}^N p_1^{(j)},
\ee
with $p_1^{(j)}$ given in \eqref{eq:p1HSPS}, which quickly diminishes as $N$ is increased. 
Ultimately, to go beyond early demonstrations and enable scalability, source multiplexing to create near-deterministic SPSs ($p_1 \sim 1$) will be necessary \cite{meyer2020single,fischer2026frequency}.

\section{Performance of Bell-state generators under generic loss model}\label{sec:lumpedHBSG}
To gain analytical traction on the structure of these HBSG circuits and the impact loss has on them, we first consider an idealized loss model dictated by three efficiency parameters. Namely, we model source loss using the amplitude-damped model of \eqref{eq:simple_rho1} with efficiency $\eta_s$, optical circuit loss via an efficiency $\eta_c$ quantifying propagation and component loss throughout the circuit, and heralding detector efficiency $\eta_d$. These losses are taken to be ``lumped,'' i.e., the same for each photon, mode, and heralding detector, respectively. 

\begin{figure}[h!]
    \includegraphics[width=\linewidth]{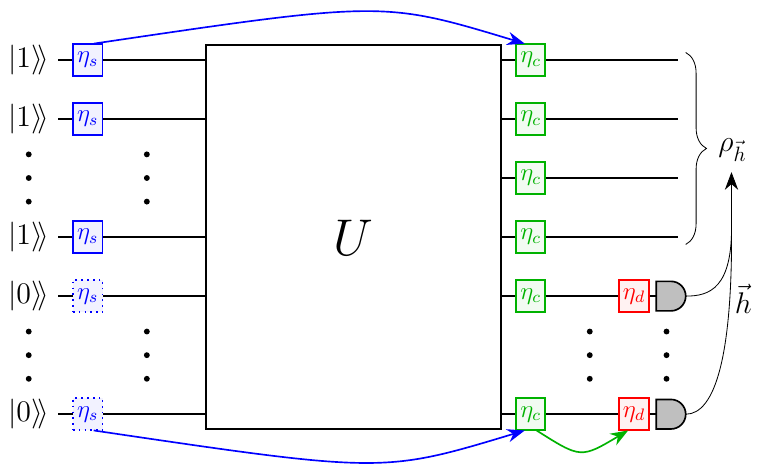}
    \caption{Lumped photon loss model, wherein loss occurs at the photon sources, within the circuit, and at the detectors with efficiencies $\eta_s$, $\eta_c$, and $\eta_d$, respectively.
    In this model one can shift the loss to the end of the circuit 
    using properties of photon loss channels.} 
    \label{fig:lumped-loss-circuit-schematic}
\end{figure}

This situation is illustrated in \figref{fig:lumped-loss-circuit-schematic}. Crucially, because of the properties of AD channels discussed in \secref{sec:photonLoss} one can further simplify the model. As the vacuum is the fixed point of a nontrivial ($\g > 0$) AD channel, one can apply the source AD to the $M-N$ vacuum-initialized modes of a HBSG circuit (denoted via the dashed blue rectangles in \figref{fig:lumped-loss-circuit-schematic}). Then because uniform loss commutes with linear optical unitaries we can effectively move the source loss through the circuit $U$. Finally, by using the composition rule for single-mode AD channels, that their efficiencies multiply, we can combine the cascaded efficiencies into a single effective efficiency experienced by each mode. The four output modes that are to encode the Bell state on a successful herald each undergo an AD channel with efficiency $\eta_b = \eta_s \eta_c$, and the heralding mode(s) undergo additional damping due to detector loss resulting in an overall efficiency $\eta_h = \eta_s \eta_c \eta_d = \eta_b \eta_d$.

Given a HBSG scheme, we want expressions for the heralding probability and Bell state fidelity as functions of the efficiencies $\eta_b$ and $\eta_h$. 
Under this loss model, one need simply compute the ideal full output state of each scheme before measurement, which can be expressed as a superposition over all click patterns $\hvec$ on the heralding modes:
\be
    \ket{\psi_{\rm HBSG}} = \sum_{\hvec} \sqrt{p_{\hvec}} \ket{\psi_{\hvec}}_b \ketm{\hvec}_h
\ee
as can straightforwardly be computed using Perceval, a computer algebra system, or (albeit tediously) by hand. Then, this state is passed through the appropriate single-mode AD channels, yielding $\rho_\textrm{HBSG} = \cE_{\eta_b, \eta_h}\p{ \kb{\psi_{\rm HBSG}}{\psi_{\rm HBSG}} }$, and an ideal heralding measurement is performed. 

For each circuit, given a target heralding outcome $\hvec^* \in C_h$ (which occurs with probability $p_{\hvec^*}$ yielding the output state $\rho_{\hvec^*}$), one can compute 
\be
    \tilde\rho_{\hvec^*} \equiv p_{\hvec^*} \rho_{\hvec^*} = \Tr_h\p{ \rho_\textrm{HBSG} \kbm{\hvec^*}{\hvec^*} }
\ee
and thus
\subalign{
    p_{\hvec^*} &= \Tr\p{ \tilde\rho_{\hvec^*} }, \\
    \cF_{\hvec^*} &= \bra{\cB_{\hvec^*}} \rho_{\hvec^*} \ket{\cB_{\hvec^*}}.
}

\subsection{Illustration for 4P5M scheme}
The full 4P5M output state is of the form
\be\label{eq:psi4P5M}
    \ket{\psi_{4P5M}} = \sum_{j=0}^4 \sqrt{p_j} \ket{\psi_j}_b \ketm{j}_h
\ee
with $\{p_j\} = \left\{ 31, 14, 6, 2, 1 \right\}/54$ the probabilities of measuring $j=0, 1, \dots, 4$ photons in the herald mode. In this loss model, a desired heralding event, $\hvec^* = (2)$ here, can either occur genuinely where no photons are lost and one gets the desired Bell state output or as a false positive where there were additional photons on the heralding modes that were lost, leading to the detection event masquerading as a successful herald.
Here one finds
$\ket{\psi_2} = -i \ket{\Phi^-}$ (the target Bell state), 
$\ket{\psi_3} = -\p{ \ketm{0100} + \ketm{0001} }/\sqrt{2}$, and 
$\ket{\psi_4} = -i \ketm{0000}$ 
such that 
\be 
    \tilde\rho_{(2)} = \sum_{j=2}^4 p_j \binom{j}{2} \eta_h^2 \g_h^{j-2}  \cE_{\eta_b}\!\p{\kbm{\psi_j}{\psi_j}}
\ee
and thus
\subalign{
    p_{(2)} 
    &= p_2 \eta_h^2 \p{1 + \g_h + \g_h^2 }, \\ 
    \cF_{(2)} &= \frac{p_2 \eta_h^2 \eta_b^2}{p_{(2)}}
    = \frac{\eta_b^2}{ 1 + \g_h + \g_h^2 }.
}

\subsection{General results}
Repeating this process for each HBSG circuit one finds that the heralding probabilities and Bell state fidelities are all of the form 
\begin{subequations}
    \label{eq:lumpedLossGeneral}
    \begin{align}
        p_{\hvec^*}(\g_h) &= p_{\hvec^*,\textrm{ideal}} \eta_h^{N_h} f(\g_h), \\
        \cF_{\hvec^*}(\g_b, \g_h) &= \frac{\eta_b^2}{f(\g_h)}, \label{eq:lumpedLossFidelity} 
    \end{align}
\end{subequations}
where $N_h = N - 2$ is the number of photons detected on the herald modes needed to flag success and 
\be\label{eq:circuitPolynomial}
    f(\g_h) = 1 + \mu_1 \g_h + \mu_2 \g_h^2
\ee
with $\mu_{1,2}$ circuit-specific coefficients \footnote{In the heralded generation of larger $n$-qubit states the corresponding polynomial will be order $n$.}. Given \eqref{eq:lumpedLossGeneral} and the $p_\textrm{ideal}$ values of Table \ref{tab:BSG_circuit_overview}, it suffices to tabulate these coefficients for each scheme as done in Table \ref{tab:circuit_mus}.
The product of these quantities,
\be\label{eq:lumpedLossYield}
     \cF_{\hvec^*} p_{\hvec^*} \equiv Y_{\hvec^*}(\g_b, \g_h) = p_{\hvec^*,\textrm{ideal}} \eta_h^{N_h} \eta_b^2,
\ee
takes on a simple, $f$-independent form. We deem this quantity the \textit{yield} as it jointly quantifies the rate (via $p_{\hvec^*}$) and quality ($\cF_{\hvec^*}$) of a HBSG scheme. We note that the relative yield $Y_{\hvec^*}/p_{\hvec^*,\textrm{ideal}}$ is of the same form for schemes with equal $N$. 

\begin{table}[ht!]
\begin{tabular}{ccc}
    Scheme & $\mu_1$ & $\mu_2$ \\ \hline
    4P5M & 1 & 1 \\
    4P6M & 2 & 1.5 \\ 
    4P8M & 2 & 1 \\
    5P5M & 0 & 4 \\
    6P6M$+$  & 0 & 6.5 \\ 
    6P6M$-$ & 0 & 2 \\ 
    6P6M average & 0 & 4.25 
    \end{tabular}
    \caption{The circuit-specific coefficients in the polynomials of \eqref{eq:circuitPolynomial} that characterize how the heralding probability and fidelity behave in the lumped-loss model.} 
    \label{tab:circuit_mus}
\end{table}

For each circuit, the $j=0$ term in $f(\g_h)$ corresponds to true positives while the $j=1,2$ terms correspond to false positives with respective probabilities of 
$p_{t+} = p_{\hvec^*,\textrm{ideal}} \eta_h^{N_h}$ and
$p_{f+} = p_{\hvec^*} - p_{t+}$.
Thus, the fidelity of the output state, \eqref{eq:lumpedLossFidelity}, can be understood simply as a degradation due to 
false positives
and 
loss on the Bell modes:
\be
   \cF_{\hvec^*} = \frac{p_{t+}}{p_{t+} + p_{f+}} \times \eta_b^2.
\ee
In particular, when uniformly amplitude damped with efficiency $\eta$, the fidelity of a dual-rail $n$-qubit state is degraded to $\cF_n = \eta^n$, which gives the $\eta_b^2$ factor in the Bell-state fidelity above.

For the 6P6M circuit we report two cases depending on whether the heralding click pattern was a 4-photon detection on one detector and zero on the other, (04) or (40), or a 3-photon detection on one detector and 1 on the other, (13) or (31), which herald the output states $\ket{\chi^+}$ and $\ket{\chi^-}$, respectively (hence the $\pm$ labeling). 
This is necessary for the 6P6M scheme as the resulting coefficients, $\mu_j$, are click-pattern dependent. Namely, the only herald-mode outputs that can lead to false positives are (06), (60), (24), and (42), such that a (04) false positive can occur in two ways, via (06) or (24), whereas a (13) false positive can only occur via (24) [likewise for (40) and (31) false positives]. Accordingly, $\mu_2$ is lower in the ``$-$'' case for the (13) and (31) click patterns. Further, the total probability of heralding and the weighted-average fidelity of these two cases follows the form of \eqref{eq:lumpedLossGeneral} and is also reported.
Henceforth we will regard these two cases as effectively different schemes.

Ostensibly such click-pattern dependent behavior could have also occurred for the 4P8M circuit, for which $|C_h| = 6$, yet the circuit has a symmetric structure such that each outcome is identically impacted by loss and thus it suffices to tabulate a single value of each coefficient.
Note, however, that this circuit comes with the caveat that different detection patterns herald $n_\textrm{states} = 3$ different Bell-like states. Namely, as implemented here the states $\ket{\Phi^+}, \ket{\Psi^-}$, and $\ket{\chi^+}$ are each heralded by two click patterns that occur with probabilities $1/32$. This is potentially troublesome in the case of heralding $\ket{\chi^+}$ for which the output qubit states are not even defined on the same modes as in the other two cases. Within a given implementation, depending on whether it suffices to keep track of which state has been generated (and adjust subsequent operations as needed) as well as the feasibility of doing feed-forward based mode switching, i.e., for permuting $\ket{\chi^+}$ to obtain $\ket{\Phi^+}$, it might be preferred to simply deem the generation of $\ket{\chi^+}$ as a failure such that the circuit would only succeed when generating $\ket{\Phi^+}$  or $\ket{\Psi^-}$ with $p_\textrm{ideal} \ra 1/8 = 12.5\%$. Accordingly, one should heed that $p_\textrm{ideal}$ may thus be reduced. 
In \secref{sec:frequencyHBSG}, where we analyze loss induced by frequency beamsplitters, we self-consistently capture the impact of loss in the frequency mode swap to map $\ket{\chi^+}$ to $\ket{\Phi^+}$.

\subsection{Impact of photon loss on fidelity}
\begin{figure}[ht!]
    \includegraphics[width=\linewidth, clip=true, trim = 15mm 0 0 0]{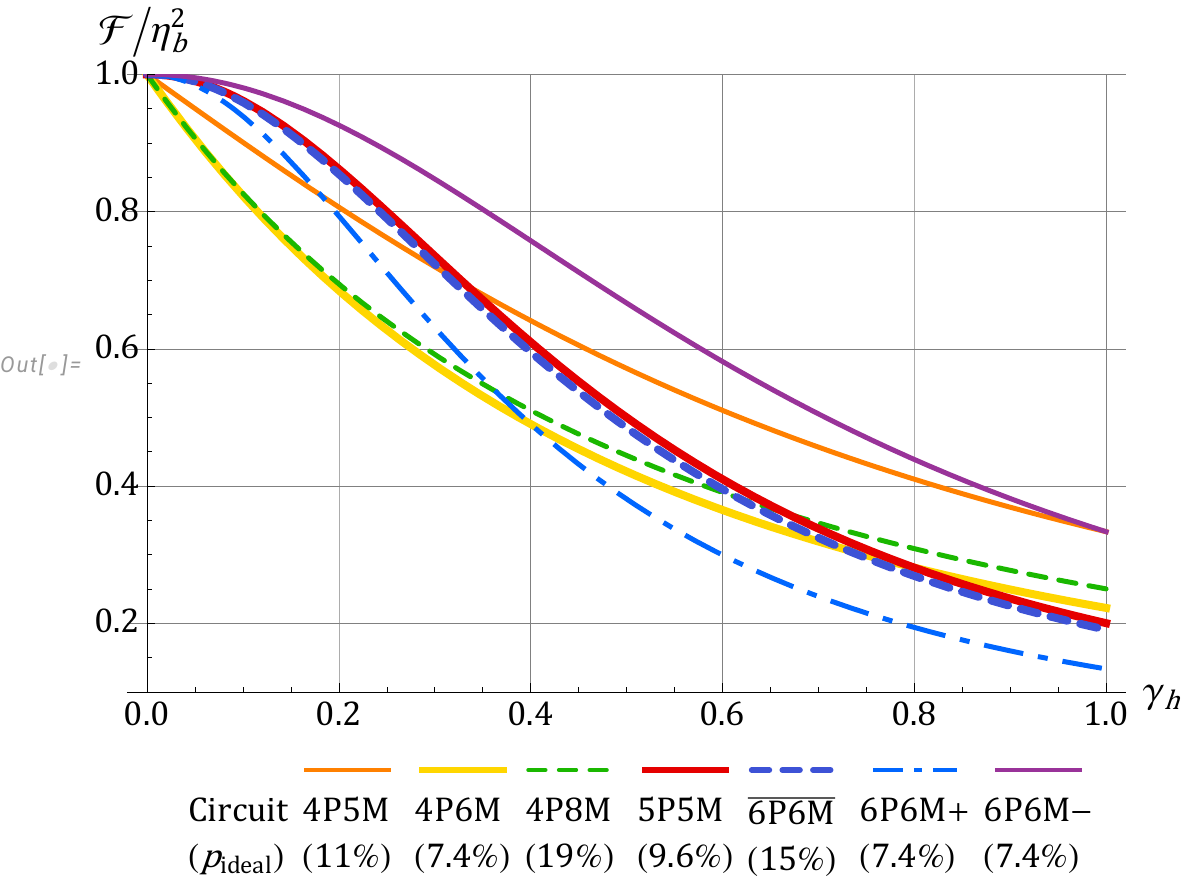} 
    \caption{Plot of $\cF/\eta_b^2 = f^{-1}$ versus $\g_h$ for each HBSG circuit considered here (see text for details).}
    \label{fig:fInverseVsGammah}
\end{figure}

In \figref{fig:fInverseVsGammah} we plot the reciprocals of the corresponding polynomials, $f^{-1}(\g_h) = \cF_{\hvec^*}/\eta_b^2  = p_{t+}/p_{\hvec^*}$, finding that the 6P6M circuit performs the best in terms of fidelity, specifically in the ``$-$'' case when heralding on a (13) or (31) click pattern. That is, 
\be
    \max(\cF_\textrm{lumped}) = \cF_\textrm{6P6M$-$}
\ee
for $0 < \g_h < 1$, where the maximum is taken over the HBSG schemes shown in Table \ref{tab:circuit_mus}. Meanwhile, compared to the average 6P6M output (denoted via an overbar) one finds that $\cF_\textrm{5P5M} \geq \cF_{\overbar{\textrm{6P6M}}}$ and
\be
    \max(\overbar{\cF}_\textrm{lumped}) = \begin{cases}
        \cF_\textrm{5P5M}, & \g_h < 1/3 \\
        \cF_\textrm{4P5M}, & \g_h > 1/3
    \end{cases}.
\ee

The robustness of the 5P5M and 6P6M circuits' Bell fidelities to small amounts of loss, $\g_b \leq \g_h \ll 1$, is owed to the fact that $\mu_1 = 0$ for these circuits, which leads to a quadratic as opposed to linear departure of $f^{-1}$ from unity: $f^{-1}(\g_h) = 1 - \mu_1 \g_h + (\mu_1^2 - \mu_2) \g_h^2 + \cO(\g_h^3)$. That $\mu_1 = 0$ for these circuits can be understood as a consequence of the \emph{zero-transmission law} of \refref{tichy2010zero}, which, at least in part, characterizes multiphoton interference effects that generalize the eponymous 2-photon Hong--Ou--Mandel interference effect to linear optical DFT circuits. For our purposes, we note that this law forbids 
(i) 4-photon contributions on any single mode in the output of the 5P5M DFT circuit and 
(ii) 2-photon contributions on any mode after the 3-mode DFTs in the 6P6M circuit, thus suppressing 5-photon outputs on the detected modes (as 5 is not in the sumset of $\{0,1,3\}$ with itself). Thus, both circuits are robust to single-photon loss events as false positives require two-photon loss events (in the lumped-loss model). 
Meanwhile, out of the $N=4$ photon schemes, the 4P5M one, which exhibits genuine $4$-photon interference \cite{shaw2023errors,fldzhyan2021compact}, is the most robust to photon loss.

The output Bell fidelities, as plotted in \figref{fig:fInverseVsGammah} relative to $\eta_b^2$, depart significantly from unity in the presence of photon loss. In part, this is because the model can be applied for any $0 \leq \g_b \leq \g_h \leq 1$, whereas we must clearly target the $\g_h \ll 1$ regime for high-quality generation.
To assess the severity of the fidelity degradation we can compare it to the fidelity of the input state, which is $\cF_{\vec{N}} = \cF_1^N = \eta_s^N$.
In the case of only source loss, $\g_s > 0$ and $\g_c = \g_d = 0$, one finds that $\cF_\textrm{Bell} > \cF_{\vec{N}}$ for each HBSG scheme. That is, if the only imperfection is a lossy input state of the form of \eqref{eq:NPhotonSourceA}, then the heralding acts to purify the Bell state relative to the input state.
Meanwhile, in the presence of circuit and detector loss ($\g_c, \g_d > 0$) this purification only holds for $\g_s > \g_s^*$ with $\g_s^*$ a circuit dependent threshold that increases with additional loss.

\section{HBSG with heralded biphoton sources}\label{sec:HBSGwHSPSs}
Now we will analyze and contrast the performance of the various HBSG schemes when using HSPSs for the input state generation. 
Such sources exhibit multiphoton errors resulting from false-positive heralds (see \secref{sec:lossyBiphotonSources}), which can then lead to further false positives in the post-circuit heralding.
To analyze the impact of such HSPS multiphoton errors in addition to photon loss, we start with a lumped circuit loss model, which is widely applicable as a reasonable starting point for HSPS-based implementations across a variety of encodings including spatial, polarization, and some frequency-bin implementations.
Later we extend this model, tailoring it to an on-chip frequency-domain implementation, wherein frequency beamsplitter loss leads to nonuniform AD channels that cannot be entirely commuted to the beginning and end of the circuit (see \secref{sec:frequencyBinHBSG}). 

\subsection{Model details}
This model corresponds to the lumped-loss model shown in \figref{fig:lumped-loss-circuit-schematic} with the only difference being that the source, previously of the form in \eqref{eq:NPhotonSourceA}, is now of the slightly more complicated form in \eqref{eq:NPhotonSourceB}. 
As before, we can shift $\eta_s$ to the end of the circuit and again summarize the circuit with two efficiencies $\eta_b = \eta_s \eta_c$ and $\eta_h = \eta_b \eta_{d}$. 
To minimize the number of parameters we need to vary, we impose the following relations:
\begin{enumerate}
    \item We take the sources to have identical efficiencies, $\eta_s$ and $\eta_{id}$.
    \item We choose each $\lambda_j$ to maximize the HSPS yield, i.e., the fidelity-probability product (see \appref{app:HSPSQuality}). Along with relation 1 this entails $\lambda_j = \lambda~\forall j$. 
    \label{item:relation2}
    \item Loss on the heralding idler modes and their detectors, $\g_{id} = 1- \eta_{id}$, is independent from $\eta_b$ and $\eta_h$. Nonetheless, one would typically expect $\eta_{id} \leq \eta_d$ assuming detectors of the same quality for source and post-circuit heralding. Thus, for ease of analysis, we replace $\eta_d$ by $\eta_{id}$ (perhaps somewhat degrading it). \label{item:relation3}
\end{enumerate}
One could easily relax these relations as desired. For instance, for 1, in a given experimental implementation, the photons may be generated by sources of varying quality and/or at different times leading to different propagation losses.
For 2, one could reduce the $\lambda$'s to increase the fidelity at the expense of heralding probability and thus generation rate \footnote{As the state of a given lossy HSPS only depends on $\g_s$ and $\xi_i = |\lambda|^2 \g_{id}$, see \eqref{eq:all_HSPS_coefficients}, a different choice of $\lambda$ will simply modify the effective $\g_{id}$ for a fixed $\xi_i$, so one can translate to other uniform $\lambda$ values.}. 

With HSPSs, if there is any heralding loss, $\eta_{id} < 1$, multiphoton contributions appear in the input state that vastly increase the size of the state space and complexity of calculating the relevant heralding probabilities $p_{\hvec}$, output states $\rho_{\hvec}$, and fidelities $\cF_{\hvec}$. (This complexity only increases when we include nonuniform beamsplitter loss in \secref{sec:frequencyBinHBSG}.)
Although analytical calculations are possible in specific instances, they quickly become unwieldy and would require case by case evaluation. 
Accordingly, we opt to perform strong linear-optical Fock-state simulation using Perceval \cite{heurtel2023perceval}. This allows us to develop a single simulation framework that can easily be applied to each of the different schemes and loss models.

To implement such simulations we must impose some cutoff in terms of which Fock state contributions we include from the lossy HSPSs of \eqref{eq:NPhotonSourceB}.
Namely, if there is loss in the idler-mode heralding, then each heralded single-photon state, $\varrho_1$, will have multiphoton contributions, i.e., $c_n>0~\forall n$ in \eqref{eq:heraldedSPS}. In the case of a single HSPS this can simply be handled by introducing a cutoff for the maximum number of included photons.
However, for multiple sources we should be careful and systematically impose a cutoff based on the most significant populations in the full state's density matrix. 

From the HSPS coefficients of \eqref{eq:HSPS_coefficients} we see that, to leading order, the contribution of the $n^{\rm th}$ term is $\cO(\g^{|n-1|})$, assuming $\g_s \sim \g_{id} \sim \g$. Similarly, for multiple sources the contribution of the Fock basis state $\ketm{\nvec}$ is $\cO(\g^{||\nvec - \vec{N}||_1})$, where $\vec{N}$ is the desired $N$-photon state and $||\cdot||_1$ is the taxicab ($L^1$) distance.
Accordingly, we will truncate the full state at some maximum power of $\g$ across all $N$ sources, imposing
\be\label{eq:nvecCutoffCriteria}
    ||\nvec - \vec{N}||_1 \leq n_\textrm{extra}
\ee
as the cutoff criterion \footnote{Alternatively, one could simply use a numerical cutoff based on the population of each Fock basis state.}. 
This limits the maximum number of photons while consistently accounting for the fact that vacuum contributions are suppressed to $\cO(\g)$.

In \appref{sec:convergenceWithNExtra} we include some convergence analysis, finding that to obtain fidelities to within the nearest percent it suffices to go to $n_\textrm{extra} = 3$ for $\g_{id} \leq 0.05$ or $n_\textrm{extra} = 4$ for $\g_{id} \leq 0.10$. As one would expect, additional idler mode heralding loss tends to require the inclusion of higher-order multiphoton contributions. These findings inform the truncation parameter fixed in subsequent results.

\subsection{Results}
\begin{figure}[ht!]
    \captionsetup[subfloat]{captionskip=-3pt}
    \subfloat[(a) $\cF_\textrm{max}$]{
        \stackinset{r}{12pt}{t}{-9pt}{\includegraphics[width=0.3\linewidth]{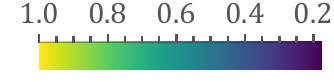}}
        {\includegraphics[width=0.96\linewidth, clip=true, trim = 3mm 1mm 7mm 2mm]{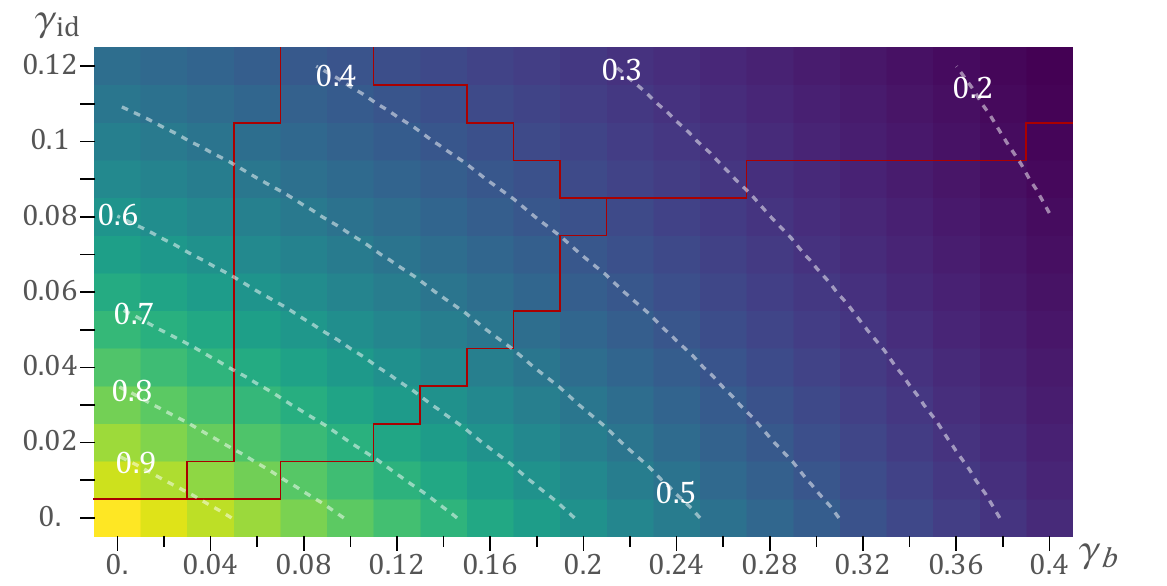}}
    } 
    \\
    \subfloat[
        (b)
        {\protect\textcolor[RGB]{255, 128, 0}{4P5M}},
        {\protect\textcolor[RGB]{255, 214, 0}{4P6M}},
        {\protect\textcolor[RGB]{26, 184, 0}{4P8M}},
        {\protect\textcolor[RGB]{230, 0, 0}{5P5M}},
        {\protect\textcolor[RGB]{0, 102, 255}{6P6M$+$}},
        {\protect\textcolor[RGB]{153, 51, 153}{6P6M$-$}}
    ]{\includegraphics[width=0.98\linewidth, clip=true, trim = 4mm 0 5mm 4mm]{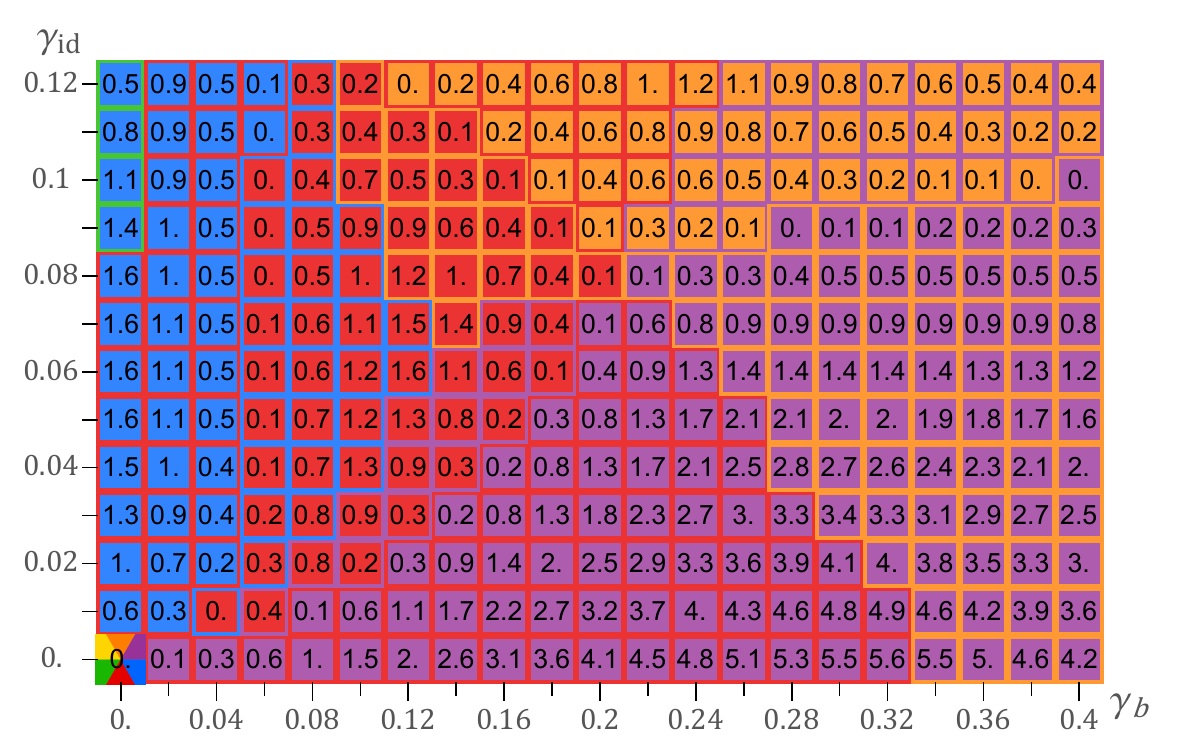}}
    \caption{
        Comparison of the HBSG circuits of \figref{fig:BSG_circuits} in the presence of HSPS multiphoton errors and lumped circuit photon loss.
        (a) Array plot of the maximum fidelity out of each HBSG scheme $s$, $\cF_\textrm{max} = \max_s(\cF_s)$, for each point $(\g_b, \g_{id})$ in the shown region; here $n_\textrm{extra} = 4$.
        We overlay contours from a linearly interpolated grid (dashed white) and lines separating regions with different ``winning'' (max) schemes (dark red).
        (b) Winner and margin plot highlighting---for each $\g_b$, $\g_{id}$---which circuit has the highest fidelity (indicated via the central pixel color in colors consistent with the subcaption and \figref{fig:fInverseVsGammah}) and by what margin, i.e., how much better it does than the second best circuit as a fidelity difference. The margin value is overlaid (rounded to the nearest $0.1\%$) and the second best circuit is indicated via the bounding color of each pixel
        (e.g., for $\g_b = 0.04$ and $\g_{id} = 0.03$ the 5P5M circuit in red wins by a margin of $0.8\%$ over the 6P6M $+$ circuit in blue).
        For $\g_{id} = \g_b = 0$ all the circuits perform ideally yielding a unit fidelity Bell state and thus ``tie'' as illustrated via the color wheel.
    }\label{fig:lossyHSPSsLumpedCircuitLossFidelityComparison}
\end{figure}

Imposing the aforementioned relations and model assumptions, one can reduce the comparison of HBSG schemes to varying two parameters, $\g_b$ and $\g_{id}$. 
This construction allows us to show the behavior, specifically the output Bell-state fidelity of \eqref{eq:averagedBellFidelity}, of the individual schemes, compare them, and highlight which performs best (in terms of fidelity and by how much) across various parameter regimes as shown in \figref{fig:lossyHSPSsLumpedCircuitLossFidelityComparison}.

Note that two of the six schemes, the 4P6M and 4P8M ones, never ``win,'' i.e., perform the best in terms of fidelity.
Out of the four winning schemes, the 6P6M$-$ (purple) wins the most in the shown parameter space and does so in regimes dominated by signal mode source and circuit photon loss, $\g_b$,
as is consistent with it always winning in the simplified source, lumped circuit model (see \figref{fig:fInverseVsGammah}).
For appreciable multiphoton errors, $\g_{id}$, and relatively small $\g_b \lesssim 2.5 \g_{id}$, other schemes outperform the 6P6M$-$ one. Namely, as $\g_b$ increases in this regime, the winner shifts from 6P6M$+$ (blue) to 5P5M (red) and then to 4P5M (orange). 
Thus, we see that higher-order interference tends to benefit a scheme's robustness to multiphoton errors as well as photon loss. 
However, precisely which scheme performs the best depends on the amount of each type of error.
Plots of the fidelity and average photon number of the output state for each scheme are given in \appref{app:relativeYieldWithHSPSs}. 
Therein, we also explore a trade-off between fidelity and probability of heralding similar to what we found for the simpler source model considered in \secref{sec:lumpedHBSG}.

Note that if one injects an extra photon into any of the modes in the 6P6M circuit, in the absence of photon loss, the probability that it leads to a $+$ case false positive [(40) or (04) detection] is $5.35\%$, whereas it is $7.82\%$ in the $-$ case [(31) or (13) detection], which is almost 1.5 times larger. 
This explains, at least heuristically, why the 6P6M$+$ scheme handles multiphoton errors better than its $-$ counterpart in the vicinity of $\g_b = 0$. 
Additional aspects of the schemes' robustness to multiphoton errors, including this heuristic, are discussed in \appref{app:higherNFlex}. Therein, we find that the 5P5M and 6P6M schemes possess valid Bell-heralding inputs with one additional photon on any input mode. Thus, if the HSPS detectors can resolve two-photon detections, certain known multiphoton emissions can be accepted as resource-generating events rather than discarded, increasing the ideal success probability by roughly a factor of two. Moreover, these higher-$N$ input configurations imply robustness against a specific second-order error process in which a single extra-photon emission is followed by the loss of one photon from the post-circuit heralding modes.

\section{Frequency-bin HBSG with architecture-dependent beamsplitter loss}
\label{sec:frequencyBinHBSG}
As demonstrated in Secs.~\ref{sec:lumpedHBSG} and \ref{sec:HBSGwHSPSs}, the lumped-loss model simplifies the analysis of HBSG performance, and in many cases even enables analytic expressions for key metrics. Often, especially in spatial-mode-encoded qubits, each mode experiences a comparable amount of loss, in which case the lumped-loss model is a very good approximation. However, there are cases where it yields a poor approximation, with a primary example being implementations where the linear-optical components themselves are lossy and applied to certain modes more than others, resulting in nonuniform loss across the photonic circuit. As a result, in this section, we extend our analysis to such architecture-dependent loss models. 

To make things concrete, we focus on HBSG with frequency-encoded qubits because, as explained below, nonuniform loss across beamsplitters is quite common in this context. 
As discussed in the introduction, frequency-bin quantum information processing has emerged as a promising means of reducing devices' hardware footprints by allowing native access to high-dimensional Hilbert spaces within a single spatial mode \cite{lukens2017frequency,lu2023frequency,clementi2023programmable,myilswamy2025chip,lukens2026paradigm,congia2026fully}.
This analysis serves to 
demonstrate the flexibility and extensibility of our model 
as well as to aid in identifying underlying component requirements for the generation of frequency-encoded photonic resource states.

\subsection{Frequency-domain linear optics}\label{sec:linearOpticsOverview}
We will first discuss frequency-domain linear optics generally, wherein the corresponding active, driven beamsplitters can be the dominant loss source. We then proceed to introduce a more refined loss model for a specific electro-optic coupled-microring frequency beamsplitter implementation, wherein the loss depends on the beamsplitter angle. This beamsplitter loss model is then employed to compare the performance of on-chip HBSG implementations in the frequency domain.

\subsubsection{Frequency beamsplitter implementations}\label{sec:freqBSs}
A beamsplitter-like interaction between disjoint modes $i$ and $j$, in the form of \eqref{eq:generalBS_matrix}, can be induced by engineering an interaction Hamiltonian of the form $g \big( e^{-i \phi} \hc{a}_i a_j +  e^{i \phi} a_i \hc{a}_j \big)$ with interaction duration $\theta/g$. 
For frequency beamsplitters, $i$ and $j$ correspond to different bins, $\red{\om_i}$ and $\blue{\om_j}$. 
Hence the inter-modal coupling, $g$, must be mediated by some drive, which could be a variety of classical (strong, undepleted) bosonic fields that induce the effective exchange $\red{\om_i} \leftrightarrow \blue{\om_j}$. 
There has been significant recent progress in such frequency beamsplitter inducing processes 
including via
microwave field interactions mediated by an electro-optic phase modulator (EOM) \cite{lukens2017frequency,lu2018electro}, 
optical wave-mixing such as Bragg-scattering four-wave mixing based approaches driven by two laser pump fields \cite{joshi2020frequency,oliver2025n},
and
phonon-mediated interactions in opto-mechanical and acousto-optic devices \cite{fan2016integrated,lukens2026paradigm}.
	
One of the primary challenges in implementing such frequency beamsplitters is that they tend to be quite lossy. Accordingly, engineering devices to minimize loss and designing protocols that are robust to loss are tasks of central importance.
We account for such photon loss via a lossy transfer matrix description, where, as compared to the ideal mapping of \eqref{eq:generalBS_matrix}, we implement the transformation
\be\label{eq:lossyTransferMatrix}
	\Xi(\theta, \phi; \eta) 
	= \sqrt{\eta} \begin{pmatrix}
		\cos{\theta} & i e^{i \phi} \sin{\theta} \\
		i e^{-i \phi} \sin{\theta} & \cos{\theta}
	\end{pmatrix}
\ee 
with efficiency $0 \leq \eta \leq 1$ and angle $\theta \in [0, \pi/2]$. As is standard, we model the non-unitary nature of \eqref{eq:lossyTransferMatrix}, i.e., photon loss, by AD channels. 

The advent of the EOM-based quantum frequency processor helped solidify the potential of frequency-encoded quantum information processing \cite{lukens2017frequency,lu2023frequency}.
For a single EOM the best efficiency one can achieve for a 50:50 beamsplitter is $\eta \approx 60\%$ \footnote{
    This assumes a sinusoidal modulation inducing Bessel-weighted sidebands about a central frequency. Then the best efficiency is $\eta = 2 J_0^2(z^*) \approx 0.60049$, where $J_n(z)$ is the Bessel function of the first kind of order $n$ and $z^* \approx 1.4347$ is the smallest positive solution to $|J_0(z)| = |J_{\pm 1}(z)|$, the condition needed for 50:50 splitting. 
}.
In principle, however, one can use an EOM, pulse shaper, EOM sequence to achieve a near-deterministic frequency beamsplitter (as well as entire multi-mode circuits \cite{lu2022high}).
In practice, this potential loss reduction is often negated by insertion loss, especially in free-space implementations, so going on-chip poses a promising though challenging path forward \cite{lukens2017frequency,myilswamy2025chip,congia2026fully}. 

\subsubsection{Driven coupled-microring frequency beamsplitters}
Another promising route to on-chip frequency beamsplitters that operate on specific bins is by leveraging EOMs in consonance with resonant microring devices. Then the resonators fix which bins can be acted upon and the EOMs mediate the coupling of such bins without appreciably affecting the other non-resonant bins. Accordingly, we will focus on such driven coupled-microring frequency beamsplitters to get a specific beamsplitter loss model, though many of the underlying principles should apply more generally to other lossy beamsplitter implementations.

Such devices have been experimentally demonstrated by \refref{hu2021chip}, requiring nontrivial integrated photonics engineering. Reference \citenum{munoz2026modeling} proceeded to theoretically analyze and develop a transfer matrix description of such devices. Ultimately, by carefully engineering and tuning the device parameters, such a device can be made to operate as a frequency beamsplitter in the form of \eqref{eq:lossyTransferMatrix} acting on the device's normal-mode frequencies with efficiency  
\be\label{eq:etaOmegaFull}
    \eta(\Omega, \theta) = 1 - \frac{2}{1+ \sqrt{1 + \Omega^2 \csc^2\theta}} 
\ee
with $\Omega$ a dimensionless modulation amplitude.
Because the underlying device engineering and physics is not a contribution of this work, but rather of Refs.~\citenum{hu2021chip} and \citenum{munoz2026modeling}, we opt to simply leverage this transfer matrix prescription of \refref{munoz2026modeling} here, deferring discussion of the underlying implementation to \appref{sec:twoRingFreqBS-Decomp}.

For each $0 < \theta \leq \pi/2$ this efficiency monotonically increases as a function of $\Omega$ from $0$ at $\Omega=0$ to $1$ as $\Omega \ra \infty$,
so to minimize loss our task, in large part, is to increase $\Omega$. 
Moreover, for a given $\Omega$, the beamsplitter efficiency, $\eta_{\sin^2{\theta}}$, is a monotonically decreasing function of the beamsplitter angle, $\theta$. 
For our beamsplitter loss model, we suppose there is some maximum reasonably achievable modulation amplitude $\Omega_\textrm{max}$ that dictates how efficient a $\theta$-beamsplitter can be.

\subsubsection{Assumptions}\label{sec:beamsplitterAssumptions}
To naturally leverage this transfer matrix description and for the simplicity of the strong linear-optical simulations in Perceval, we make the best-case assumption that the photons are fully indistinguishable apart from their frequency bins and that this is not changed by the frequency beamsplitters \cite{munoz2026modeling}. This assumption is consistent with our focus on HSPSs for which near-unity source purity and indistinguishability can be achieved and multiphoton contamination tends to be a larger concern \cite{wiesner2024influence}. 
Additionally, we leverage the flexibility of frequency beamsplitters to act on non-adjacent modes, assuming that the necessary inter-ring couplings and modulation frequencies can be achieved. In particular, we suppose the frequency beamsplitters can act on bins separated by up to four bin spacings as needed for the all-frequency-beamsplitter implementation of the 4P8M circuit of Fig.~\ref{fig:BSG_circuits}(e).

\subsection{HBSG in the frequency domain}\label{sec:frequencyHBSG}
\begin{figure}[h]
    \captionsetup[subfloat]{captionskip=0pt}
    \includegraphics[width=0.95\linewidth]{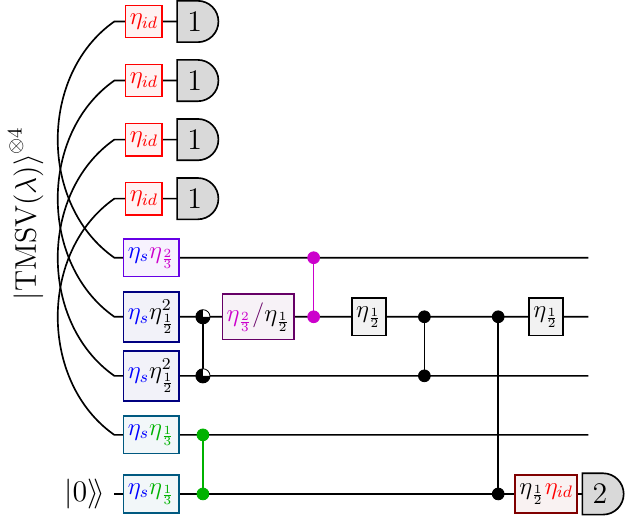}
    \caption{
        Lossy HSPS and frequency beamsplitter model for the 4P5M scheme.
        The top four modes are heralding idler modes that are entangled with the top four signal modes via TMSV states as indicated via the linking arcs.
        In addition to heralding detector loss, $\red{\eta_{id}}$, and signal mode source loss, $\blue{\eta_s}$, we account for beamsplitter loss 
        with efficiencies      
        $\textcolor{forestgreen}{\eta_{\scriptscriptstyle \frac{1}{3}}} > \eta_{\scriptscriptstyle \frac{1}{2}} > \textcolor{lightpurple}{\eta_{\scriptscriptstyle \frac{2}{3}}}$.
        Using the properties of photon loss channels (see \secref{sec:photonLoss}), we shift said beamsplitter loss to the source to the extent possible, leaving four single-mode loss channels (two internal to the circuit and two after it) that we account for using the beamsplitter loss model.
    }\label{fig:4P5M_HBSG_shifting_loss}
\end{figure}

Returning to HBSG, these lossy frequency beamsplitters will lead to nonuniform AD channels that cannot generally be mapped onto the lumped-loss model.
As the frequency bins are propagating in a common bus waveguide, the post-circuit signal-mode heralding requires the unloading of specific bins. This could be accomplished using resonant microrings followed by detection. Accordingly, we will take the corresponding AD efficiencies to be the same as for the HSPS idler-mode readout, $\eta_{id}$ (making relation \ref{item:relation3} further justified in this context).
In \figref{fig:4P5M_HBSG_shifting_loss}, we illustrate this ultimate lossy model for the 4P5M scheme of \figref{fig:BSG_circuits}(a). The methodology is analogous for the other schemes. 

\subsubsection{Distinctive aspects of 4P8M circuit in the frequency domain}\label{sec:4P8MFrequencyImplementation}

\begin{figure}[ht!]
    \includegraphics[width=\linewidth]{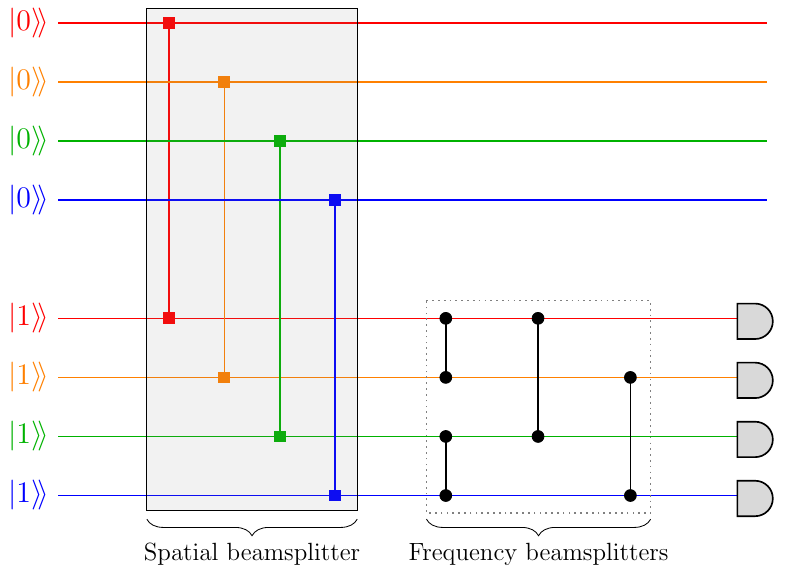} 
    \caption{Spatial-frequency hybridized 4P8M HBSG circuit.
	The beamsplitters with square mode labels act between identical frequency bins across different spatial modes and can be implemented via a single spatial beamsplitter. 
    As in Fig.~\ref{fig:BSG_circuits}, the beamsplitters with circle mode labels act on frequency bins. Here color denotes the different frequency bins; each beamsplitter is 50:50.
    }\label{fig:hybrid_4P8M_HBSG}
\end{figure}

Especially in early experiments we expect frequency beamsplitter loss to be the dominant error mechanism. Accordingly, it is worthwhile to minimize the number of frequency beamsplitters when possible.
Fortunately, some heralded frequency-encoded photonic entanglement generation circuits, including the 4P8M HBSG circuit of Fig.~\ref{fig:BSG_circuits}(e), naturally admit a separation across distinct spatial modes (i.e., different waveguides or fibers). Therein, the frequency-encoded resource states can be produced on a single waveguide in a less lossy manner by leveraging an additional waveguide---linked to the first by a spatial beamsplitter---whose output is used for the generation heralding.
The spatial beamsplitter splits photons across two spatial modes without altering other DoFs such as the frequency or wave-packet shape (at least within some finite bandwidth that is typically much larger than the frequency-bin mode spacings we will be concerned with) and can thus be used to perform a single transformation to all pairs of identical frequency bins across two spatial modes simultaneously.

This ``hybrid'' nature of the 4P8M HBSG circuit is illustrated in Fig.~\ref{fig:hybrid_4P8M_HBSG}. Other elements in this class of spatial-frequency hybridizable photonic circuits include the infinite subclass of $n\geq 2$-qubit GHZ generation circuits of  \refref{gimeno2016towards} as well as the effectively identical variants of \refref{chin2024heralded} (see \appref{app:hybrid_spat_freq}). Moreover, one can easily see that the boosted type-II fusion circuits of \refref{ewert20143} can naturally leverage this spatial-frequency hybridization.

\subsubsection{Bell-state fidelities}

\begin{figure*}[ht!]
    \includegraphics[width=0.99\linewidth, clip=true, trim = 1em 1em 1em 1em]{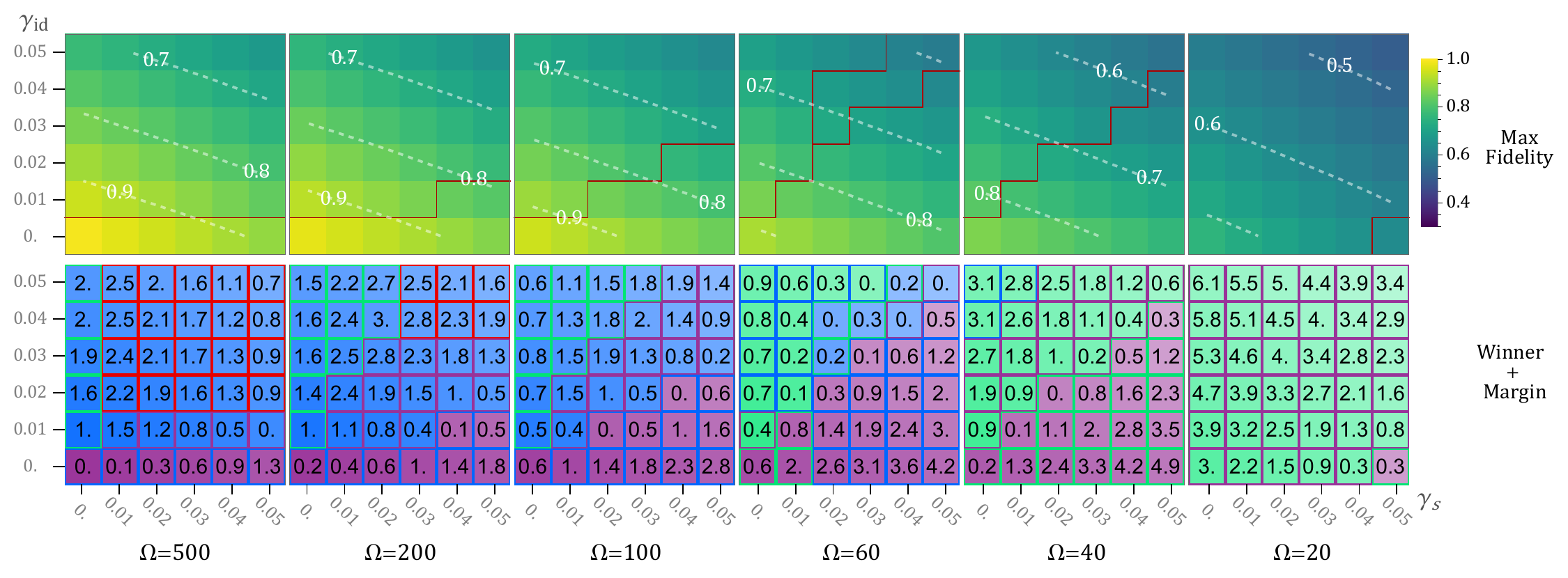}
    \caption{
    Comparison of the HBSG circuits of \figref{fig:BSG_circuits} (and variants thereof) across a variety of parameter regimes with lossy HSPSs and lossy electro-optic microring frequency beamsplitters with $n_\textrm{extra} = 3$.
    (top panel) For each $\Omega \in \{ 20, 40, 60, 100, 200, 500 \}$, we construct an array plot of the maximum fidelity of the output Bell states over the considered schemes
    as a function of $\g_s$ and $\g_{id}$.
    (bottom panel) 
    Winner and margin plot that highlights which circuit ``wins'' in terms of having the highest fidelity (indicated via pixel color) and by what margin (overlaid value) relative to the second best circuit (indicated via the pixel border color). 
    This construction is analogous to \figref{fig:lossyHSPSsLumpedCircuitLossFidelityComparison} and uses the same color scheme: 
        {\protect\textcolor[RGB]{255, 128, 0}{4P5M}},
        {\protect\textcolor[RGB]{255, 214, 0}{4P6M}},
        {\protect\textcolor[RGB]{26, 184, 0}{4P8M}},
        {\protect\textcolor[RGB]{230, 0, 0}{5P5M}},
        {\protect\textcolor[RGB]{0, 102, 255}{6P6M$+$}},
        {\protect\textcolor[RGB]{153, 51, 153}{6P6M$-$}},
        with the addition of 
        {\protect\textcolor[RGB]{4, 230, 117}{4P8M hybrid}}.
    The corresponding fidelities and heralding probabilities for each scheme is given in \appref{app:freqImplementationHeraldingProbabilities}.
    Note that for the 5P5M and 6P6M$\pm$ schemes we consider two variants of the circuits: (i) and (ii), as defined in \appref{app:circuitVariants}.
    }\label{fig:modestOmega_fullComparison_WinnerAndMargins}
\end{figure*} 

\begin{figure}[ht!]
    \captionsetup[subfloat]{farskip=1pt, captionskip=-3pt}
    \subfloat[(a) Average $\cF$ for $\Omega=20$]{\includegraphics[width=0.9\linewidth]{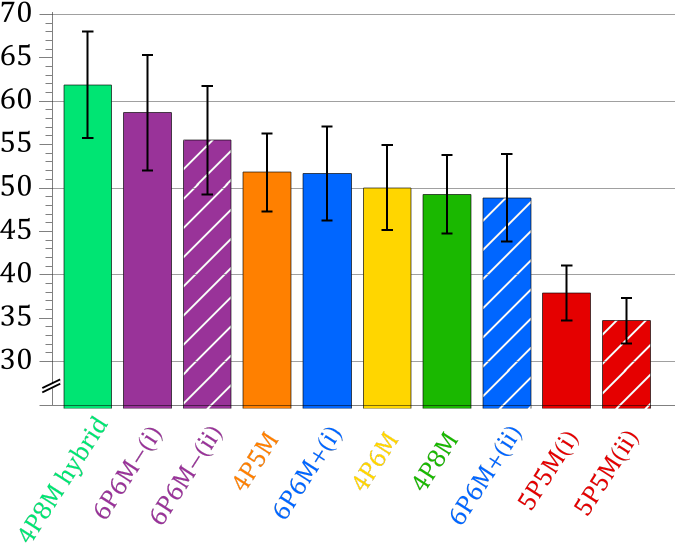}} \\ 
    \subfloat[(b) Average $\cF$ for $\Omega=100$]{\includegraphics[width=0.9\linewidth]{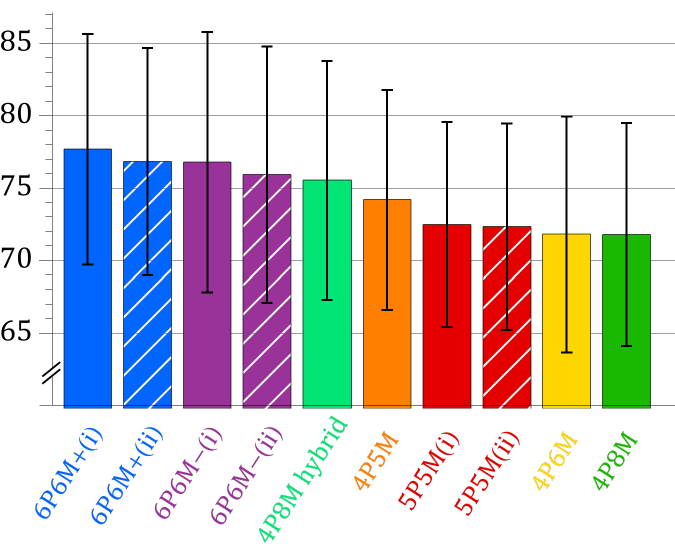}} 
    \caption{Bar charts contrasting the overall fidelity performance of each circuit and their variants for (a) $\Omega = 20$ and (b) $\Omega=100$.
    In each case, for each circuit we plot the average and standard deviation of the fidelities across the $36$ data points corresponding to the values of $\g_s$ and $\g_{id}$ swept over in each density plot of \figref{fig:modestOmega_fullComparison_WinnerAndMargins}. 
    }\label{fig:freqModel_fidelity_Barchart}
\end{figure}

The comparison of the HBSG schemes in this coupled-microring frequency-domain implementation, as shown in \figref{fig:modestOmega_fullComparison_WinnerAndMargins}, is analogous to that of \figref{fig:lossyHSPSsLumpedCircuitLossFidelityComparison} except that circuit loss (previously lumped into $\eta_b = \eta_s \eta_c$) is now taken to be dominated by nonuniform beamsplitter loss as parameterized by $\Omega$. 
Which scheme performs best in terms of fidelity is sensitive to not just how much loss occurs but where it occurs. Overall we find that there are three winning schemes depending on what parameter regime one is operating in.
\begin{enumerate}
    \item The hybrid 4P8M scheme wins when dominated by frequency-beamsplitter loss, especially in regimes with significant multiphoton errors. 
    \item The 6P6M$-$ scheme wins when dominated by nearly uniform source and circuit loss across all modes, as is consistent with its preeminence in the lumped circuit model.
    \item The 6P6M$+$ scheme wins when dominated by multiphoton errors and the frequency-beamsplitters are quite efficient. 
\end{enumerate}
For 1, we note that for small $\Omega$, when dominated by frequency-beamsplitter loss, it naturally is beneficial to minimize the number of such beamsplitters. In \figref{fig:freqModel_fidelity_Barchart}(a) we show the averaged circuit performance for $\Omega = 20$, which is comparable to what has been achieved experimentally \footnote{
    In the experimental demonstration of \refref{hu2021chip}, the strongest electro-optic modulation they use is $\epsilon \approx 25$~dBm which we estimate corresponds to $\epsilon/2\pi \approx 2.72$~GHz~\cite{munoz2026modeling} and the internal loss rate of the device with the largest splitting was reported to be $\kint/2\pi \approx 0.17$~GHz. Accordingly, the maximum dimensionless modulation strength used is  $\Omega = \epsilon/\kint \sim 16$. This can be increased by increasing $\epsilon$ (modulating harder) and reducing ring loss $\kint$.
}.
Therein, we see that the best performing scheme on average, the 4P8M hybrid one, has the smallest ratio of the number of frequency beamsplitters to modes, $n_\textrm{fBS}/M = 1/2$, and the worst performing schemes, the 5P5M ones, have the highest ratio, $n_\textrm{fBS}/M = 2$ (see \appref{app:circuitVariants}). However, the other schemes---for which this ratio is closer, $5/6 \leq n_\textrm{fBS}/M \leq 7/6$---do not follow this heuristic. 

The relative performance of the various schemes changes considerably for more efficient beamsplitters. In particular, as $\Omega$ is increased the hybrid 4P8M scheme gets surpassed by the 6P6M$-$ scheme, especially when photon loss dominated. Then as $\Omega$ is further increased the 6P6M$+$ scheme starts outperforming both the hybrid 4P8M and 6P6M$-$ schemes [see \figref{fig:freqModel_fidelity_Barchart}(b)], especially for appreciable multiphoton errors (larger $\g_{id}$) as was found in \figref{fig:lossyHSPSsLumpedCircuitLossFidelityComparison} for lumped circuit loss.

In terms of generation rate, the $N=5,6$ photon schemes are clearly more expensive when using probabilistic HSPSs, with each additional photon making the generation $25\%$ as probable at best in a given direct generation attempt. Additionally, they require detectors with more stringent PNR capabilities.
Accordingly, it is worth contrasting the $N=4$ photon schemes; to which, we note that 
between $\Omega = 100$ and $150$ the all-frequency-beamsplitter 4P8M scheme starts outperforming the 4P6M scheme. Moreover, 
between $\Omega = 200$ and $500$ the 4P5M scheme starts winning on average in terms of the $N=4$ schemes, even versus the hybrid 4P8M, especially when photon loss dominated as compared to multiphoton errors. 

\subsection{Extensions}
Here we considered HBSG schemes from the literature as written in terms of beamsplitters operating on pairs of modes, which is one reason we focus on driven coupled-microring devices that are amenable to implementing two-mode beamsplitters.
However, a unique resource of frequency-bin processing is the ability to natively do multi-bin operations \cite{lukens2017frequency,lu2022high,oliver2025n,munoz2026modeling,lukens2026paradigm}.
For instance, the subcircuit emphasized in the dashed region of Fig.~\ref{fig:hybrid_4P8M_HBSG} is a rotated Hadamard matrix of order $n=4$, $H_4$, decomposed in terms of four beamsplitters \footnote{Rotated in that it is decomposed in terms of symmetric Loudon beamsplitters, in the form of \eqref{eq:generalBS_matrix} with $\phi=0$. If candid Hadamard beamsplitters in the form of \eqref{eq:twoByTwoHadamard} were used, one would obtain the standard real-entried $H_4$ (with the physicist normalization).}. As is shown in \refref{munoz2026modeling}, this circuit can be implemented natively in the frequency domain with a single device consisting of four coupled, driven microring resonators.

Meanwhile, the 5P5M DFT circuit, for instance, could be implemented directly using a quantum frequency processor \cite{lu2022high}, which is an alternating concatenation of EOMs and pulse shapers, whereas for us its implementation required an expensive decomposition in terms of 10 two-mode frequency beamsplitters (see \appref{app:circuitVariants}).
Accordingly, analysis of how to best leverage such multi-mode operations for frequency-encoded resource-state generation, and contrasting them against more conventional approaches such as that pursued here, is an interesting, potentially fruitful area of study. For instance, one could imagine performing numerical optimization over circuits similar to \refref{hartnett2026automated} with the inclusion of native frequency operations.

We focused on the impact of multiphoton errors as, beyond photon loss, they tend to be the dominant error mechanism for biphoton HSPSs \cite{wiesner2024influence}. Nevertheless, in general one should account for the possibility that the photons being interfered are partially distinguishable (especially when considering solid-state sources \cite{Note3}), which will degrade the quality of the interference orchestrated by the photonic circuit \cite{menssen2017distinguishability,randles2025interference} and ultimately of the heralded resource states as analyzed in \refref{shaw2023errors}. Accordingly, natural extensions of the work in this section would include analyses of the wavepacket distortion and timing errors induced by frequency beamsplitters, and how to minimize them, as well as of the corresponding impact of distinguishability in frequency-bin resource-state generation.

Analyses of nonuniform circuit loss in other implementations, e.g., due to optical switching or insertion loss, could also be performed in this modeling framework.
Moreover, this framework could be used to assess 
the impact of static or fluctuating system parameters (e.g., the beamsplitter angles $\theta$ and $\phi$),
the implications of more intricate detector models (see \appref{app:detectorPOVMs}),
and 
the source-multiplexing requirements needed to achieve target resource-state generation rates at fidelities above a specified threshold.

\section{Discussion and conclusion}\label{sec:conclusion}
In this paper, we analyzed the performance of five heralded Bell-state generation schemes from the literature, with a focus on on-chip implementations using SFWM-based heralded single-photon sources (HSPSs). 
We further developed a frequency-domain implementation model, demonstrating how the broader modeling methodology can be tailored to architectures with distinct physical constraints and advantages.

Overall, we find that schemes exhibiting higher-order multiphoton interference tend to be more robust to both photon loss and multiphoton errors in terms of the fidelity of the generated Bell state. References \citenum{shaw2023errors} and \citenum{fldzhyan2021compact} similarly found this to be the case in related settings with the former further finding that higher-order multiphoton interference benefits the schemes' robustness to partial distinguishability errors (though without considering the 6P6M scheme).
Namely, the 6P6M and 5P5M schemes exhibit genuine 6- and 5-photon interference, respectively, and for nearly uniform intra-circuit loss channels they are the most robust to photon loss and multiphoton errors. This robustness is due, at least in part, to the zero-transmission law of \refref{tichy2010zero}, which suppresses a subset of potential false-positive heralding events. 

When using HSPSs, we find that the output Bell-state fidelities are quite sensitive to multiphoton errors, i.e., there is a relatively sharp contrast between no multiphotons ($\g_{id} =0$) and some multiphotons ($\g_{id} > 0$) for each scheme. However, how sensitive varies appreciably. For instance, within the 6P6M scheme itself we find that different detection patterns herald Bell states that are less susceptible to multiphoton errors yet more susceptible to photon loss and vice versa.
Meanwhile, the $N=4$ photon circuits are likely preferable in contexts where single photons are difficult to produce, i.e., with non-multiplexed HSPSs, and where detectors with high PNR capabilities are not available. Then the 4P5M circuit tends to perform the best across the parameter ranges studied, unless one is operating in the frequency domain in which case the 4P8M scheme can be done in a less lossy hybrid spatial-frequency fashion.

Focusing on the heralded generation of Bell states, the most modest seed states, enabled the exhaustiveness of our comparison, which, together with the analysis of \refref{shaw2023errors}, provides a largely complete picture of how the studied HBSG schemes behave in the presence of realistic error mechanisms.
Moreover, this work provides a natural framework for addressing future questions about generating larger entangled resource states of photons and using them for QIS applications.
Corresponding analyses could explore
how errors propagate under fusions when constructing larger resource states 
and 
when larger resource states should be generated directly.
As a simple example, we note that using the highest probability HBSG circuit considered here (with $p_\textrm{ideal} = 3/16$), 
the probability of making a 3-qubit GHZ state via the type-I fusion of two Bell states in a single shot is $\frac{1}{2} p_\textrm{ideal}^2 = 9/512 \approx 1.758\%$ (assuming access to the 8 single-photon inputs). Meanwhile, the 6P12M 3-qubit GHZ state generation scheme of \refref{varnava2008good} ideally succeeds with probability $1/32 = 3.125\%$ and requires two fewer single-photon inputs. Thus, in an ideal implementation the latter approach is preferable in terms of direct rate. However, by source multiplexing (say as made possible using quantum memories and optical switching) this is no longer necessarily the case. 
More generally, the determination of which approaches are preferable in a given experimental context---in terms of both quality of the output resource state and the rate of its generation---is crucial for designing and realizing photonic quantum information processing primitives.

The modeling methodology and computer simulations of this work, along with those of Refs.~\citenum{shaw2023errors} and \citenum{wiesner2024influence}, provide one direction for tackling such questions. 
Moreover, this work also provides an analytical framework for exploring such questions, developing effective models for resource-state generation and subsequent applications, and performing system-level analysis.
\newline
    
\section{Acknowledgments} 
We acknowledge support from the Sandia National Laboratories Laboratory Directed Research and Development program (EPIQ project).
We thank Lucas Cohen, Paul Davids, Michael Gehl, Andrew Landahl, Nils Otterstrom, and Kevin Thompson for helpful discussions.
Sandia National Laboratories is a multimission laboratory managed and operated by National Technology and Engineering Solutions of Sandia LLC, a wholly owned subsidiary of Honeywell International Inc.~for the U.S.~Department of Energy’s National Nuclear Security Administration under contract DE-NA0003525.
    
\bibliographystyle{unsrt} 
\bibliography{BSG_overleaf}

\newpage

\appendix
\addtocontents{toc}{\protect\setcounter{tocdepth}{1}} 
\printAppendixTOC 

\section{Beamsplitter conventions}\label{app:beamsplitter_identities}
Spatial beamsplitters tend to have fixed phases, e.g., the symmetric Loudon beamsplitter given by \eqref{eq:generalBS_matrix} with fixed $\phi=0$ is common as is the real-entried beamsplitter 
\be\label{eq:twoByTwoHadamard}
    BS_H(\theta) = \begin{pmatrix}
        \cos\theta & \sin\theta \\
        \sin\theta & -\cos\theta
    \end{pmatrix},
\ee
which for a 50:50 splitting is the $2 \times 2$ Hadamard matrix, $H_2 = BS_H(\pi/4)$. 
Similar unitaries appear as the Jones matrices for waveplates in polarization optics.

One can convert between beamsplitters in the form of \eqref{eq:generalBS_matrix} to that of \eqref{eq:twoByTwoHadamard} using phase shifters (indexing from $0$) 
as
\subalign{
    BS(\theta, 0) &= 
    PS_1(\pi/2)
    BS_H(\theta)
    PS_1(\pi/2)
    \\
    &= e^{-i \theta} H_2 PS_0(2 \theta) H_2 \\
    &= e^{i \theta} H_2 PS_1(-2 \theta) H_2, 
}
where the bottom two equalities give Mach--Zehnder interferometric forms of an arbitrary reflectivity beamsplitter. Meanwhile, one finds that phases can be ``moved through'' beamsplitters in the form of \eqref{eq:generalBS_matrix} by adjusting the beamsplitter's phase as
\subalign{
    PS_0(\phi_2) BS(\theta, \phi_1) &= BS(\theta, \phi_1 + \phi_2) PS_0(\phi_2), \\
    PS_1(\phi_2) BS(\theta, \phi_1) &= BS(\theta, \phi_1 - \phi_2) PS_1(\phi_2).
}

\section{Additional HBSG circuit details}\label{app:HBSG_extra}
First we list the specific Bell-like states heralded by each circuit using the notation
$\ket{\psi^{N\textrm{P}M\textrm{M}}_{\hvec}}$, where $\hvec$ is the heralding click pattern (counting modes starting from $0$): 
\begin{enumerate}[label=(\alph*)]
    \item $\ket{\psi^\textrm{4P5M}_{(2_4)}} = -i \ket{\Phi^-}$ with $p_\textrm{ideal} = 1/9$,
    \item $\ket{\psi^\textrm{5P5M}_{(3_4)}} = \ket{\Psi^+}$ with $p_\textrm{ideal} = 12/125$, \label{heraldStates:5P5M}
    \item $\ket{\psi^\textrm{4P6M}_{(1_2 1_3)}} = -\ket{\chi^-}$ with $p_\textrm{ideal} = 2/27$,
    \item $\ket{\psi^\textrm{6P6M}_{(0_2 4_3)}} = -\ket{\psi^\textrm{6P6M}_{(4_2 0_3)}} = i\ket{\chi^+}$ and $\ket{\psi^\textrm{6P6M}_{(1_2 3_3)}} = \ket{\psi^\textrm{6P6M}_{(3_2 1_3)}} = -\ket{\chi^-}$ each with probability $1/27$ hence $p_\textrm{ideal} = 4/27$ overall, and \label{heraldStates:6P6M}
    \item 
    $\ket{\psi^\textrm{4P8M}_{(1_4 1_5 0_6 0_7)}} = \ket{\psi^\textrm{4P8M}_{(0_4 0_5 1_6 1_7)}} = \ket{\Phi^+}$,
    $\ket{\psi^\textrm{4P8M}_{(1_4 0_5 0_6 1_7)}} = -\ket{\psi^\textrm{4P8M}_{(0_4 1_5 1_6 0_7)}} = \ket{\Psi^-}$,
    and 
    $\ket{\psi^\textrm{4P8M}_{(1_4 0_5 1_6 0_7)}} = \ket{\psi^\textrm{4P8M}_{(0_4 1_5 0_6 1_7)}} = \ket{\chi^+}$ each with probability $1/32$ hence $p_\textrm{ideal} = 3/16$ overall. \label{heraldStates:4P8M}
\end{enumerate}

\subsection{5P5M}\label{app:5P5Mdetails}
The Clements decomposition of this circuit is given in \figref{fig:5P5M_Clements_decomp} and is discussed in the surrounding \appref{app:circuitVariants} text. The corresponding DFT-based approach of \refref{paesani2021scheme} generalizes to the heralded generation circuits for $N$-photon, $d$-dimensional GHZ states. 

\subsection{4P6M}\label{app:4P6Mdetails}
The 4P6M scheme of Fig.~\ref{fig:BSG_circuits}(c) is taken from \refref{forbes2025heralded}, which is based on the circuit presented in \refref{carolan2015universal}, however, with phase conventions altered so that the final 50:50 beamsplitter can be omitted. 
Note that the prescribed $\pi$ phase on the top beamsplitter is not necessary, it simply acts to `flip' the output Bell-like state to $-\ket{\chi^-}$, whereas with the phase omitted, the scheme would yield $\ket{\chi^+}$. This adjustment will not impact the calculated heralded state fidelities and  generation probabilities. 

As emphasized in \refref{fldzhyan2021compact}, this circuit only exhibits 3-photon interference, which can be seen in that the top photon (injected onto mode 1) cannot reach the lower detector and symmetrically the bottom photon cannot reach the upper detector. Accordingly, in the absence of multiphoton errors, a 4-photon detection event will never occur for this scheme. Moreover, this circuit is separable into two subcircuits, where one could start with the topmost three photons, let them pass through the three beamsplitters entirely on those modes and be detected by the upper detector. Then only if a 1-photon detection is registered would one need to inject the additional photon onto mode 4, which would interfere with the heralded two-photon state of the previous subcircuit via the bottom two beamsplitters, and upon a 1-photon detection on the lower detector would herald the generation of the Bell-like state. That is, by including feed-forward one can make the fourth photon necessary only when the first heralded subunitary succeeds (a related, though more elaborate, concept was presented in \refref{fldzhyan2021compact}).

\subsection{4P8M}\label{app:4P8Mdetails}
This scheme, first proposed in \refref{zhang2008demonstration}, contains a significant amount of symmetry and structure [see \eqref{eq:H4_identity} and the surrounding discussion]. Said structure allows one to effectively ``shift'' the whole $H_4$ subunitary around relative to four transversal beamsplitters, denoted here as $T$.
For instance, one will also herald Bell-like states of the same form as \hyperref[heraldStates:4P8M]{(e)} 
if $H_4$ is placed before $T$ on either the top or bottom four rails, provided the 4 input photons are injected into the same modes as $H_4$. 
Such structure can also be used to show that the $n=2$ (Bell) case of the $n$-qubit GHZ state generation scheme of \refref{chin2024heralded} is closely related to this circuit (see \appref{app:hybrid_spat_freq}). 
Using related circuits, \refref{bartolucci2021creation} demonstrate additional, more elaborate schemes that, with feed-forward, can be used to increase the probability of heralding the generation of a Bell-like state to $1/4$ or above using a combination of entanglement distillation, additional photons, and a process they term ``bleeding.'' 

Moreover, under the lumped-loss model of \secref{sec:lumpedHBSG}, this scheme exhibits the property that on a successful herald the resulting output state is an identically damped Bell state 
\be
    \rho_{\hvec^*} = \cE_{\eta_e}\p{ \ketbra{\cB_{\hvec^*}}{\cB_{\hvec^*}} }
\ee
with effective efficiency 
\be
    \eta_{e} = \frac{1-\g_b}{1+\g_h}
    = \frac{\eta_b}{2-\eta_h}
\ee
as can be shown via direct computation. The resulting fidelity is $\cF_\cB = \eta_e^2$ as is consistent with \eqref{eq:lumpedLossFidelity} as
$f(\g_h) = 1 + 2\g_h + \g_h^2 = (1+\g_h)^2$ for this scheme.
This property is also exhibited by the 3-qubit GHZ state scheme of \refref{varnava2008good} (see Fig.~1 in the arXiv version) and can be understood heuristically via analogous reasoning as the type-II fusion of two Bell states in the successful heralding subspace.

\subsection{Detector PNR requirements}
The number of detectors needed for each circuit and their necessary PNR capabilities varies.  
By explicitly calculating the output states of the ideal 4P6M and 4P8M schemes one can easily compute the respective probabilities of a false positive, as induced by using threshold (as opposed to $0, 1, \geq 2$) detectors for the post-circuit heralding, to be $3/27$ and $15/128$ such that the corresponding fidelities with the target Bell state(s) become $2/5 = 0.4$ and $24/39 \approx 0.62$.
Meanwhile, the 4P5M, 5P5M, and 6P6M schemes have increasingly stringent PNR requirements, for which one should be able to discern between $0, 1, \dots, n$ and $> n$ photons with $n$ of 2, 3, and 4, respectively.

\subsection{Implementation flexibility}
There is additional freedom in exactly how each of these circuits is implemented including via permutations of the modes (e.g., to herald different Bell states) 
as well as different decompositions in some cases such as for the 5P5M and 6P6M circuits (see \appref{app:circuitVariants}).
Moreover, different arrangements of the input photons herald success in some of the circuits. Specifically, in the 4P6M and 4P8M schemes as shown in \figref{fig:BSG_circuits}, for the 50:50 beamsplitters that interfere a single photon and vacuum-initialized mode, one can inject the single photon onto either mode leading to $2^2$ and $2^4$ valid single-photon input configurations (up to known phase differences in the heralded Bell states), respectively. 
Moreover, the HBSG schemes have additional input state flexibility if we allow the number of input photons, $N$, to be increased as discussed in \appref{app:higherNFlex}, though this further heightens the PNR requirements.

\section{Simplifying lossy circuits}\label{app:lossyCircuits}
As discussed in \secref{sec:photonLoss}, photon loss is modeled using amplitude damping (AD) channels. We first define AD channels and then show that they satisfy properties \ref{item:AD_multiplicative} and \ref{item:uniformAD_commutes}.

\subsection{Amplitude damping channels}
On a single bosonic mode with creation (annihilation) operator $\hc{a}$ ($a$), the AD channel is:
\be
    \rho \ra \cE_\eta(\rho) = \sum_{l = 0}^\infty E_l \rho \hc{E}_l
\ee
with Kraus operators
\be 
    E_l = \p{\frac{\g}{\eta}}^{l/2} \frac{a^l}{\sqrt{l!}} \eta^{\hc{a}a/2}
\ee
and $\g \equiv 1- \eta$. One can easily verify that this is equivalent to the beamsplitter loss model of \eqref{eq:singleModeBSLossModel}.

\subsection{Single-mode loss is multiplicative} 
The action of an AD channel on the Fock basis is thus
\begin{align}
    \cE_\eta(\ketm{n_1}\!\bram{n_2}) 
    = &\sum_{k=0}^{m}
    \br{\eta^{n_1+n_2} \binom{n_1}{k} \binom{n_2}{k}}^{1/2} \\
    &\quad\times \p{\frac{\g}{\eta}}^{k} \ketm{n_1-k}\!\bram{n_2-k} \nonumber
\end{align}
with $m=\min(n_1, n_2)$. After a few lines of algebra one finds
\be
    \cE_{\eta_2}(\cE_{\eta_1}(\ketm{n_1}\!\bram{n_2})) = \cE_{\eta_1 \eta_2}(\ketm{n_1}\!\bram{n_2}),
\ee
which by completeness of the Fock basis verifies property \ref{item:AD_multiplicative}.
On the pure Fock state $\ketm{n}$ 
\begin{subequations}
    \begin{align}
        \mathcal{E}_{\eta}(\ketm{n}\!\bram{n}) = \sum_{k=0}^n  \varepsilon_{n,k}(\eta) \ketm{k}\!\bram{k}, \label{eq:FockStateAD} \\
        \varepsilon_{n,k}(\eta) := \binom{n}{k} \eta^k (1 - \eta)^{n-k}. \label{eq:epsCoeffFockAD}
    \end{align}
\end{subequations}

\subsection{Uniform loss commutes with linear optics} 
An $m \times m$ linear optical unitary, $U$, transforms annihilation operators as $a_i \overset{U}{\ra} \sum_{j=1}^m U_{ij} a_j$ (and similarly for creation operators).
Meanwhile, under the beamsplitter loss model, loss induces the transformation $a_i \overset{L}{\ra} \sqrt{\eta} a_i + \sqrt{\gamma} b_i$, where $\{b_i\}$ are a set of $m$ annihilation operators for the vacuum-initialized loss modes that are to be traced over.
Hence,
\be
    a_i \overset{U L\,}{\longrightarrow} \sqrt{\eta} \sum_j U_{ij} a_j + \sqrt{\gamma} b_i,
\ee 
whereas
\be
    a_i \overset{L U}{\longrightarrow} 
    \sqrt{\eta} \sum_j U_{ij} a_j + \sqrt{\gamma} b'_i
\ee 
with $b'_i = \sum_j U_{ij} b_j$ a unitarily transformed loss mode annihilation operator. 

Thus, these dilated transformations are the same on the $a$ modes, yet not on $b$.	 
However, unitary transformations to the modes being traced over, $b$, have no physical impact. Namely, by
inserting the identity $\hc{U}_b U_b = \mathbbm{1}_b$ and leveraging the cyclic property of the partial trace on the subsystem being traced over, one has
$\rho_a = \Tr_b(\rho_{ab})= \Tr_b(U_b \rho_{ab} \hc{U}_b) $. Thus, as quantum channels on the $a$ modes, uniform loss and unitary transformations commute as they only differ by a unitary on auxiliary modes, verifying property \ref{item:uniformAD_commutes}.
See App.~E of \refref{oszmaniec2016random} for a proof using a first-quantization framework.

\section{Detector models}\label{app:detectorPOVMs}
We analyze photodetection in the framework of a positive operator-valued measure (POVM), which is a set of complete measurement operators, $\{ \hat{\Pi}_n \}$, each corresponding to a distinct measurement outcome $n$. For an individual optical mode, e.g., a specific frequency bin, each of the POVM elements admits a Fock state representation
\be\label{eq:PikFockRep}
    \hat{\Pi}_n = \sum_{k=0}^\infty \varpi_{n,k} \ketm{k}\!\bram{k},
\ee
where the weights are constrained by $\sum_n \varpi_{n,k} = 1$ for each $k$ such that the POVM satisfies the completeness relation
\be
    \sum_n \hat{\Pi}_n = \sum_{k=0}^\infty \ketm{k}\!\bram{k}
    = \mathbbm{1}. 
\ee 
    
Here we detail three photodetection POVMs we consider in this work. Ideally one would have access to a set of ideal photon-number resolving (PNR) detectors that can perfectly discern the precise number of photons in a given mode with POVM elements $\hat{\Pi}_n = \ketm{n}\!\bram{n}$. However, in practice such PNR detectors have less than unit efficiencies (App.~\ref{app:PNRDs}) and, moreover, are expensive so it is worth considering cheaper threshold detectors either individually (App.~\ref{app:thresholdDets}) or in arrays (App.~\ref{app:arrayDets}).

\subsection{PNR detectors}\label{app:PNRDs}
The standard photocounting formula of an ideal PNR detector gives the probability of detecting exactly $n$ photons in a single-mode radiation field of state $\rho$ to be 
\be
    p_n = {\Tr}(\rho \hat{\Pi}_n)
\ee
with
\be\label{eq:idealPinNO} 
    \hat{\Pi}_n = : e^{-\hc{a}a} \frac{\p{\hc{a}a}^n}{n!} :,
\ee
where $\hc{a}$ and $a$ are the respective single-mode creation and annihilation operators and $:\,:$  denotes the normal ordering operator \cite{sperling2012true}.
One can immediately see that these form a complete POVM and, upon a brief calculation, can show that they have the expected Fock state representation $\hat{\Pi}_n = \ketm{n}\!\bram{n}$.
    
These POVM elements can be modified to account for PNR detectors with efficiency $0 < \eta \leq 1$ and dark counts $d \geq 0$ by modifying the number operator in \eqref{eq:idealPinNO} as $\hc{a} a \ra \eta \hc{a} a + d$ \cite{sperling2012true}, yielding
\subalign{
    \hat{\Pi}_n(\eta, d) &=~: e^{-\eta \hc{a} a - d} \frac{\p{\eta \hc{a} a + d}^n}{n!} : \\
    &= e^{-d} \sum_{j=0}^n \frac{d^{n-j}}{(n-j)!} \hat{\Pi}_j(\eta, 0).  \label{eq:nonidealPinNO}
}
Inserting the Fock state resolution of the identity, one finds
\be\label{eq:PNRD_POVM_els}
    \hat{\Pi}_j(\eta, 0) = \sum_{k=j}^\infty 
    \binom{k}{j} \eta^j (1-\eta)^{k-j} \ketm{k}\!\bram{k}.
\ee 
Thus, if $\eta < 1$, there are contributions where there were actually $k > n$ photons yet they masqueraded as an $n$-photon detection. 
This PNR detector model in the absence of dark counts, $d=0$, is equivalent to an AD channel with efficiency $\eta$ before an ideal detector, $\hat{\Pi}_n = \ketm{n}\!\bram{n}$. This property allows us to nicely exploit the properties of AD channels in consonance with measurement.

Using \eqref{eq:PNRD_POVM_els} one can put the POVM elements of \eqref{eq:nonidealPinNO} into the form of \eqref{eq:PikFockRep} with the weights 
\be
    \varpi_{n,k} \ra e^{-d} d^n \g^k
    \sum_{j=0}^{\min(n,k)}
     \frac{\binom{k}{j}}{(n-j)!} 
     \p{\frac{\eta}{d \g}}^j
\ee
and $\g = 1 - \eta$. 
For $n=0$ and $n=1$ the respective full POVM elements are 
\subalign{ 
    \hat{\Pi}_0(\eta, d) &= e^{-d} \sum_{k=0}^\infty \g^k \ketm{k}\!\bram{k}, \\
    \hat{\Pi}_1(\eta, d) &= e^{-d} \sum_{k=0}^\infty
    \g^{k-1}
    \p{ d \g +  k \eta } \ketm{k}\!\bram{k}. \label{eq:1ClickPNRDElement}
}

\subsection{Threshold detectors}\label{app:thresholdDets}
Meanwhile, a threshold detector cannot resolve photon number, it either clicks if any light is detected or does not with respective POVM elements
\begin{subequations}\label{eq:PiClickNoClick}
\begin{align}
    \hat{\Pi}_\textrm{click}(\eta) &= \sum_{n=1}^\infty \hat{\Pi}_n(\eta,d) 
    = \mathbbm{1} - \hat{\Pi}_0(\eta,d) \nonumber \\
    &= \sum_{k=0}^\infty \p{1 - e^{-d} \g^{k}} \ketm{k}\!\bram{k}, \label{eq:PiClick} \\
    \hat{\Pi}_\textrm{no-click}(\eta) &= 
    \hat{\Pi}_0(\eta, d) = e^{-d} \sum_{k=0}^\infty \g^{k} \ketm{k}\!\bram{k}. \label{eq:PiNoClick}
\end{align}
\end{subequations}

\subsection{Threshold array detectors}\label{app:arrayDets}
A class of effective detectors with intermediate photon number-resolving capabilities consists of replacing a single threshold detector by an array of $N_d$ threshold detectors. Namely, we assume the signal to be measured is split up uniformly across $N_d$ paths whose outputs are measured using threshold detectors, which are taken to be identical for simplicity. This splitting can be accomplished using a diffractive element and will come with its own loss mechanisms, which we take to be subsumed into the detector efficiencies.

Following \refref{sperling2012true}, the POVM element for detecting $n$ clicks with $N_d \geq n$ multiplexed threshold detectors with efficiencies $\eta$ and dark counts $d$ is
\be\label{eq:arrayPOVMElement}
    \hat{\Pi}_{N_d,n} = \binom{N_d}{n} : \p{ e^{-\mathfrak{W}} }^{N_d-n} \p{ \mathbbm{1} - e^{-\mathfrak{W}} }^n : 
\ee
with $\mathfrak{W} = \frac{\eta}{N_d} \hc{a}a + d$.
After some algebra, one can express this in the form of \eqref{eq:PikFockRep} with the weights 
\begin{align}
    \varpi_{n,k} \ra c^{(k)}_{N_d,n}
    = \binom{N_d}{n} &\sum_{j=0}^n \binom{n}{j} (-1)^{n-j} e^{-d(N_d-j)} 
    \nonumber \\
    &\times 
    \p{ \g + \frac{\eta j}{N_d} }^k.
\end{align}
In the case of a single detector, $N_d=1$, one simply reobtains \eqref{eq:PiClickNoClick}. 
Meanwhile, in the $N_d \ra \infty$ limit one finds
\be
    \lim_{N_d \ra \infty} \hat{\Pi}_{N_d,n}(\eta, d/N_d) 
    = \hat{\Pi}_n(\eta, d),
\ee
so for large $N_d$ one simply approaches the expression for a noisy PNR detector assuming that if present, the dark counts are  ``distributed'' in a manner such that $d_\textrm{array} = d_\textrm{PNR}/N_d$. 

In practice, the dark counts will not be distributed, so $N_d$ should not be increased without bound (even without physical and monetary considerations) as then the dark counts would dominate the measurements.
Namely, suppose we want to detect $n$ photons with such an array detector. Given a fixed $\eta$ and $d$, the optimal value of $N_d$ can be defined as the one that maximizes the $k=n$ weight, $c^{(n)}_{N_d,n}$, relative to all of the $n$-click weights.
That is, we want to maximize 
\be
    \mathcal{C}_n(x) = \frac{c^{(n)}_{x,n}}{\sum_{k=0}^\infty c^{(k)}_{x,n}},
\ee
where we replaced $N_d \ra x \in \mathbbm{R}^+$.
This can equally be understood as maximizing the ratio of true positives to overall positives or the Hilbert-Schmidt inner product between the target POVM element, $\kbm{n}{n}$, and the unit-trace normalized actual POVM element.

This optimization is analytically tractable for $n=1$ and can be done numerically more generally. 
In the $n=1$ case, which is relevant for the HSPSs we are concerned with,
\be
    \mathcal{C}_1(x) = \frac{ \eta (x-1) \br{ \g (e^d - 1) x + \eta e^d } }{ x \br{1 + (e^d-1) x} },
\ee
which is maximum at $x = x^*$ with
\be
    \frac{1}{x^*} = \sqrt{(e^d - 1) (e^d + 1 - 1/\eta)} - (e^d-1).
\ee 
Thus, within this model, one should take $N_d$ to be $x^*$ rounded to the nearest integer. 
   
\section{Biphoton source theory}\label{app:sourceTheory}

\subsection{Biphoton source Hamiltonian}\label{app:biphotonSources}
SPDC (SFWM) involves the interaction of three (four) waves within a $\chi^{(2)}$ ($\chi^{(3)}$) nonlinear medium, in which, with some probability, one (two) pump photon(s) of frequency $\om_p$ is (are) converted into two new photons, a signal and an idler of frequencies $\om_s$ and $\om_i$, respectively. These produced photons are constrained by energy conservation 
($\om_s + \om_i = \om_p$ for SPDC and
$\om_s + \om_i = 2 \om_p$ for SFWM).
In both cases, the Hamiltonian for the process is of the phenomenological form ($\hbar = 1$)
\be
    H_\textrm{bip} = i k \p{a_s a_i e^{-i \varphi} - \hc{a_s} \hc{a}_i e^{i \varphi}},
\ee
where $\hc{a}_{s,i}$ and $a_{s,i}$ are the bosonic creation and annihilation operators for the signal, idler modes. The realization of this precise form requires a tailored experimental implementation, e.g., the nonlinearity should be sufficiently weak to be in a perturbative regime, the pump field should be a strong, effectively non-depleting, coherent state, and the process should be phase matched. We assume such considerations are taken, see Refs.~\citenum{vernon2015spontaneous} and \citenum{signorini2020chip} for the underlying theory and a review of such sources, respectively.  

The corresponding time evolution operator is
$S = e^{i H_\textrm{bip} t}$, which is nothing but the two-mode squeezing operator, that can be factored into the form \cite{schumaker1985new}
\begin{align}
    S(\lambda) 
    &= \sqrt{1 - |\lambda|^2} e^{\lambda \hc{a_s} \hc{a}_i} \nonumber \\ 
    & \times e^{\p{ \hc{a}_s a_s + \hc{a}_i a_i} \ln\p{\sqrt{1 - |\lambda|^2}} } e^{- \lambda^* a_s a_i}.
\end{align}
Here $\lambda \equiv e^{i \varphi} \tanh{r}$ and $r=k t$ a squeezing strength parameter that is proportional to $\chi^{(3)} P_p$ for SFWM and to $\chi^{(2)} \sqrt{P_p}$ for SPDC with $P_p$ the pump power.
This form elucidates the action of $S$ on the vacuum, $\vac$, namely, it generates the two-mode-squeezed vacuum state of \eqref{eq:TMSW_pure}.
As shown in \secref{sec:lossyBiphotonSources}, detecting the  idler output heralds the state of the signal mode, which with some probability should be a single photon.

Then, $N$ such heralded single-photon sources (HSPSs) can be used to prepare the necessary $N$ photon input state, $\ketm{\vec{N}}$, for an $N$P$M$M HBSG circuit.
The overall state of the signal and idler modes before heralding can be generated via $N$ disjoint squeezers, $\comm{S_i}{S_j} = 0$: 
\be\label{eq:MMSV}
    \ket{\psi_0} = \Motimes_{j=1}^N S_j(\lambda_j) \vac
    = \Motimes_{j=1}^N \ket{\textrm{TMSV}(\lambda_j)}.
\ee
Alternatively, a single quantum frequency comb can be used to generate such multi-mode squeezed vacuum states  within a frequency-bin encoding \cite{qin2026high}.

\subsection{Heralded single photons from biphoton sources}\label{app:biphotonsForHSPSs}
After the multi-mode squeezed vacuum state $\rho_0 = \ketbra{\psi_0}{\psi_0}$ of \eqref{eq:MMSV} is generated, both the signal and idler modes propagate through lossy components, such as unloading microrings and output waveguides.
Then, the idler modes are measured with the goal of heralding a target state of the signal modes, $\ketm{\vec{N}}$.
The heralding succeeds if we obtain the desired ``all-click'' pattern on the detected idler modes with corresponding POVM element $\hat{\Pi}_c$.

As the efficiencies of single-mode loss channels are multiplicative, we can capture these losses via a 
single AD channel for each of the signal and idler modes, $j \in \{1, 2, \cdots, N \}$, with overall efficiencies $\eta_s^{(j)}$ and $\eta_i^{(j)}$, respectively. The precise forms of $\eta_{s,i}^{(j)}$ in terms of underlying physical parameters will depend on how the HSPSs are implemented.
Moreover, the losses on the heralding idler modes can be subsumed into the corresponding detector POVM elements by modifying their efficiencies: 
$\eta^{(j)}_d \ra \eta^{(j)}_{id} = \eta_i^{(j)} \eta^{(j)}_d$. 
Meanwhile, loss on the signal modes commutes with the heralding idler measurement and so can be applied to the state before or after, we opt for after.

We first evaluate $\tilde{\rho}_s$, the state of the signal modes after the heralding measurement with the idler loss subsumed into $\hat{\Pi}_c$, yet before the signal mode loss. Assuming the detectors are independent, $\hat{\Pi}_c$ can be decomposed in terms of a tensor product of the POVM elements for each heralding detector as 
\be\label{eq:PicTensorProduct}
    \hat{\Pi}_{c} = \Motimes_{j = 1}^{N} \hat{\Pi}^{(j)}_{c}
\ee
with 
\be\label{eq:PijcFockRep}
    \hat{\Pi}^{(j)}_{c} = \sum_{k=0}^\infty w^{(j)}_k \ketm{k_j}_i {}_i\bram{k_j}
\ee  
the POVM element for a click on idler detector $j$ expanded in the Fock basis as in \eqref{eq:PikFockRep} with $n=1$.
The weights, $w^{(j)}_k$, depend on the detector model and idler efficiencies, and are specified explicitly below.
    
Conditioned on a successful heralding, with probability
\be
    p_c = {\Tr}(\rho_0 \hat{\Pi}_c)
    = \bra{\psi_0} \hat{\Pi}_c \ket{\psi_0},
\ee
the heralded state of the signal modes is 
\be\label{eq:rhosGen}
    \tilde{\rho}_s =  \frac{ \Tr_i( \rho_0 \hat{\Pi}_c) }{p_c}.
\ee 	
By leveraging the nice tensor product structure of Eqs.~(\ref{eq:MMSV}) and (\ref{eq:PicTensorProduct}), owed to the independence of the HSPSs and detectors, respectively, one finds
\be
    \tilde{\rho}_s = \Motimes_{j=1}^N \tilde{\varrho}_j
\ee
and 
\be
    p_c = \prod_{j=1}^N p_c^{(j)}
\ee
with 
\subalign{
    \tilde{\varrho}_j &= \frac{1}{p_c^{(j)}}
    \sum_{n=0}^\infty \tilde{\sigma}^{(j)}_{n} 
    \ketm{n_j}_s {}_s\bram{n_j}, 
    \\
    p_c^{(j)} &= \sum_{n=0}^\infty \tilde{\sigma}^{(j)}_{n}, \label{eq:pcjInputStateGen}
    \\
     \tilde{\sigma}^{(j)}_{n} &= \p{1 - |\lambda_j|^2} |\lambda_j|^{2 n} w^{(j)}_n.
}

Now we can account for signal mode loss by applying an AD channel to each of the $N$ modes. As $\tilde{\rho}_s$ is separable and these channels are independent in terms of these modes, we need only determine how $\tilde{\varrho}_j$ transforms under an AD channel with efficiency $\eta_s^{(j)}$.
After some algebra and using \eqref{eq:FockStateAD}, the result is simply a change in the coefficients $\tilde{\sigma}_n^{(j)} \ra \sigma_n^{(j)}$, namely,
\be
    \rho_s = \Motimes_{j=1}^N \varrho_j
\ee
with
\subalign{
    \varrho_j &= \cE_{\eta_s^{(j)}}\!\p{ \tilde{\varrho}_j } = \frac{1}{p_c^{(j)}}
    \sum_{n=0}^\infty \sigma^{(j)}_{n} 
    \ketm{n_j}_s {}_s\bram{n_j},
    \\
    \sigma_n^{(j)} &= \sum_{k=n}^\infty \tilde{\sigma}^{(j)}_{k}  
     \varepsilon_{k,n}(\eta_s^{(j)}), \label{eq:sigmanRhoCoeff}
}
where $\varepsilon_{k,n}$ is defined in \eqref{eq:epsCoeffFockAD}.

\subsection{HSPS quality}\label{app:HSPSQuality}
For each type of detector considered in App.~\ref{app:detectorPOVMs} we now specify the weights $w^{(j)}_k$ of \eqref{eq:PijcFockRep}, then evaluate $p_c^{(j)}$ and $\sigma_n^{(j)}$ of Eqs.~(\ref{eq:pcjInputStateGen}) and (\ref{eq:sigmanRhoCoeff}), respectively.
For brevity, we omit the idler mode index $j$,
taking $p_c^{(j)} \ra p_1$,
and use the shorthands
$\g_{id} = 1-\eta_{id}$, $\g_s = 1 - \eta_s$,
$\xi_i = |\lambda|^2 \g_{id}$,
$\xi_s = |\lambda|^2 \g_s$,
and
$\xi = |\lambda|^2 \g_s \g_{id}$.
Then for the  $n=1$ and/or click POVM elements of Eqs.~(\ref{eq:1ClickPNRDElement}), (\ref{eq:PiClick}), and (\ref{eq:arrayPOVMElement}), respectively, we have 
\begin{widetext}
\begin{subequations}\label{eq:HSPSPNRFull}
    \begin{align}
    w_k^{(\textrm{PNR})} &= e^{-d} \g_{id}^{k} \p{ d + k \frac{\eta_{id}}{\g_{id}} }, \\
    p_1^{(\textrm{PNR})}
    &= e^{-d} \frac{1 - |\lambda|^2}{(1-\xi_i)^{2}} \br{
     d (1-\xi_i)
     + \frac{\eta_{id}}{\g_{id}} \xi_i
    }, \\
    \sigma_n^{(\textrm{PNR})} 
    &= e^{-d} \p{1 - |\lambda|^2} \p{\frac{\eta_s}{\g_s}}^n
    \frac{\xi^n}{(1-\xi)^{n+2}}
    \br{ d (1-\xi) + \frac{\eta_{id}}{\g_{id}} (n+\xi) },
    \end{align}
\end{subequations}
\subalign{
    w_k^{(\textrm{threshold})} &= 1 - e^{-d} \g_{id}^{k}, \\
    p_1^{(\textrm{threshold})} 
    &= 1 - e^{-d} \frac{1 - |\lambda|^2}{1-\xi_i}, \\
    \sigma_n^{(\textrm{threshold})} 
     &= \p{1 - |\lambda|^2} \p{\frac{\eta_s}{\g_s}}^n
     \br{
         \frac{\xi_s^n}{(1-\xi_s)^{n+1}} 
        - e^{-d}  \frac{\xi^n}{(1-\xi)^{n+1}}
     }, 
}   
and
\begin{subequations}\label{eq:HSPS_arrayFull}
    \begin{align}
         w_k^{(\textrm{array})} &= N_d e^{-N_d d} 
    \br{ e^{d} \p{ \g_{id} + \eta_{id}/N_d }^k - \g_{id}^k }, \\
    p_1^{(\textrm{array})} 
    &= N_d e^{-N_d d} \p{1 - |\lambda|^2} \br{
        \frac{e^{d}}{1-\xi_i - |\lambda|^2 \eta_{id}/N_d }
        - \frac{1}{1-\xi_i}
    }, \\
    \sigma_n^{(\textrm{array})}
    &= N_d e^{-N_d d}
    \p{1 - |\lambda|^2} \p{\frac{\eta_s}{\g_s}}^n
    \br{
        e^{d} \frac{(\xi + \eta_{id} \xi_s/N_d)^n}{(1-\xi - \eta_{id} \xi_s/N_d)^{n+1}}
        - \frac{\xi^n}{(1-\xi)^{n+1}}
    }.
    \end{align}
\end{subequations} 
\end{widetext}

\begin{figure}[ht!]
    \includegraphics[width=0.9\linewidth, clip=true, trim = 15mm 0 0 0]{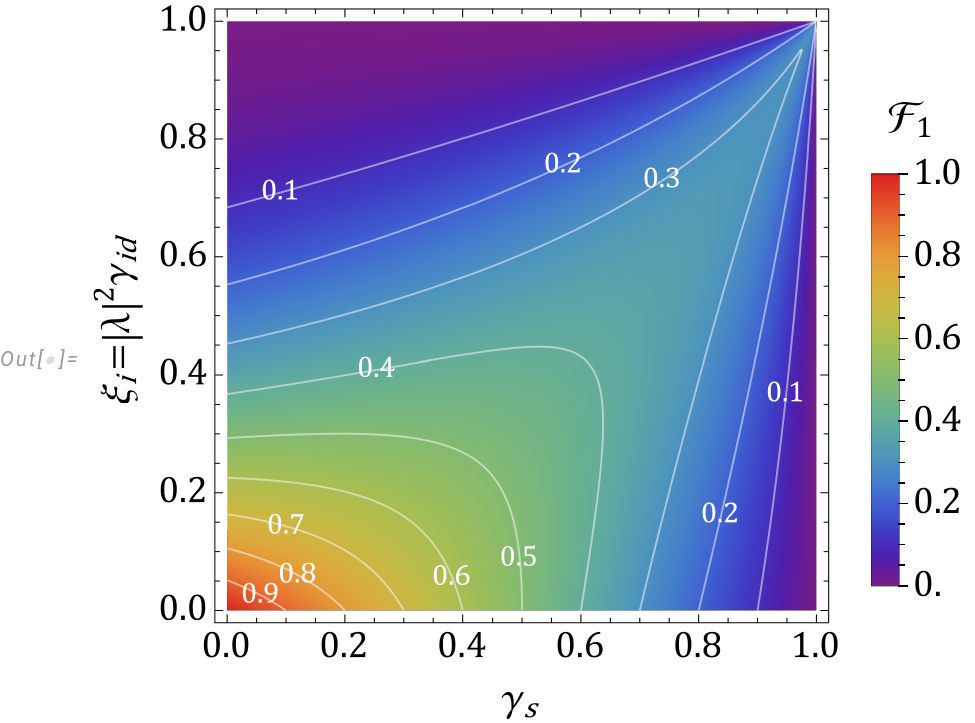} 
    \caption{HSPS fidelity of \eqref{eq:HSPS_fidelity}.
    }\label{fig:HSPS_fidelity}
\end{figure}

Focusing on the case of PNR detectors with $d=0$, from \eqref{eq:HSPSPNRFull} one finds the HSPS state considered in the main text, \eqref{eq:heraldedSPS}. The corresponding single photon fidelity is
\be\label{eq:HSPS_fidelity}
    \cF_1 = \bram{1} \varrho_1 \ketm{1}
    = c_1
    = \eta_s \frac{ \p{1 - \xi_i}^2 \p{1 + \xi } }{ \p{1-\xi}^{3} },
\ee
which is plotted in \figref{fig:HSPS_fidelity}.
One can further calculate the average number of photons in the heralded state to be 
\be
    \bar{n}_{\varrho_1} = \sum_{n=0}^\infty n c_n 
    = \eta_s \frac{1 + \xi_i}{1 - \xi_i},
\ee
which is lowered by signal loss, $\eta_s < 1$, and raised by idler loss via $\xi_i = |\lambda|^2 \g_{id}$, 
so evidently one can simply decrease $|\lambda|$ to reduce multiphoton errors.

For reasonably small photon loss rates ($\g_s < 1/2$), $\cF_1$ is a monotonically decreasing function of $|\lambda|$ yet the probability of heralding, $p_1$, is $0$ at $|\lambda| = 0$ and has a maximum value of $1/4$ at the intermediate value $|\lambda| = 1/\sqrt{2 - \g_{id}} \geq 1/2$. 
Thus, there is a tradeoff between fidelity and rate. One simple way of compromising is to find the value of $|\lambda|$ that maximizes the product $\cF_1 p_1$ (if one cares more about $\cF_1$ or $p_1$ this can be adjusted by changing the relative powers of each in the product).
The corresponding maximum of this product occurs at 
\be\label{eq:lambdaSqYieldMax}
    |\lambda|^2 = \p{ 1 - 2 \g_{id} \g_s + \sqrt{1+3 \g_{id}^2 \g_s^2} }^{-1}.
\ee	

\subsection{Implications of dark counts}
In-depth analysis of dark counts and the potential for array detectors is beyond the scope of this work, though we have set up the basic machinery. We do, however, note that the dominant effect of dark counts depends on where they occur.

Dark counts in the HSPS detectors lead to one thinking there was a click when there was not, thus leading to an increased probability of injected vacuum into the circuit and thus acting similar to signal mode loss.
Namely, as in \eqref{eq:heraldedSPS} we can expand the Fock expansion coefficients for the noisy heralded single photons for small $\g_{id}, \g_s \sim \g$. To include dark counts while accounting for the fact that they are rare compared to photon loss we take $d \sim \g^2$, which likely overestimates their impact, yet allows us to heuristically understand their effects.
For PNR detectors, with $c_n \ra \sigma_n^{(\textrm{PNR})} /p_1^{(\textrm{PNR})}$ from \eqref{eq:HSPSPNRFull}, under this model the only change to the coefficients of \eqref{eq:HSPS_coefficients} to $\cO(\g^2)$ is to increase the vacuum coefficient $c_0$ by $d/|\lambda|^2$ and correspondingly reduce $c_1$.

Meanwhile for threshold array detectors, if one takes the number of detectors to be large, $N_d \sim 1/\g$ (such that $N_d d \sim \g$), $c_n \ra \sigma_n^{(\textrm{array})} /p_1^{(\textrm{array})}$ can be computed from \eqref{eq:HSPS_arrayFull} to leading order as
\subalign{
    c_0 &= \g_s + \frac{N_d d}{|\lambda|^2} + \cO(\g^2), \\
    c_1 &= 1 - \sum_{n \neq 1} c_n, \\
    c_2 &= 2 \xi_i + \frac{|\lambda|^2}{N_d} + \cO(\g^2), \\
    c_{n \geq 3} &= \cO(\g^{n-1}).
}
Thus, for array detectors dark counts lead to an increased rate of false-positive vacuum injection by $N_d$ as well as multiphoton contributions, suppressed to $\cO(1/N_d)$, due to cases where multiphoton events are not split among different detectors in the array. 

Conversely, dark counts in the post-circuit heralding will lead to false positives inducing multiphoton contributions. For instance, suppose we are in the branch of the 4P5M scheme where a single photon is to reach the heralding signal mode detector yet a dark count occurs leading to us registering a 2-photon detection event. Then the output Bell modes will acquire a 3-photon contribution  (in the absence of other losses). 

\section{HBSG with heralded single photons}

\subsection{Convergence in terms of $\texorpdfstring{\bm{n}}{n}_\textrm{extra}$}\label{sec:convergenceWithNExtra}

\begin{figure*}[ht!]
    \captionsetup[subfloat]{captionskip=-5pt}
    \subfloat[(a) $\g_{id} = 0.01$]{\includegraphics[width=0.33\linewidth]{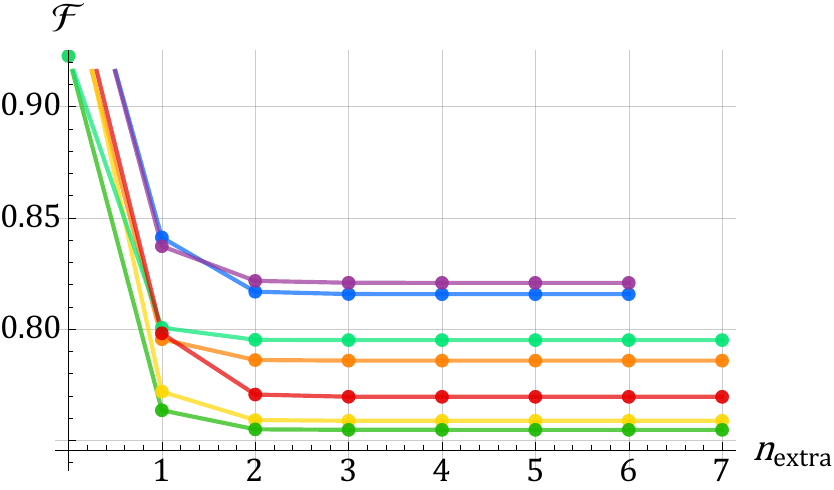}} \hfill
    \subfloat[(b) $\g_{id} = 0.05$]{\includegraphics[width=0.33\linewidth]{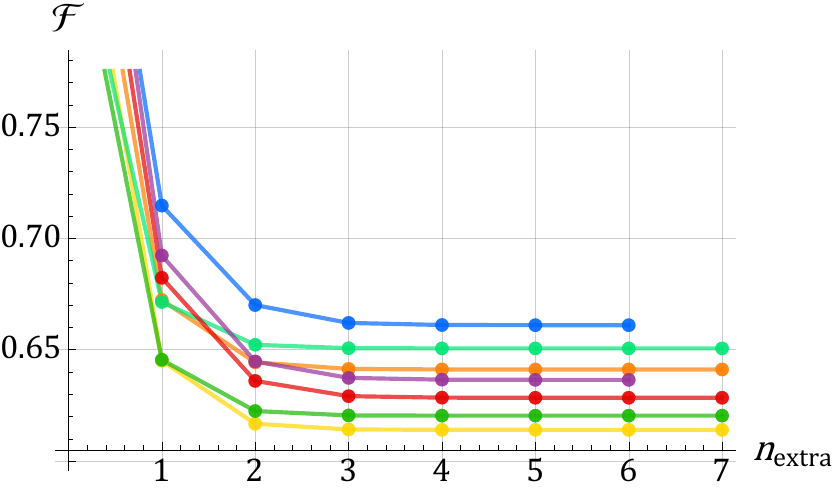}} \hfill
    \subfloat[(c) $\g_{id} = 0.10$]{\includegraphics[width=0.33\linewidth]{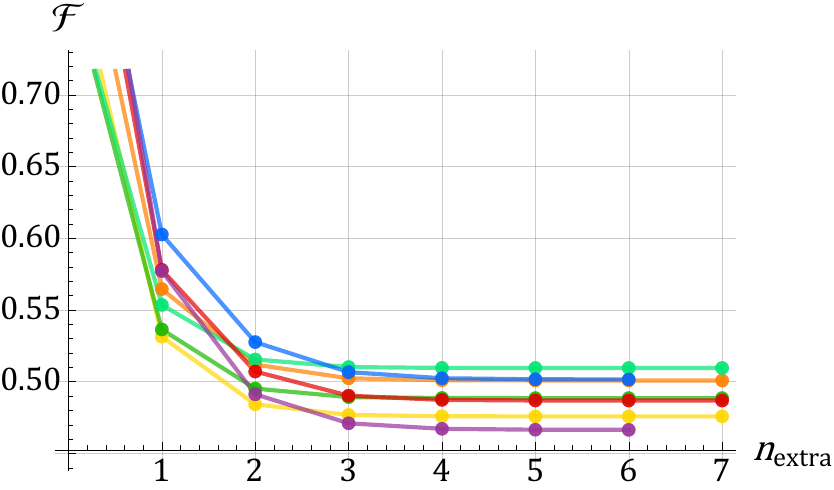}} 
    \\
    \captionsetup[subfloat]{farskip=0pt, captionskip=-5pt}
    \subfloat[]{\includegraphics[width=0.33\linewidth]{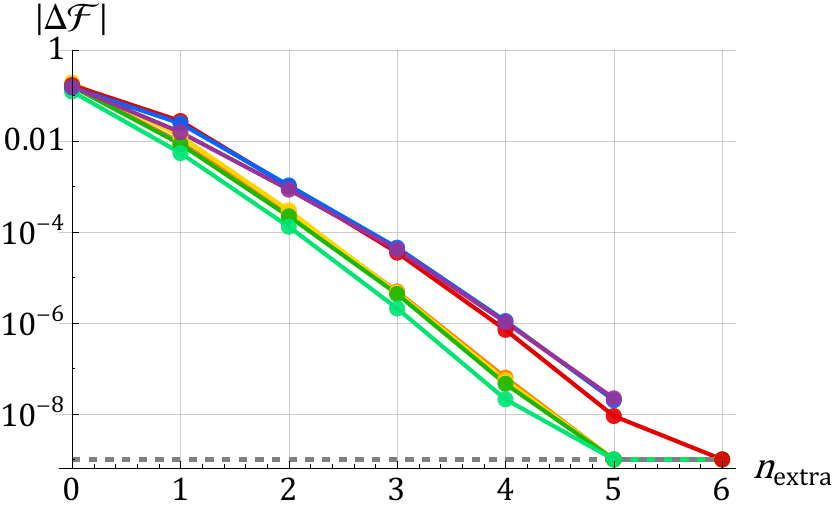}} \hfill
    \subfloat[]{\includegraphics[width=0.33\linewidth]{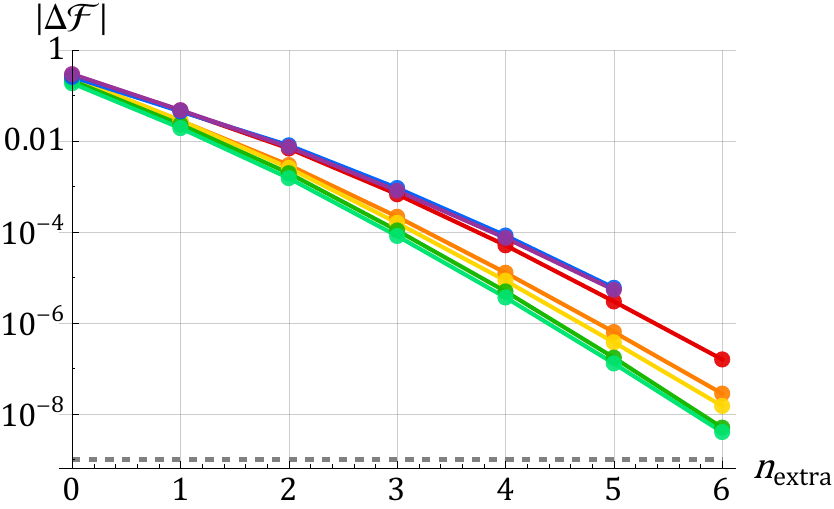}} \hfill
    \subfloat[]{\includegraphics[width=0.33\linewidth]{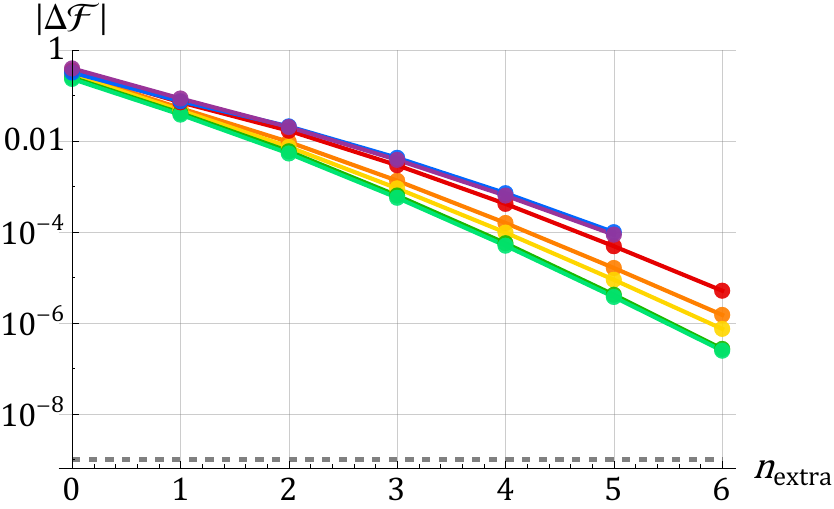}} 
    \\ \vspace{3mm}
    \subfloat[(d)]{\includegraphics[width=0.33\linewidth]{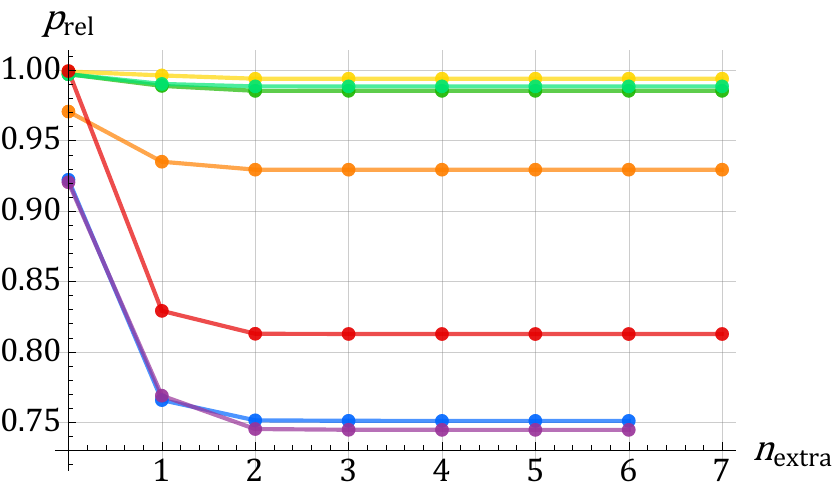}} \hfill
    \subfloat[(e)]{\includegraphics[width=0.33\linewidth]{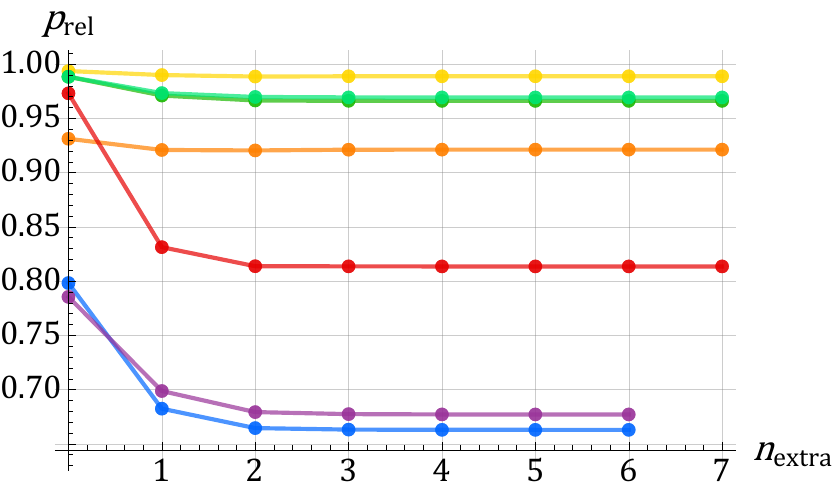}} \hfill
    \subfloat[(f)]{\includegraphics[width=0.33\linewidth]{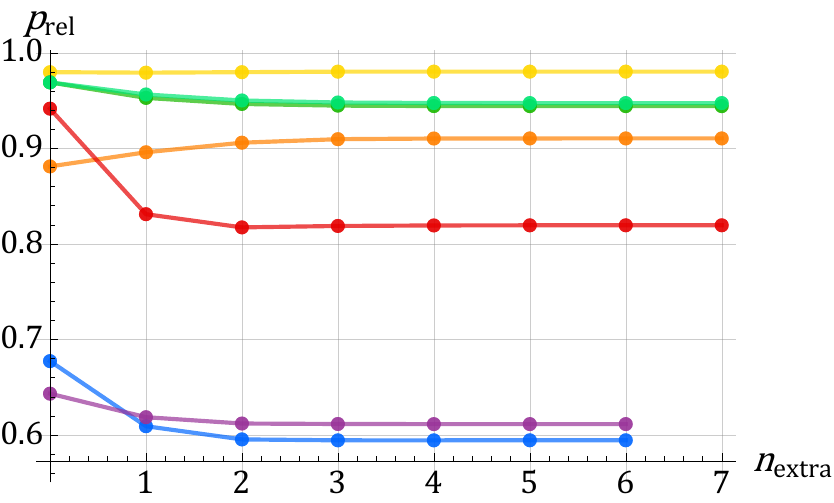}} 
    \\
    \captionsetup[subfloat]{farskip=0pt, captionskip=-5pt}
    \subfloat[]{\includegraphics[width=0.33\linewidth]{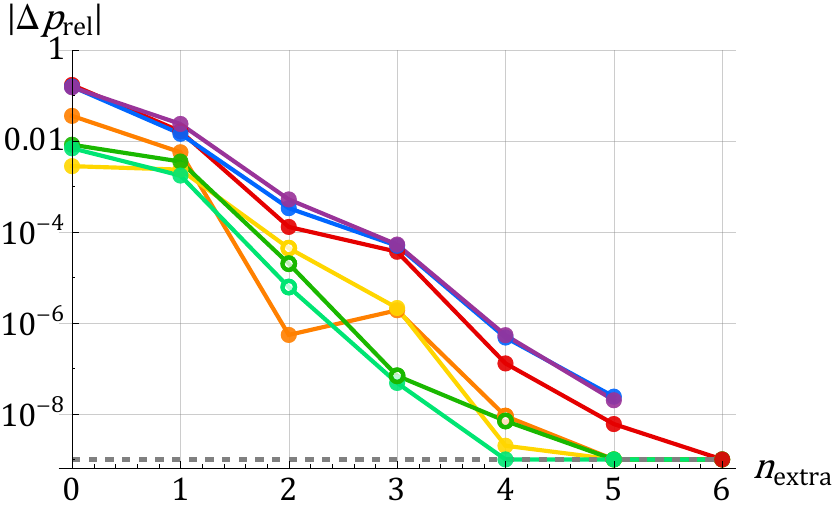}} \hfill
    \subfloat[]{\includegraphics[width=0.33\linewidth]{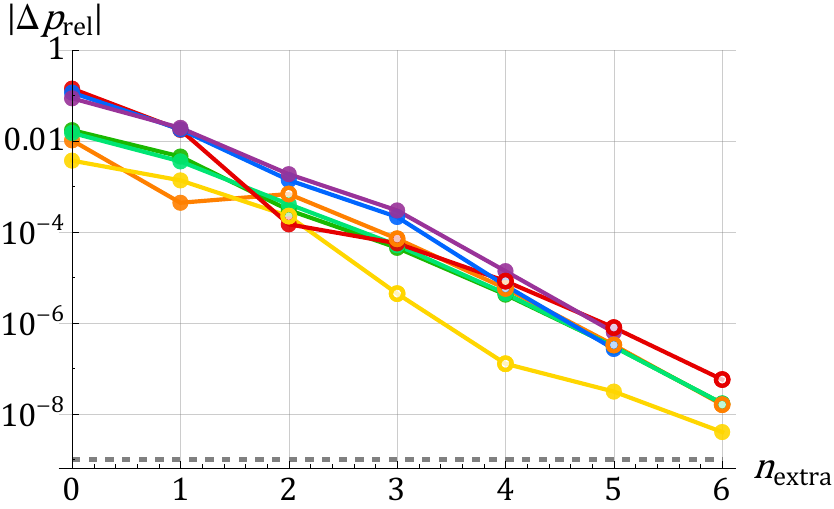}} \hfill
    \subfloat[]{\includegraphics[width=0.33\linewidth]{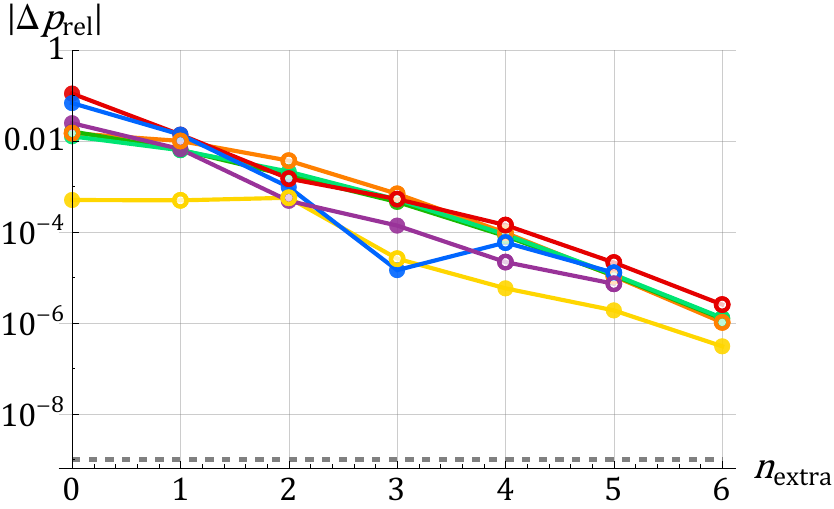}} 
    \\ \vspace{3mm}
    (g)\; \raisebox{10pt}{\subfloat[]{\includegraphics[width=0.7\linewidth, clip=true, trim = 15mm 0 0 0]{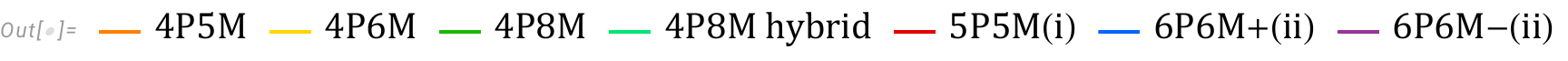}}}
    \caption{
       Plot illustrating the convergence of (a,b,c) the Bell state fidelity and (d,e,f) the relative heralding probability for each circuit shown in \figref{fig:BSG_circuits} as $n_\textrm{extra}$ is increased for three different multiphoton regimes, dictated by $\g_{id}$ values of $0.01$, $0.05$, and $0.10$, respectively.
       Under each fidelity (probability) plot we give a plot of the magnitude of the differences between the subsequent and current fidelities (probabilities) using solid disks and open circles to denote negative and positive differences, respectively. 
       These values were only exported to nine decimal places, so differences of zero, which occur for $\g_{id} = 0.01$, are replaced with that $10^{-9}$ precision (dashed gray line).
       (g) Scheme legend.
    }\label{fig:nExtraConvergence}
\end{figure*}

Here we investigate how the circuits behave as the truncation parameter $n_\textrm{extra}$ of \eqref{eq:nvecCutoffCriteria} is increased as is needed to capture the impact of higher-order multiphoton contributions.
The simulations get more expensive in terms of memory and runtime as $n_\textrm{extra}$ is increased. Accordingly, for the $N=6$ schemes we go up to $n_\textrm{extra}$ of 6, whereas we go to 7 for the $N=4,5$ schemes. Moreover, to keep things manageable, we focus on the frequency-domain implementation of \secref{sec:frequencyHBSG} and fix the values: $\Omega = 100$ and $\gamma_s = 0.03$ as a reasonably representative case.
The resulting fidelity convergence is shown in \figref{fig:nExtraConvergence} for several values of $\gamma_{id}$ ($1\%$, $5\%,$ and $10\%$), which dictates the multiphoton error parameter $\xi_i = |\lambda|^2 \g_{id}$, where $\lambda$ is fixed based on \eqref{eq:lambdaSqYieldMax} as in the main text.

Therein, we also plot the magnitude of the fidelity difference 
$\Delta\cF(n_\textrm{extra}) = \cF(n_\textrm{extra}+1) - \cF(n_\textrm{extra})$, 
which more precisely shows the extent to which the fidelity is converging. In particular, we see that $|\Delta\cF|$ decreases approximately linearly on the log plots, which implies exponential convergence of $\cF$ in terms of $n_\textrm{extra}$. 
We see that with additional multiphotons the convergence occurs more slowly just as one would expect, i.e., as $\gamma_{id}$ is increased it is necessary to increase $n_\textrm{extra}$ to retain a given level of accuracy because the corresponding multiphoton contributions are more significant, see \eqref{eq:HSPS_coefficients}. Additionally, the higher $N$ schemes tend to converge more slowly.

We also plot the corresponding relative probabilities $p_\textrm{rel} = p_\textrm{her}/p_\textrm{ideal}$ and their differences $\Delta p_\textrm{rel}$, defined analogously to $\Delta\cF$, which  are monotonically decreasing. In some cases, we see that $|\Delta p_\textrm{rel}|$ is not strictly decreasing as $n_\textrm{extra}$ is increased. For instance, in \figref{fig:nExtraConvergence}(f) we see that for the 6P6M$+$ scheme $|\Delta p_\textrm{rel}|$ decreases up until $n_\textrm{extra} = 3$ whence it is $\cO(10^{-5})$, then it increases to $\cO(10^{-4})$ in a manner such that $p_\textrm{rel}$ oscillates.
Further analysis would be needed to diagnose this effect. We suspect it is due to some combination of Perceval's working precision and potentially genuine interference from the higher-order multiphoton contributions.
Bounding the impact of such multiphoton truncation on output fidelities and probabilities is a valuable topic for future analyses.

\subsection{Fidelity, average photon number, and relative yield}\label{app:relativeYieldWithHSPSs}
\begin{figure*}[ht!]
    \subfloat[(a) Bell-state fidelity, $\cF$]{
        \stackinset{r}{-10pt}{c}{}{\includegraphics[height=4cm]{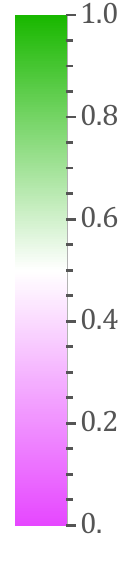}}
        {
        \includegraphics[width=0.95\linewidth, clip=true, trim = 27mm 3mm 25mm 8mm]{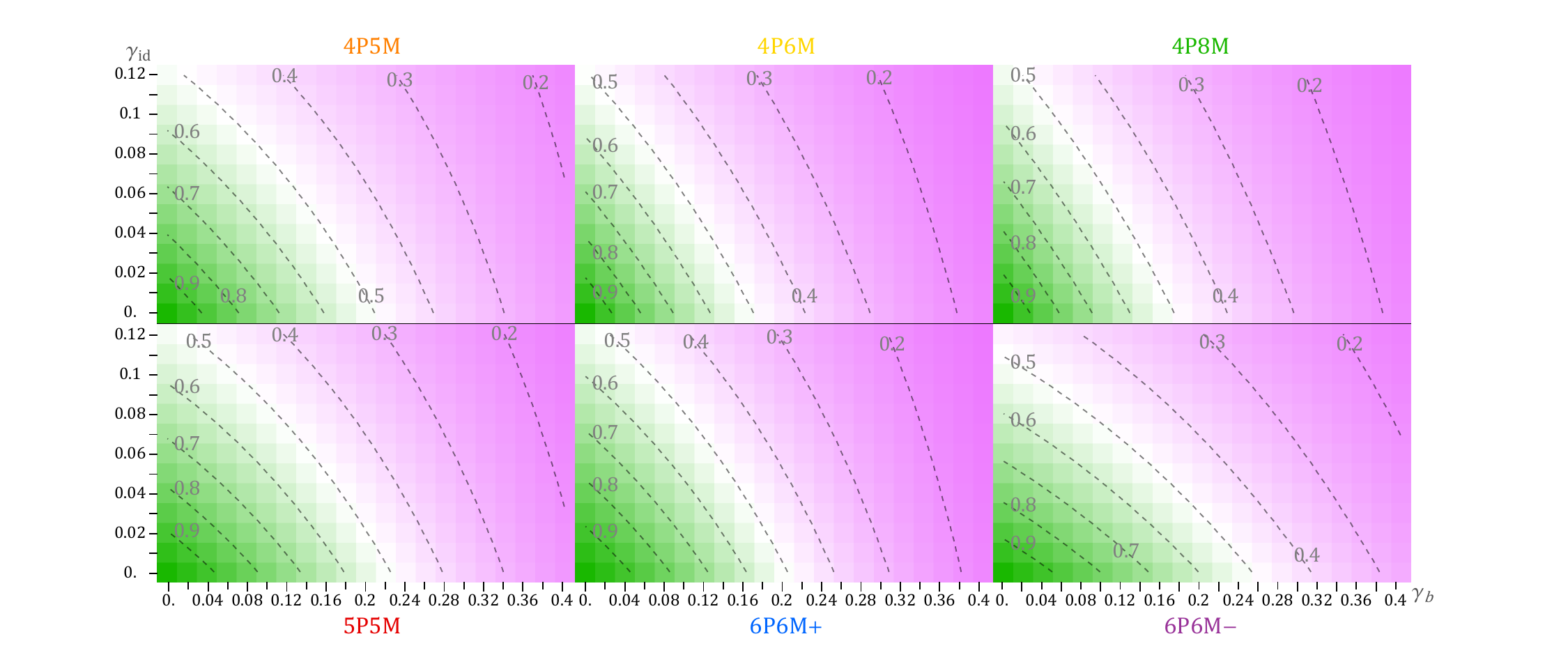}} 
    } 
    \\
    \subfloat[(b) Bell-state average photon number, $\overbar{n}$]{
        \stackinset{r}{-10pt}{c}{}{\includegraphics[height=4cm]{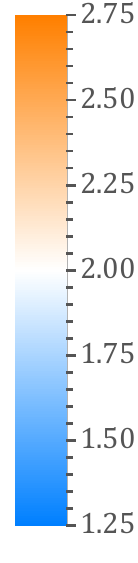}}
        {
        \includegraphics[width=0.95\linewidth, clip=true, trim = 27mm 3mm 25mm 8mm] {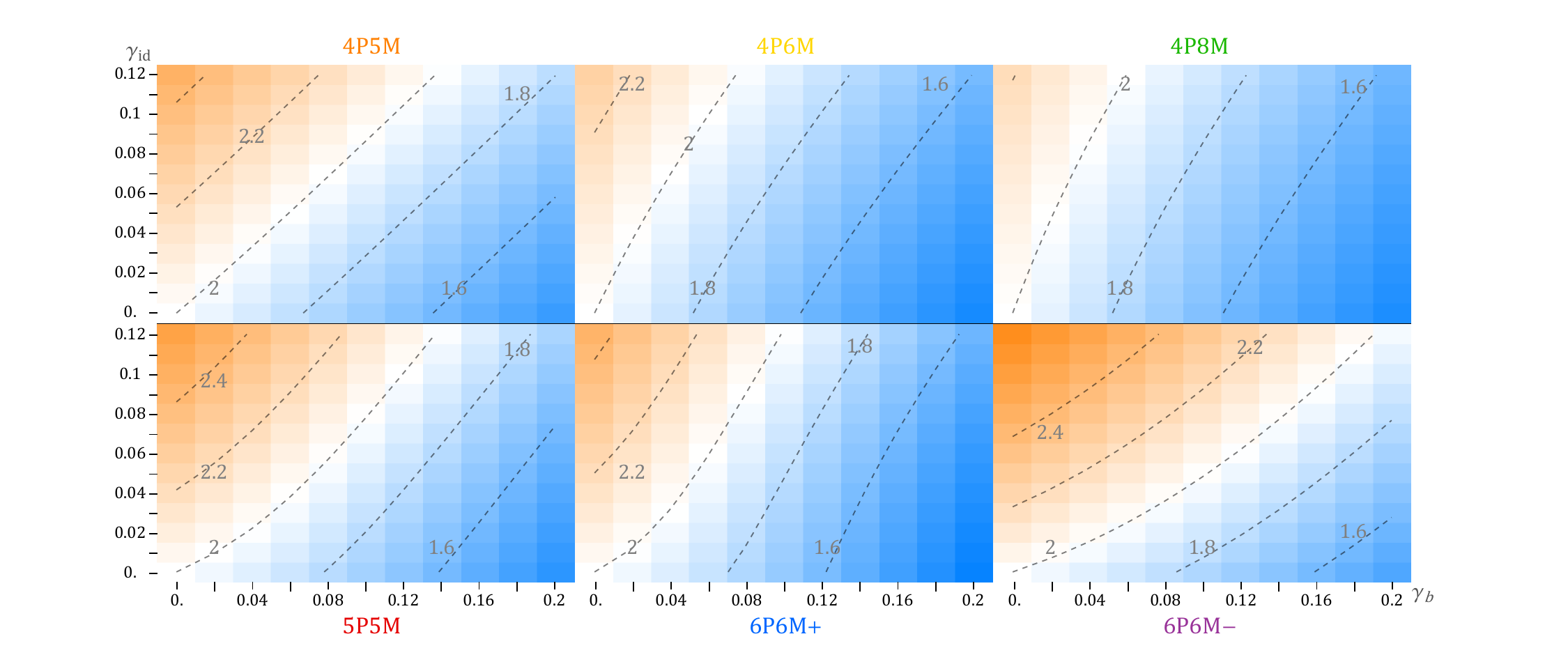}
        }
    } 
    \caption{
        For each HBSG scheme, we construct a density plot of the (a) fidelity and (b) average photon number of the output Bell state, as a function of multiphoton errors, $\g_{id}$ (vertical axis), and Bell mode loss, $\g_b$ (horizontal axis; note the different domains). 
    }\label{fig:lumped_circuit_fidelity_and_nBar} 
\end{figure*}

In \figref{fig:lumped_circuit_fidelity_and_nBar} we 
report the fidelities for each HBSG scheme using HSPSs under the lumped circuit loss model. The corresponding data was used to generate \figref{fig:lossyHSPSsLumpedCircuitLossFidelityComparison}.
We also report the average photon number of the output Bell state,
\be
    \overbar{n} = \Tr\bigg( \rho_\tout \sum_j \hc{a}_j a_j \bigg)
\ee
with the sum over the output modes, as an overall metric characterizing the state's photon number makeup. 

We see that although the fidelities have similar shapes across the various HBSG schemes, the corresponding photon number distributions vary appreciably. Fidelity degradation is ultimately due to false positive detections (as well as direct amplitude damping of the output state). We see that some of the schemes, 4P8M and 4P6M, preferentially filter out false-positives that lead to multiphoton errors in which there are more than the expected two photons in the output state. Meanwhile, other schemes, 6P6M$-$ and 5P5M, have larger (orange) regions in which there are significant multiphoton errors in the output state, e.g., $\overbar{n} > 2$.
We note that $\overbar{n}$ is a limited metric in that it does not inform one of the state's structure as a mixture of different photon number sectors nor across the output modes.  
Further study of this underlying structure and of how multiphoton contributions cascade under operations like fusions and, together with loss, lead to future false positives is left to future works.

\begin{figure*}[htb]
    \subfloat[(a) $\g_b = 0$]{\includegraphics[width=0.375\linewidth]{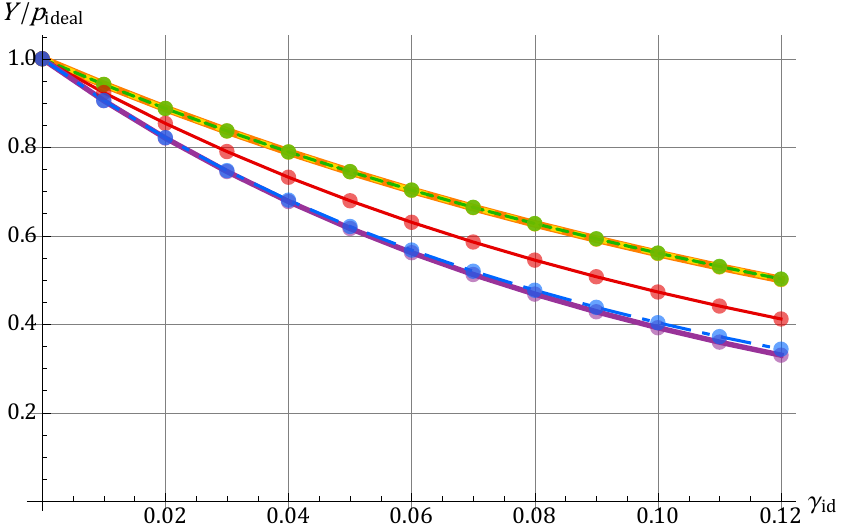}} \hfill
    \subfloat[(b) $\g_b = 0.06$]{\includegraphics[width=0.375\linewidth]{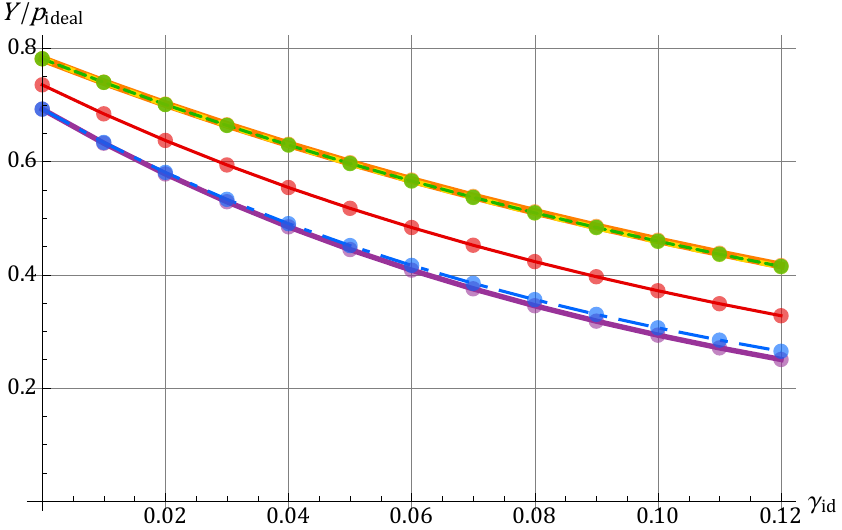}} \hfill
    \subfloat[(c)]{\includegraphics[width=0.225\linewidth]{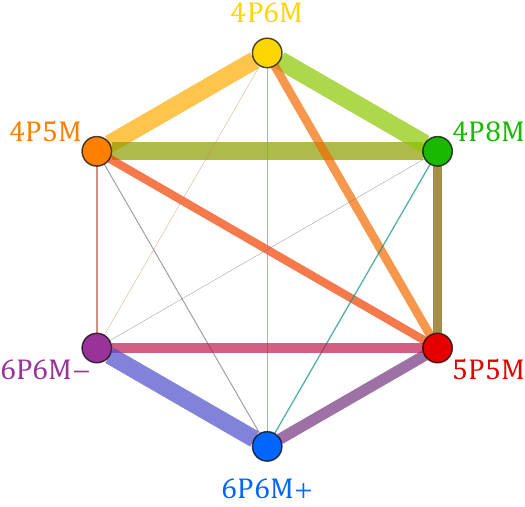}}
    \caption{
        Cross-sections of the relative yield, $Y(\g_b, \g_{id})/p_\textrm{ideal}$, of each circuit for (a) $\g_b = 0$ and (b) $\g_b = 0.06$. 
        (c) A weighted graph with vertices representing the various HBSG circuits and the weights, illustrated via edge thickness, indicating how similar the relative yields are for each pair of circuits. Thicker (thinner) edges indicate circuit pairs with similar (well separated) relative yield landscapes. This similarity is measured by the square root of the sum of squares of the differences in the relative yields between each circuit across the $21 \times 13$ $(\g_b, \g_{id})$ grid values considered in \figref{fig:lossyHSPSsLumpedCircuitLossFidelityComparison}:
        $\mathcal{M}_{jk} = \sqrt{ \sum_{(\g_b, \g_{id}) \in \textrm{grid}} \br{ \frac{ Y_{j}(\g_b, \g_{id}) }{ p_{j\textrm{-ideal}} } - \frac{ Y_{k}(\g_b, \g_{id}) }{ p_{k\textrm{-ideal}} } }^2 }$ for circuits $j,k$.
        In particular, the edge thicknesses are given by a linearly decreasing function of $\mathcal{M}_{jk}$ with the maximum thickness (minimum $\mathcal{M}_{jk}$) corresponding to the 4P6M $\leftrightarrow$ 4P8M edge.
        The circuits are color coded as indicated in (c), consistent with previous figures.
    }\label{fig:relative_yields_HSPS_lumpedCircuit} 
\end{figure*}

For the amplitude damped single-photon source and lumped circuit loss model we found a precise tradeoff between the probability of post-HBSG-circuit heralding and the fidelity of the output Bell state as quantified by their product, the yield $Y$, see \eqref{eq:lumpedLossYield}. 
Such a tradeoff appears to also be present when using HSPSs in the lumped circuit model.
Therein, the functional structure of the yield relative to the ideal probability of heralding, $Y/p_\textrm{ideal}$, is likewise nearly consistent for circuits with the same number of input photons, $N$. Said relative yield reduces with additional photons, and the probability of HSP generation is not included in the yield as defined, so this reduction is not simply an artifact of HSPSs being probabilistic.

One can see this behavior, at least heuristically, by examining the corresponding relative yield surfaces, $Y(\g_b, \g_{id})/p_\textrm{ideal}$, directly and comparing them, analogous to the fidelity plots of \figref{fig:lumped_circuit_fidelity_and_nBar}(a). For ease of presentation, in \figref{fig:relative_yields_HSPS_lumpedCircuit} we simply give two cross-sections for fixed $\g_b$ values seeing that the relative yield varies little within circuits of a fixed $N$ but appreciably for different $N$. This property appears to hold across all the $\g_b$ considered as is illustrated in \figref{fig:relative_yields_HSPS_lumpedCircuit}(c). 

\subsection{Higher-$\texorpdfstring{\bm{N}}{N}$ input flexibility and multiphoton-error robustness}\label{app:higherNFlex}
As alluded to in \appref{app:HBSG_extra}, if we do not restrict to the prescribed number of input photons, $N$, for each circuit, there are additional inputs $\ketm{\vec{N}}$ that can be passed through these circuits to herald the generation of Bell-like states.
For each circuit we search exhaustively over all possible input configurations with $N \leq 10$ and all potentially valid heralding click patterns for Bell-like states using Perceval. We find that there are many such inputs, which we list below in the form $(\ketm{\vec{N}}, p_\textrm{ideal}) \times n_\textrm{perms}$, where $p_\textrm{ideal}$ is the ideal probability of heralding a Bell-like state given input $\ketm{\vec{N}}$ (rounded to the nearest $0.01\%$), and $n_\textrm{perms}$ is the total number of Bell-heralding permutations of the mode occupation numbers in $\vec{N}$ (omitted if 1). 
\\

\textbf{4P5M}:
$(\ketm{11110}, 11.11\%)$,
$(\ketm{11115}, 0.96\%)$,
$(\ketm{22221}, 1.80\%)$ 

\textbf{4P6M}:
$(\ketm{011110}, 7.41\%) \times 4$ 

\textbf{4P8M}:
$(\ketm{00001111}, 18.75\%) \times 16$,
$(\ketm{00002222}, 1.85\%) \times 16$,
$(\ketm{00001117}, 0.04\%) \times 64$, 
$(\ketm{00001135}, 0.03\%) \times 192$, 
$(\ketm{0001 2232}, 0.08\%) \times 192$, 
$(\ketm{0011 1133}, 0.12\%) \times 96$ 

\textbf{5P5M}:
$(\ketm{11111}, 9.60\%)$,
$(\ketm{11112}, 3.84\%) \times 5$,
$(\ketm{22222}, 2.58\%)$

\textbf{6P6M}:
$(\ketm{111 111}, 14.81\%)$,
$(\ketm{111 112}, 4.32\%) \times 6$,
$(\ketm{112 112}, 1.65\%) \times 9$, 
$(\ketm{111 222}, 0.78\%) \times 2$, 
$(\ketm{111 223}, 1.21\%) \times 6$, 
$(\ketm{112 222}, 0.52\%) \times 6$ 
\\

We omit the corresponding heralding click patterns for brevity, though they can easily be found (e.g., using Perceval) and necessitate detecting $N-2$ photons on the herald mode(s). We do not analyze how the resulting heralded Bell states are impacted by loss.
We see that, at least up to $N =10$, the 4P6M circuit only has the prescribed $N=4$ input (up to permutations swapping $\ketm{0}$ and $\ketm{1}$ on modes joined by an initial 50:50 beamsplitter).
The 4P5M and 4P8M circuits have multiple additional Bell-heralding inputs (several hundred for the 4P8M circuit), however, they each require at least twice the number of input photons, making them far less practical than the original implementations.
Meanwhile, the 5P5M and 6P6M circuits each have the property that a single additional photon on any one of the modes can also herald the generation of Bell-like states (with the 6P6M circuit having valid inputs for each checked $N \geq 6$, i.e., up to and including $N=10$).
This property is worth emphasizing for two reasons: 
(i) it bolsters the circuits' achievable overall success probability when using biphoton sources with heralding detectors that can resolve $2$-photon clicks,
and
(ii) it makes these circuits robust against the second-order error process in which one source produces an extra photon and one photon is subsequently lost from the heralding mode(s).

\underline{5P5M details.}
Suppose the state $\ketm{11112}$ (or, in fact, any of the 5 permutations) is passed through the 5P5M circuit, the output state is of the form
\be 
    \ket{\tilde{\psi}^{(11112)}_{5P5M}} = \sum_{j=0}^6 \sqrt{p_j} \ket{\tilde{\psi}_j}_b \ketm{j}_h.
\ee 
Notably $\ket{\tilde{\psi}_4}_b = \ket{\Psi^+}$ is the same state as in the original $N=5$ implementation,  \hyperref[heraldStates:5P5M]{(b)}, with $p_4 = 24/625 = 3.84\% = \tfrac{2}{5} p^{5P5M}_\textrm{ideal}$. 
Thus, for (i) we note that if such source events are identified and accepted, this freedom could be leveraged to increase the ideal overall success probability
by $ 5 p_4 (1-|\lambda|^2)^5 (|\lambda|^2)^6$ 
from $p^{5P5M}_\textrm{succ,0} =  p^{5P5M}_\textrm{ideal} [ (1-|\lambda|^2) |\lambda|^2 ]^5$
to $ p^{5P5M}_\textrm{succ} = (1 + 2 |\lambda|^2 ) \times p^{5P5M}_\textrm{succ,0}$. For $|\lambda|^2 = 1/2$ this precisely doubles the success probability from $0.009375\%$ to $0.01875\%$ (including the next valid input, $\ketm{22222}$, only marginally increases this doubling factor: $2 \ra 2.0084$).

Suppose one attempts to prepare the input $\ketm{11111}$ yet a single multiphoton error occurs leading to an extra photon on one mode (or a mixture over these possibilities).
Without photon loss here the herald $\ketm{3}_h$ will lead to a false positive in which the output state, $\ket{\tilde{\psi}_3}_b$, will have an extra photon compared to the target Bell state with $p_3 = 12/125 = 9.6\%$. 
However, if photon loss occurs on the herald mode, $\ketm{4}_h \ra \ketm{3}_h$, one will still get the desired Bell state, $\ket{\tilde{\psi}_4}_b$; this is the second order, $\cO(\xi_i \times \g_h)$, 
robustness of (ii).

\underline{6P6M details.}
Suppose the state $\ketm{111 112}$ or any of the 6 permutations is passed through the 6P6M circuit, the output state is of the form
\be
    \ket{\psi^{(111 112)}_{6P6M}} = \sum_{(j,k):~j+k \leq 7} \sqrt{p_{jk}} \ket{\tilde\psi_{jk}}_b \ketm{j k}_h.
\ee
Here $\ket{\tilde\psi_{05}}_b = \ket{\tilde\psi_{50}}_b = \ket{\chi^+}$
with $p_{05} = p_{50} = 1.54321\% = 5/324$
and $\ket{\tilde\psi_{23}}_b = \ket{\tilde\psi_{32}}_b = \ket{\chi^-}$
with $p_{23} = p_{32} = 0.61728\% = 1/162$,
which are the same states as in the original $N=6$ implementation, \hyperref[heraldStates:6P6M]{(d)}. 
Thus, for (i) we note that with HSPSs one can increase the ideal overall success probability 
from $p^{6P6M}_\textrm{succ,0} = p^{6P6M}_\textrm{ideal} [ (1-|\lambda|^2) |\lambda|^2 ]^6$
to $ p^{6P6M}_\textrm{succ} = (1 + \tfrac{7}{4} |\lambda|^2 ) \times p^{6P6M}_\textrm{succ,0}$. For $|\lambda|^2 = 1/2$ this increases the success probability from $0.00362\%$ to $0.00678\%$ increasing it by $1.875$ times. If the valid 8-photon inputs are included this $1.875$ factor goes to $2.125$ (it goes up slowly from there, slightly surpassing 2.18 by $N=10$).

Now suppose we try to prepare $\ketm{111 111}$ and a single multiphoton error occurs. We see that the $\pm$ cases both exhibit the second-order loss robustness of (ii). Namely, if a single multiphoton error occurs and one herald mode photon is lost, then the target click patterns will yield intended states:
$\ket{05}_h \ra \ket{04}_h$ (or $\ket{50}_h \ra \ket{40}_h$) will herald $\ket{\chi^+}$
and
$\ket{23}_h \ra \ket{13}_h$ (or $\ket{32}_h \ra \ket{31}_h$) will herald $\ket{\chi^-}$.
If no photon loss occurs false positives can occur for heralding
$\ket{\chi^+}$ due to a $(04)$ or $(40)$ click pattern with $p_{04} = p_{40} = 2.6749\% = 13/486$
or
$\ket{\chi^-}$ due to a $(13)$ or $(31)$ click pattern with $p_{13} = p_{31} = 3.90947\% = 19/486$. 
The ratio of these probabilities $p_{13}/p_{04} = 19/13 \approx 1.46$ largely accounts for why the $+$ case of 6P6M scheme handles multiphoton errors better than the $-$ case. 

\begin{figure}[ht!]
    \includegraphics[width=\linewidth]{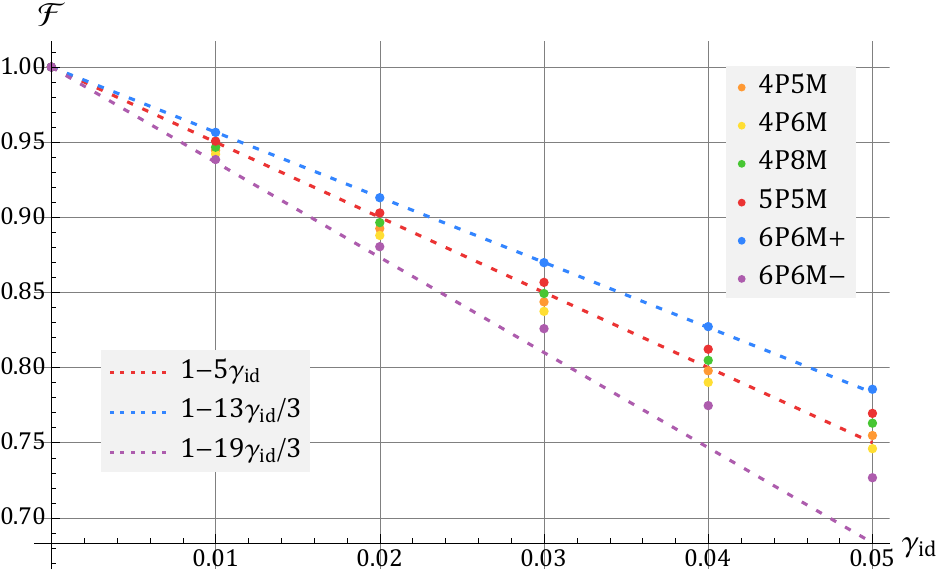} 
    \caption{Multiphoton error impact on heralded Bell-state fidelities for each scheme. The data points are the same as the $\g_b = 0$ values in \figref{fig:lossyHSPSsLumpedCircuitLossFidelityComparison}(a).
    The dashed lines correspond to the fidelity approximation under the mixture model to linear order in $\g_{id}$.
    }\label{fig:multiphoton_FidDeg}
\end{figure}

In \figref{fig:multiphoton_FidDeg} we plot the behavior of each scheme's fidelity in the presence of small multiphoton errors (including consistent heralding detector loss, see relation \ref{item:relation3}). 
The differences in the schemes' multiphoton robustness can be seen explicitly using the mixture model: 
\be
    \rho \approx \frac{p_\textrm{mp} 2 N \xi_i \ketbra{\psi_\textrm{mp}}{\psi_\textrm{mp}} + p_{\hvec} (1 - 2 N \xi_i) \ketbra{\cB_{\hvec}}{\cB_{\hvec}}}{p_\textrm{mp} 2 N \xi_i + p_{\hvec} (1 - 2 N \xi_i)}, \nonumber
\ee 
which captures the leading-order effect of multiphoton errors for $\xi_i \ll 1$ in the absence of other errors. Namely, under this approximating model one either generates the desired Bell state, $\ket{\cB_{\hvec}}$, or the multiphoton-contaminated 3-photon state $\ket{\psi_\textrm{mp}}$ (e.g., $\ket{\tilde\psi_3}$ for the 5P5M scheme) with the scaling following from \eqref{eq:c2HSPS} and the $N$ HSPSs being independent. Here $p_\textrm{mp}$ and $p_{\hvec}$ are the respective probabilities that the multiphoton and target branches yield the accepted heralding pattern.
Based on the model assumption \ref{item:relation2}, with $\g_b = 0$ here we take $|\lambda|^2 \ra 1/2$ such that $\xi_i = \g_{id}/2$.

The resulting Bell state fidelities for the 5P5M and 6P6M$\pm$ schemes under this mixture model are shown as the dashed lines in \figref{fig:multiphoton_FidDeg} (as we have already characterized the corresponding leading order multiphoton contributions). This could likewise be done for the $N=4$ schemes via a weighted average over the multiphoton contribution being on different input modes.

This simple mixture model serves as a useful starting point in understanding multiphoton errors in HBSG. Developing more sophisticated models that account for higher-order multiphoton errors together with photon loss, as required for the second-order robustness of (ii), is an interesting direction for future work. Such models could help explain more systematically why certain resource-state-generation schemes are more robust in different error regimes.

\section{Coupled-ring frequency beamsplitters}\label{sec:twoRingFreqBS-Decomp}
Here we elaborate on the driven coupled-microring frequency beamsplitters of \secref{sec:frequencyBinHBSG}.

\subsection{Physical setup}
\begin{figure}[ht!]
    \subfloat[(a)]{\includegraphics[width=0.95\linewidth, clip=true, trim = 0 0 0 8mm]{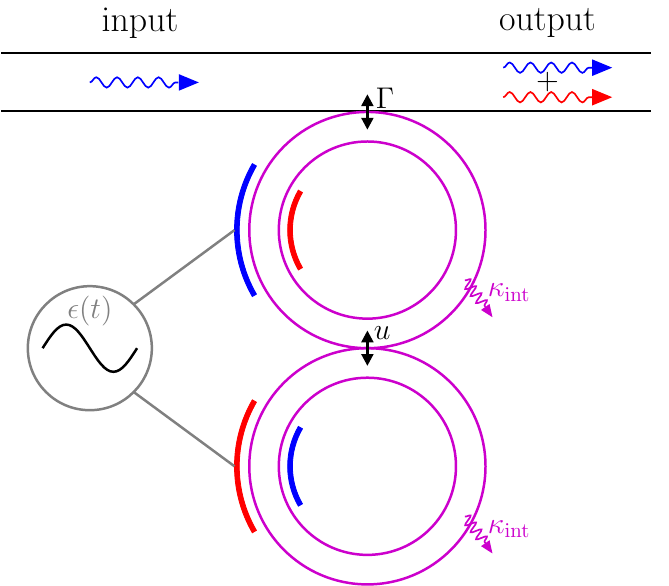}} \\
    \subfloat[(b)]{\includegraphics[width=0.7\linewidth]{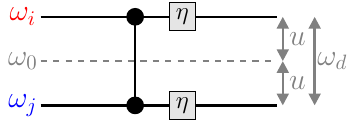}} 
    \caption{(a) A pair of strongly coupled microring resonators driven by an electro-optic phase modulator can be used as a frequency beamsplitter 
    \cite{hu2021chip,munoz2026modeling}. 
    (b) Corresponding lossy frequency beamsplitter circuit schematic. The loss channels, shown via their efficiencies as in \eqref{eq:singleModeAD}, can be taken to both be before or after the beamsplitter as uniform loss channels commute with linear optics (see property \ref{item:uniformAD_commutes}). 
    }\label{fig:freq_BS_illustration}
\end{figure}

Using a pair of strongly coupled microring resonators each driven by an EOM, one can engineer frequency beamsplitters that act on a pair of frequency bins on the bus waveguide that one of the rings is coupled to \cite{hu2021chip,munoz2026modeling}. 
A schematic illustration of the device is given in Fig.~\ref{fig:freq_BS_illustration}(a), where (using the notation of \refref{munoz2026modeling})  $u$ is the inter-ring coupling rate and $\Gamma$ is the coupling rate between the bus waveguide and top ring. The rings are assumed to be identical, each with central frequency $\om_0$ and internal loss rate $\kint$. 

The coupled rings support two normal modes at frequencies $\red{\om_i} = \om_0 - u$ and $\blue{\om_j} = \om_0 + u$ as shown in Fig.~\ref{fig:freq_BS_illustration}(b). An excitation exchange interaction between these normal modes is induced by modulating each ring with drives $\epsilon \cos(\om_d t + \phi)$ resonantly, $\om_d = \blue{\om_j} - \red{\om_i} = 2 u$, but with opposite signs [indicated via the arcs of opposite colors in Fig.~\ref{fig:freq_BS_illustration}(a)].
We assume the frequency bins are well-separated compared to the total resonance linewidth of the hybridized normal modes, $u \gg \kappa = \Gamma/2 + \kint$, as is necessary for the condition $[\red{a_{\om_i}}, \blue{\hc{a}_{\om_j}}] = 0$ to hold.

\subsection{Transfer matrix}
We will now show that the transfer matrix derived in \refref{munoz2026modeling}: 
\[
    \Xi = \begin{pmatrix}
        1 - \frac{\Gamma \kappa}{\kappa^2 + \epsilon^2} & i e^{i\phi} \frac{\Gamma \epsilon}{\kappa^2 + \epsilon^2} \\
        i e^{-i\phi} \frac{\Gamma \epsilon}{\kappa^2 + \epsilon^2} & 1 - \frac{\Gamma \kappa}{\kappa^2 + \epsilon^2}
    \end{pmatrix}
    \equiv \begin{pmatrix}
        \Xi_{11} & \Xi_{12} \\
        -\Xi^*_{12} & \Xi_{11}
    \end{pmatrix}
\]
can be put into the form of \eqref{eq:lossyTransferMatrix} with the efficiency as given by \eqref{eq:etaOmegaFull}. We note that $-1 \leq \Xi_{11} \leq 1$ and $|\Xi_{12}| \leq 1$.
To decompose this transfer matrix into the desired form,
\[
    \sqrt{\eta} \begin{pmatrix}
        s \cos{\theta} & i e^{i \phi} \sin{\theta} \\
        i e^{-i \phi} \sin{\theta} & s \cos{\theta}
    \end{pmatrix}
\]
with $s = \textrm{sign}(\Xi_{11}) = \pm 1$ (or $0$ for a full swap),
it will help to work in terms of the dimensionless couplings $\Omega = \epsilon/\kint$ and $\cC = \Gamma/(2\kint)$ 
such that 
\subalign{
    \Xi_{11} &= \frac{\Omega^2 + 1 - \cC^2}{\Omega^2 + (1+\cC)^2}
    = s \sqrt{\eta} \cos{\theta}, \\
    \Xi_{12} &= i e^{i\phi} \frac{2\Omega \cC}{\Omega^2 + (1+\cC)^2}
    = i e^{i \phi} \sqrt{\eta} \sin{\theta}.
}

\begin{figure}[ht]
    \includegraphics[width=\linewidth, clip=true, trim = 15mm 0 0 0]{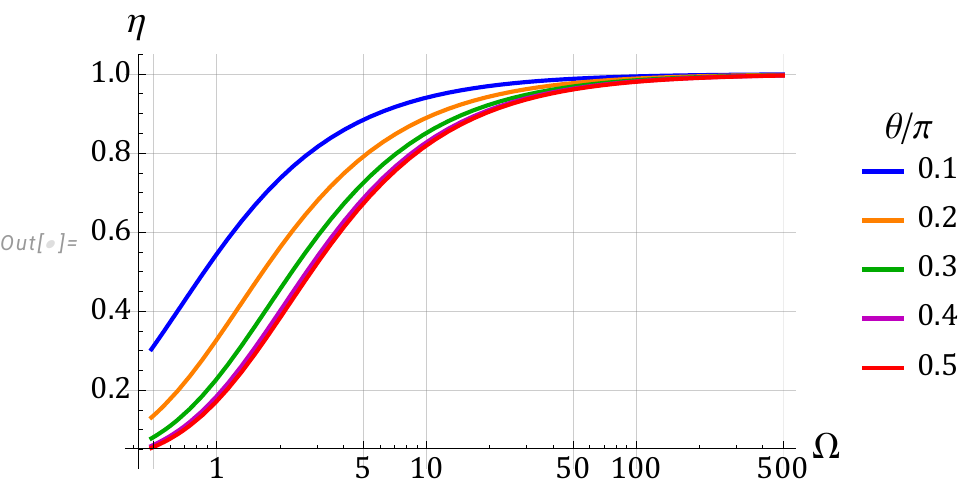}
    \caption{Plot of the frequency beamsplitter efficiency, $\eta(\Omega, \theta)$ of \eqref{eq:etaOmegaFull}, versus $\Omega$ for various $0 < \theta \leq \pi/2$.}
    \label{fig:etaVsOmega_various_theta}
\end{figure}

These simultaneous equations can be used to solve for $\eta$ and $\theta$ in terms of $\Omega$ and $\cC$ as
\be\label{eq:etaOmegacC}
    \eta(\Omega, \cC) = \Xi_{11}^2 + |\Xi_{12}|^2 
    = 1 - \frac{4 \cC}{\Omega^2 + (1+\cC)^2}
\ee
and 
\be
    \tan\br{\theta(\Omega, \cC)} = \frac{|\Xi_{12}|}{|\Xi_{11}|} 
    = \frac{2 \Omega \cC}{| \Omega^2 + 1 - \cC^2 |}.
\ee
The latter expression can be inverted to find the two solutions
\be\label{eq:cCsCooperativity}
    \cC_s(\Omega, \theta) = \sqrt{1 + \Omega^2 \csc^2\theta} - s \Omega \cot\theta,
\ee
which coincide for $s=0$ ($\theta=\pi/2$). Thus, by engineering the waveguide-ring coupling to take on one of the two corresponding values, $\Gamma_s = 2 \kint \cC_s$, one obtains $\eta(\Omega, \theta)$ as given by \eqref{eq:etaOmegaFull} and plotted in Fig.~\ref{fig:etaVsOmega_various_theta}. Note that the $s=-1$ beamsplitter can be put into the form of \eqref{eq:lossyTransferMatrix}, up to a global phase, by mapping $\phi \ra \phi + \pi$, i.e., by flipping the modulation signs, which we assume is done when necessary. 

For large $\Omega$ one finds that 
\be 
    \cC_s = \Omega \br{\tan(\theta/2)}^s + \frac{\sin{\theta}}{2 \Omega} + \cO(\Omega^{-2})
\ee
such that 
\subalign{
    \Gamma_\pm(\Omega, \theta) &= 2 \kint \br{\sqrt{1 + \Omega^2 \csc^2\theta} \mp \Omega \cot\theta} \\
    &\approx  2 \epsilon \br{\tan(\theta/2)}^{\pm1},
}
where $0 \leq \tan(\theta/2) \leq 1$ for $0 \leq \theta \leq \pi/2$. That is, for large modulation amplitudes, $\epsilon \gg \kint$, $\Gamma_\pm$ are nearly proportional to $\epsilon$. Moreover, for appreciable $\theta$ values (not near 0) the proportionality factor is order one, $\Gamma_+ \sim \epsilon$ and $\Gamma_- \gtrsim \epsilon$, e.g.,
for a 50:50 beamsplitter, $\theta = \pi/4$, $\Gamma_+ \approx 0.8 \epsilon$ and $\Gamma_- \approx 5 \epsilon$. Meanwhile for small $0 < \theta \ll 1$,
$\Gamma_+ \ll \epsilon$ and $\Gamma_- \gg \epsilon$.
	
Accordingly, one wants
(i)
$\epsilon \gg \kint$ to minimize loss, and
(ii)
$(\blue{\om_j} - \red{\om_i})/2 = u \gg \kappa = \Gamma/2 + \kint$ to have well-separated frequency bins compared to the total linewidth.
Because $\Gamma \gtrsim \epsilon$ (unless $0 < \theta \ll 1$), from (i) we have that $\Gamma \gg \kint$ such that with (ii) one finds $u \gg \Gamma/2$.
Note that it will typically be preferable to operate at the smaller $\Gamma_+$---both as it is easier to realize smaller couplings and as otherwise one has $\Gamma_- \geq 2 \epsilon \gg \kint$ constraining $\epsilon$ from above and below---for which the desired hierarchy of couplings is
\be\label{eq:couplingHierarchy}
    u \gg \Gamma_+ \sim \epsilon   \gg \kint.
\ee
In practice, one would not be able to easily control the precise value of $\Gamma$ when fabricating a device. Once the actual value is known, the modulation amplitude, $\epsilon$, should be fine-tuned to the appropriate operating point.

\section{Frequency-domain HBSG}\label{app:extraFreqImplemenationDetails}

\subsection{Fidelities and heralding probabilities}\label{app:freqImplementationHeraldingProbabilities}

\begin{figure*}[ht!]
    \includegraphics[width=0.95\linewidth]{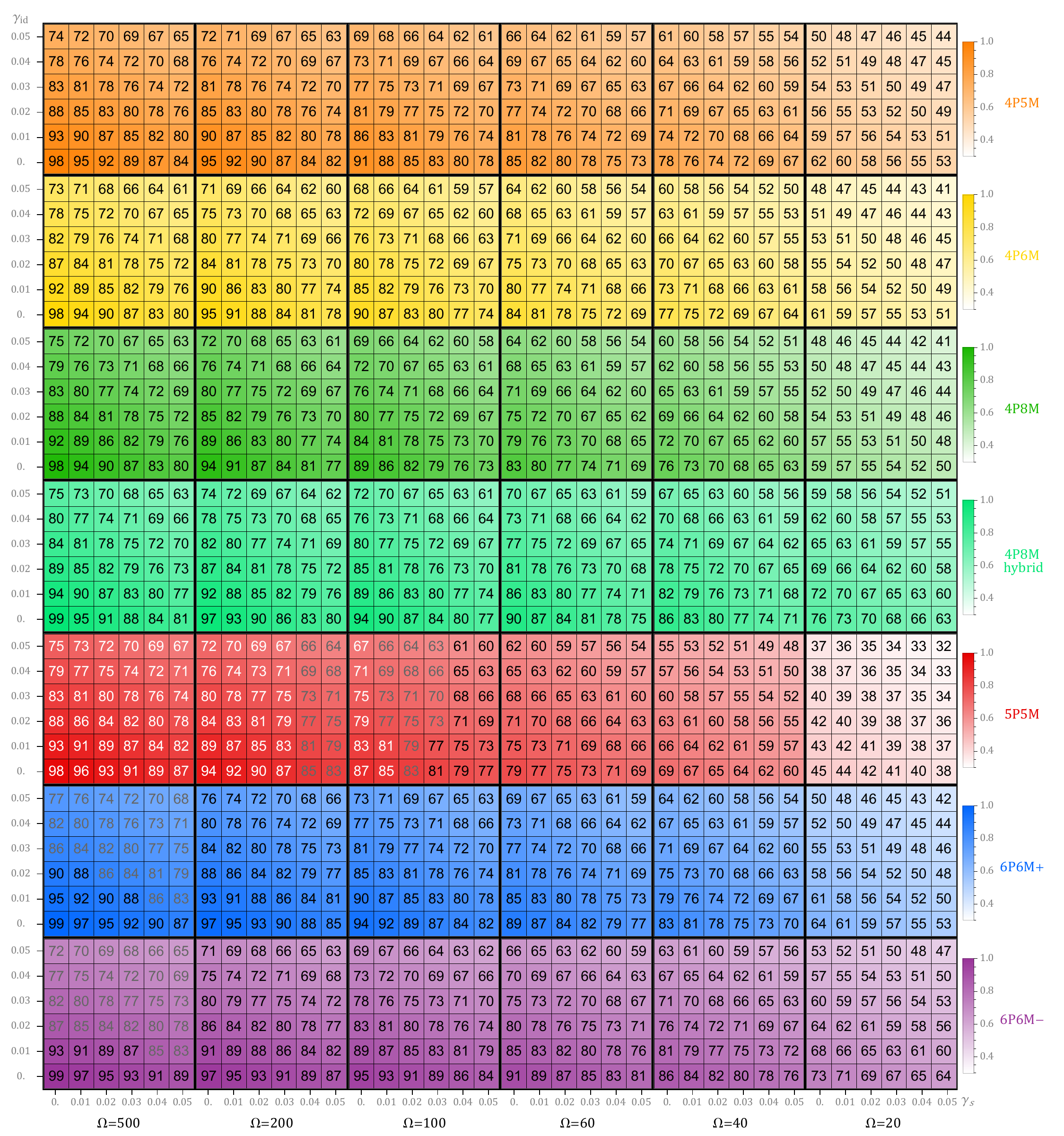}
    \caption{
    For each circuit and value of $\Omega \in \{ 20, 40, 60, 100, 200, 500 \}$, we construct a density plot of the (averaged) fidelity of the output Bell state(s) as a function of $\g_s$ and $\g_{id}$.
    We arrange these density plots into a larger grid varying the circuit vertically (indicated by color, labeled on the far right) and the value of $\Omega$ horizontally with smaller values, corresponding to lossier frequency beamsplitters, going right.
    For the 5P5M and 6P6M$\pm$ schemes we consider two variants of the circuits: (i) and (ii), as defined in \appref{app:circuitVariants}. 
    Rather than including density plots for each variant, we report the fidelity of the better performing variant.
    In each case, variant (i) is found to win significantly (by up to a few percent) for relatively small $\Omega$ and perform comparably for large $\Omega$ (see \figref{fig:variant_histograms}). This behavior is coarsely indicated via the color of the overlaid fidelity percentage with black, gray, and white indicating that (i) wins by greater than $0.2\%$, (i) barely wins by $0$-$0.2\%$, and (ii) wins yet never by more than a $0.1\%$, respectively.
    }\label{fig:freq_full_fidelity_grid} 
\end{figure*}

Now we present all of the fidelities and heralding probabilities for the frequency-domain implementation across each circuit and for each value of $\Omega$, $\g_s$, and $\g_{id}$ (with $n_\textrm{extra} = 3$).
The corresponding varying of 4 quantities (3 numerical values and 1 label) can be done using the ``grid of grids'' construction shown in \figref{fig:freq_full_fidelity_grid} for the fidelities (from which \figref{fig:modestOmega_fullComparison_WinnerAndMargins} derives).
Note that the results of \figref{fig:lumped_circuit_fidelity_and_nBar}(a) are formally equivalent to taking the perfect beamsplitter, $\Omega \ra \infty$, limit in a single row of the outer grid in \figref{fig:freq_full_fidelity_grid} yet with an extended $\eta_s \ra \eta_b$ axis to account for lumped circuit loss.

\begin{figure*}[p]
    \subfloat[(a)]{\includegraphics[width=0.9\linewidth, clip=true, trim = 1mm 0 1mm 5mm]{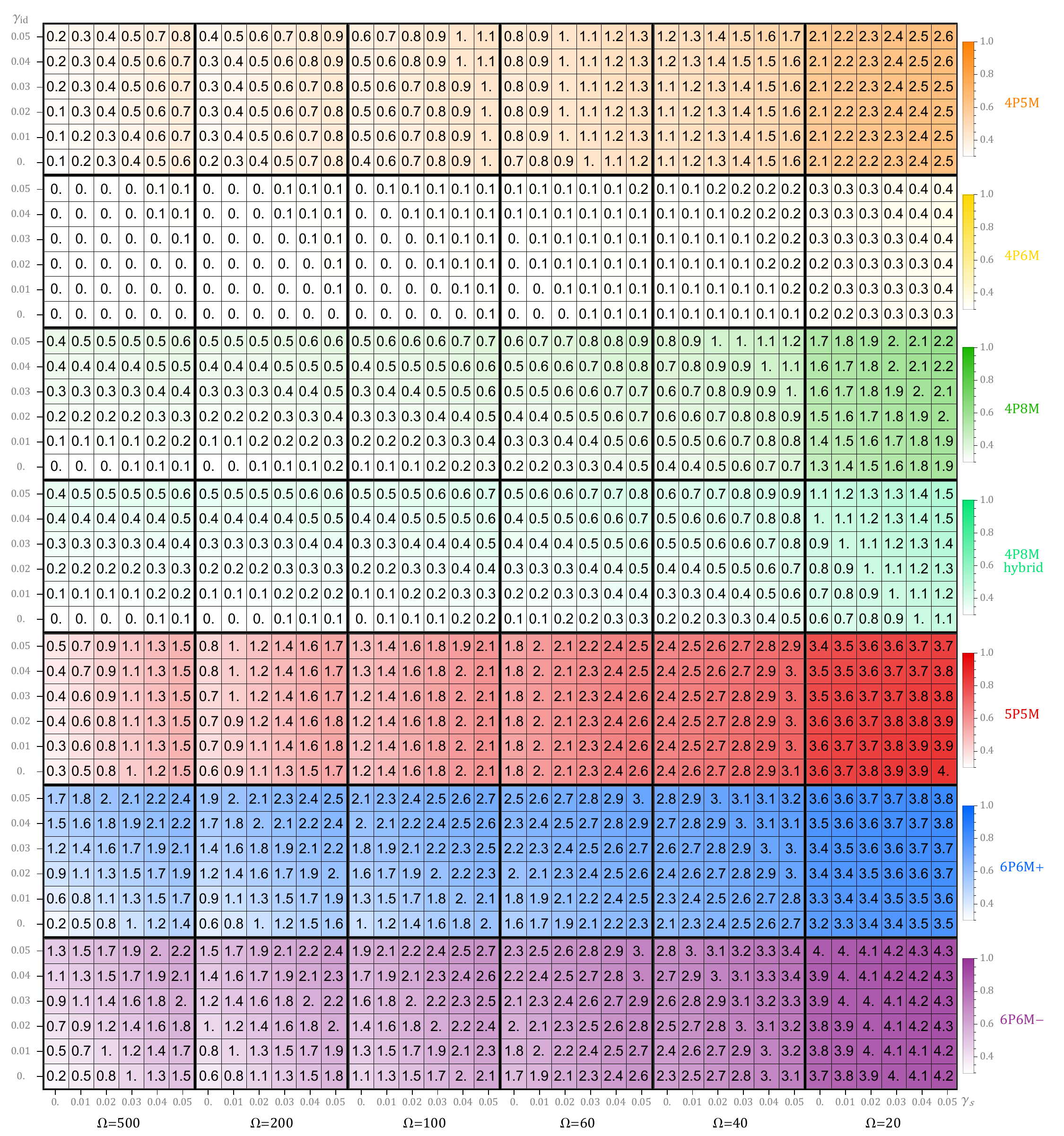}}
    \\
    \captionsetup[subfloat]{captionskip=-5pt}
    \subfloat[(b)]{\includegraphics[width=0.6\linewidth, clip=true, trim = 15mm 12mm 0 4mm]{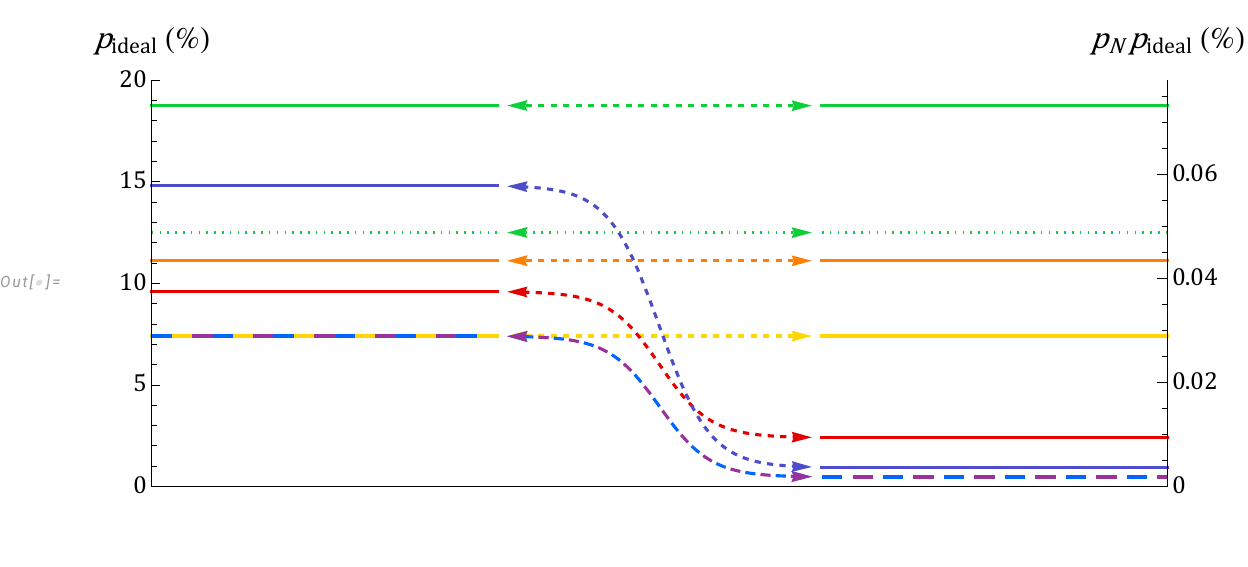}}
    \caption{(a) Comparison of the HBSG schemes heralding probabilities relative to their ideal values, $p_\textrm{ideal}$, given in Table \ref{tab:BSG_circuit_overview}. 
    In each density plot we are plotting the value of $p_\textrm{ideal} - p_\textrm{her}$ and we annotate the grid points with their values rounded to the nearest $0.1\%$.
    However, here we only consider the (i) variants of the 5P5M and 6P6M circuits which are better performing in terms of fidelity, see \figref{fig:variant_histograms}.
    (b) Plot of $p_\textrm{ideal}$ for each scheme (left) and the corresponding \emph{maximum} single-shot probability of success (right), $p_N p_\textrm{ideal}$, which accounts for the HSPSs being probabilistic, $p_N \leq 1/4^N$. We use the same color scheme as (a), though we merge the 4P8M schemes (green) as they have the same heralding probabilities and also plot the probabilities if the permuted Bell output, $\protect\ket{\chi^+}$, is not accepted (dotted green). The combined 6P6M scheme probabilities (dark blue) are simply twice the 6P6M$\pm$ ones.
    }\label{fig:fullComparison_PHer_diff}
\end{figure*}

In \figref{fig:fullComparison_PHer_diff} we show the grid of grids construction for the heralding probability degradation, $p_\textrm{ideal} - p_\textrm{her}$, of each circuit for the same parameter ranges as the fidelity grids of \figref{fig:freq_full_fidelity_grid}.
We see that in each case this probability difference is positive (or near zero), so the various errors act to decrease the observed heralding probability, $p_\textrm{her}$, relative to its ideal value. Moreover, this decrease tends to be more significant for the higher $N$ schemes.
Similar to \appref{app:relativeYieldWithHSPSs}, there appears to be some tradeoff between the fidelity $\cF$ and relative heralding probability $p_\textrm{her}/p_\textrm{ideal}$. This can be seen quite directly in \figref{fig:nExtraConvergence} and more generally by plotting the relative yields and seeing that they tend to coalesce for schemes with the same $N$. However, this pattern appears less well-defined here than in the lumped-loss models.

\subsection{Hybrid spatial-frequency encodings}
\label{app:hybrid_spat_freq}
\begin{figure*}[ht!]
    \hfill
    \subfloat[(a)]{\includegraphics[height=7.5cm]{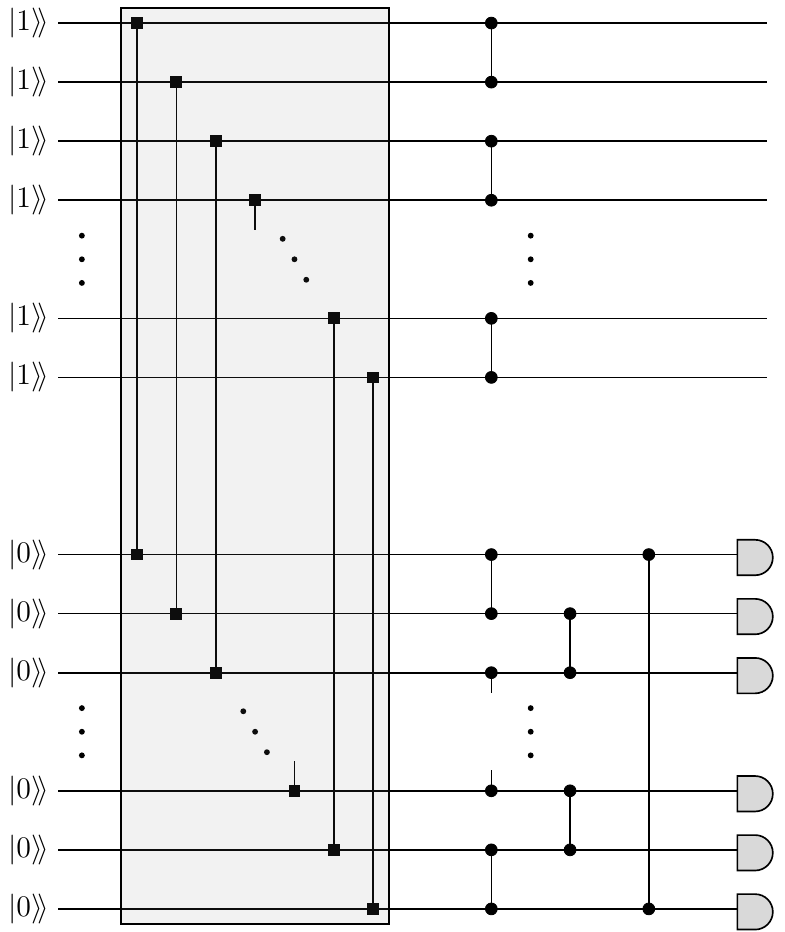}}
    \hfill
    \subfloat[(b)]{\includegraphics[height=7.5cm]{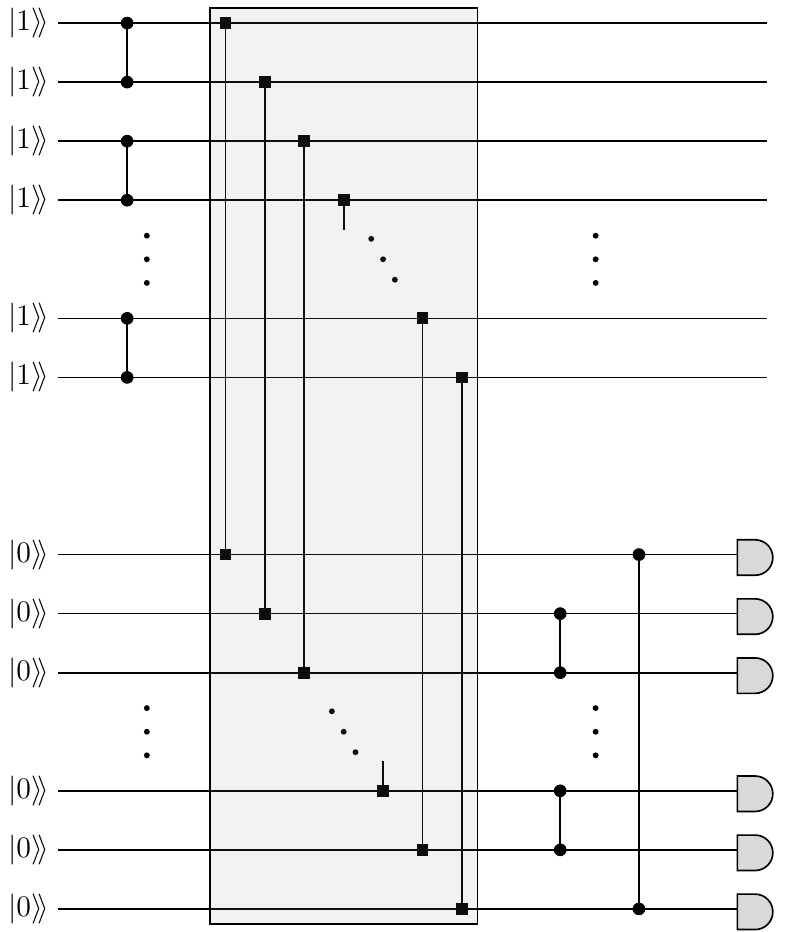}} 
    \hfill
    \caption{Photonic circuits for the heralded generation of dual-rail $n$-qubit GHZ states. They are presented here to emphasize that a hybrid spatial-frequency implementation is natural because the $2n$ transverse beamsplitters (those adorned with square mode indicators) can be implemented with a single spatial beamsplitter across two waveguides as in \figref{fig:hybrid_4P8M_HBSG}.
    (a) $n$-GHZ scheme of \refref{gimeno2016towards} and (b) the closely related variant of \refref{chin2024heralded}. 
    }\label{fig:hybrid_circuits}
\end{figure*}

As noted in \secref{sec:4P8MFrequencyImplementation}, several heralded schemes for frequency-encoded resource-state generation can be spatially hybridized. This was illustrated for the 4P8M HBSG scheme in \figref{fig:hybrid_4P8M_HBSG} and is illustrated for certain GHZ generation schemes in \figref{fig:hybrid_circuits}.
Our motivation for such hybridization is that spatial beamsplitters are both significantly less lossy than the corresponding frequency beamsplitters and, moreover, they enable a certain multimode transformation across all the frequency bins at once. 
The resulting benefit can be seen in the strong performance of the hybrid 4P8M HBSG scheme in \secref{sec:frequencyHBSG}. Accordingly, further analyses of how to leverage such hybridization are potentially fruitful.

In each case, there are twice as many modes as photons, $M = 2N$, and the circuit can be naturally bifurcated in a manner such that there are $N$ beamsplitters linking distinct modes in each half. Accordingly, we are able to trade $N$ ``transverse'' frequency beamsplitters---that operate across several frequency bins---for a single spatial-mode beamsplitter.
Having $N$ frequency bins per spatial mode is important as we want the final $N$-photon entangled resource state to be encoded in the frequency domain within a single guided spatial mode. 

In \figref{fig:hybrid_circuits}(a) and (b) we schematically show two closely related implementations of a class $2n$-photon, $4n$-mode schemes for generating dual-rail photonic GHZ states on $n \geq 2$ qubits,
\be\label{eq:nQubitGHZ}
    \ket{n\textrm{-GHZ}} = \p{\ket{0}^{\otimes n} + \ket{1}^{\otimes n}}/\sqrt{2},
\ee
as found in Refs.~\citenum{gimeno2016towards} and \citenum{chin2024heralded}, respectively.
These schemes are directly translated from the polarization encoding (see Fig.~4.11 of \refref{gimeno2016towards} and Fig.~3 of \refref{chin2024heralded}) to the abstracted mode representation used throughout the text that could equally correspond to frequency, spatial, or another encoding. 
We find it instructive to present the full translated schemes here, especially as doing so reveals that the circuits are naturally hybridizable.  
To check that the translation was performed correctly, we  verified in Perceval that these schemes indeed herald the desired GHZ states in the aforementioned manner up to $n=6$ (for 12 photons across 24 modes).

In both cases, for a given $n$, the corresponding scheme succeeds with probability 
\be\label{eq:nGHZpIdeal}
    p_\textrm{ideal}(n\textrm{-GHZ}) = \frac{1}{2^{2n-1}}.
\ee
Success is heralded by detecting a single photon on each of the $n$ pairs of modes joined by one of the final beamsplitters, i.e., on the mode pairs 
\begin{align}
    \label{eq:nGHZLastLayer}
    \{ (2n+2j+1, 2n + (2j+2 \!\mod 2n) ) \\
    : j \in \{0,1,\cdots,n-1\}  \}, \nonumber
\end{align}
where the $\textrm{mod } 2n$ accounts for the long beamsplitter across all $2n$ of the bottom heralding modes.
The output state is of the form of \eqref{eq:nQubitGHZ} up to known relative phase factors 
(determinable based on the heralding click pattern) with dual-rail qubits defined as in \eqref{eq:dualRailQubitDef} on adjacent modes. 

By expressing the $n$-GHZ schemes in this manner, one can easily see that the schemes of \refref{chin2024heralded} are effectively identical to that of \refref{gimeno2016towards} via the $H_4$ circuit identity
\be\label{eq:H4_identity}
    \raisebox{-0.45\totalheight}{ \includegraphics[width=0.38\linewidth]{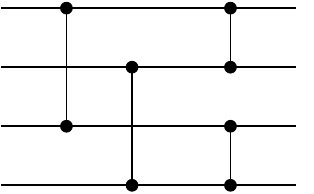}}
    = 
    \raisebox{-0.45\totalheight}{ \includegraphics[width=0.38\linewidth]{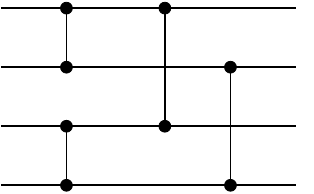}}.
\ee
Namely, it follows that in \figref{fig:hybrid_circuits}(a) one can commute the layer of $2n$ adjacent mode beamsplitters to the left through the transversal beamsplitters, yet one can omit the beamsplitters on the bottom heralding modes as all-vacuum-initialized modes are invariant under linear optics. Moreover, as noted in \refref{gimeno2016towards}, their $n$-GHZ schemes generalize the 4P8M HBSG scheme of \refref{zhang2008demonstration} and the 3-GHZ scheme of \refref{varnava2008good}. 

For HBSG, taking $n=2$ in \figref{fig:hybrid_circuits}(a), the resulting circuit is slightly different than that of \refref{bartolucci2021creation}, our \figref{fig:BSG_circuits}(e), because of the beamsplitters on the $2n \ra 4$ upper output modes (the apparent difference in the $H_4$ implementation is simply a mode permutation). 
These two beamsplitters can be understood simply as single dual-rail qubit Hadamard gates, so for $n=2$ they simply map Bell states to Bell states.
Moreover, by removing (or equivalently adding due to involution) these two beamsplitters one maps output states of the form $(\ketm{2000} + \ketm{0200} + \ketm{0020} + \ketm{0002})/2$---as heralded by two disjoint detections on a pair of modes joined by the beamsplitters in 
\eqref{eq:nGHZLastLayer} (and none detected on the other pair)---onto the permuted Bell state $\ket{\chi^+}$. This accounts for the difference in $p_\textrm{ideal}$ from the $1/8$ of \eqref{eq:nGHZpIdeal} to the reported $3/16$ when $\ket{\chi^+}$ is accepted.

\subsection{Circuit decomposition variants}\label{app:circuitVariants}
For the frequency-domain implementation of \secref{sec:frequencyHBSG}, the place where the beamsplitter loss occurs matters, we must pick specific instantiations of the circuits.
For the 4P5M, 4P6M, and 4P8M (as well as its hybrid variant) we simply use the circuits of \figref{fig:BSG_circuits}.
For the 5P5M and 6P6M circuits, we note that an arbitrary $m$-mode unitary can be decomposed into $m(m-1)/2$ two-mode beamsplitters (10 and 3 for the $m=5$ and 3 mode DFTs, respectively) using either the Reck (triangular) or Clements (rectangular) decompositions of Refs.~\citenum{reck1994experimental} and \citenum{clements2016optimal}, respectively. 
This assumes the beamsplitters have tunable phase freedom as is the case in this coupled-microring implementation, see \eqref{eq:lossyTransferMatrix}, otherwise additional internal phase shifters will be needed. Moreover, a series of phase shifters on each mode, at either the beginning or end of the circuit, are necessary to exactly match the unitary. However, in our cases we can omit the corresponding prefatory phases acting on the single photon inputs as these phases simply act to change the phase of the output Bell states in a well-defined manner.

\begin{figure}[ht!]
    \includegraphics[width=0.9\linewidth]{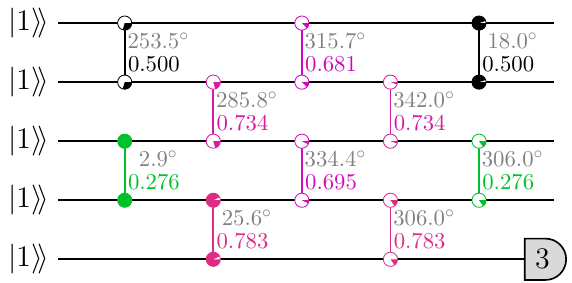}
    \caption{Clements decomposition of $5 \times 5$ DFT circuit. Rounded values of $\phi$ and $\sin^2\theta$ are reported to the right of each beamsplitter as shown in gray (above) and in colors consistent with the legend of Fig.~\ref{fig:BSG_circuits} with color blending (below), respectively.}
    \label{fig:5P5M_Clements_decomp}
\end{figure}

\begin{figure}[h]
    \includegraphics[width=0.7\linewidth]{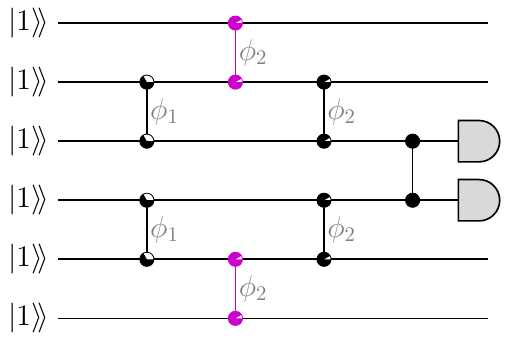}
    \caption{Alternate 6P6M circuit implementation with the 3 DFT circuits ``flipped'' 
    relative to \figref{fig:BSG_circuits}(d).}
    \label{fig:6P6M_flipped_3DFTs_BSG}
\end{figure}

As noted in \figref{fig:modestOmega_fullComparison_WinnerAndMargins}, for the 5P5M and 6P6M schemes we consider two variants, denoted (i) and (ii).
For the 5P5M case, we focus on the Clements decomposition shown in \figref{fig:5P5M_Clements_decomp} to distribute the beamsplitter loss more evenly over the modes and reduce optical depth relative to the Reck decomposition. 
However, we consider either
\begin{enumerate}[label=(\roman*)]
    \item detecting on the bottom mode as shown in \figref{fig:BSG_circuits}(b) and \figref{fig:5P5M_Clements_decomp}, 
    or 
    \item detecting on the central mode that is ``touched'' by the most beamsplitters (5 here; as are the inner non-central modes), 
    which also heralds $\ket{\Psi^+}$ with the qubits defined on the adjacent modes, $(0,1)$ and $(3,4)$. 
\end{enumerate}
For the 6P6M case, there is freedom in terms of whether the 3 DFTs are implemented in a ``$\blacktriangle$ fashion'' with one beamsplitter touching the top mode and two the bottom or vice versa in a ``$\blacktriangledown$ fashion.'' We consider two corresponding options
\begin{enumerate}[label=(\roman*)]
    \item $\blacktriangle\blacktriangledown$ pairing shown in \figref{fig:6P6M_flipped_3DFTs_BSG},
    or
    \item the $\blacktriangledown\blacktriangle$ pairing shown in \figref{fig:BSG_circuits}(d).
\end{enumerate}

\begin{figure}[htp!]
    \captionsetup[subfloat]{farskip=0pt, captionskip=-15pt}
    \subfloat[\quad (a) 5P5M]{\includegraphics[width=0.85\linewidth]{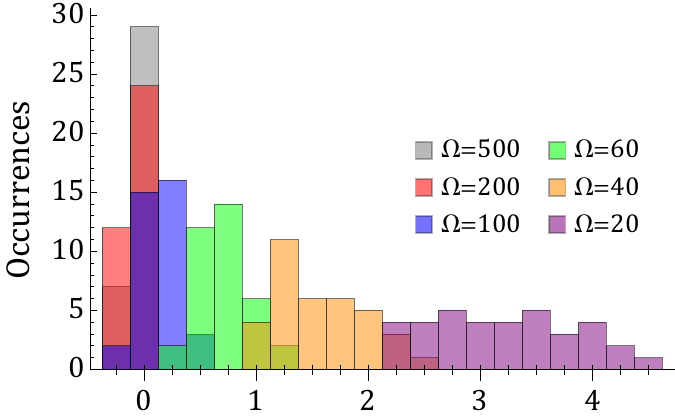}} \\ 
    \subfloat[\quad (b) 6P6M$+$]{\includegraphics[width=0.85\linewidth]{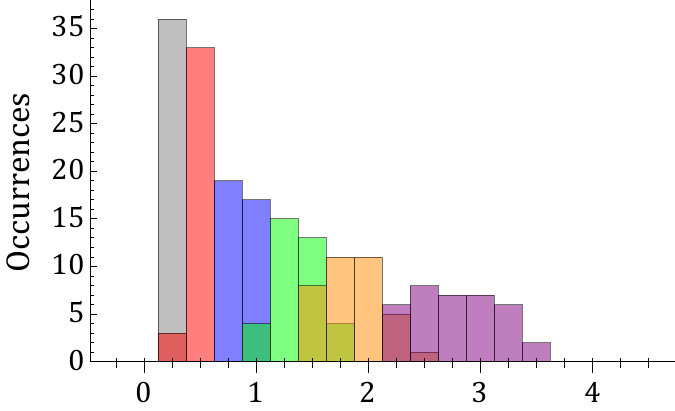}} \\
    \subfloat[\quad (c) 6P6M$-$]{\includegraphics[width=0.85\linewidth]{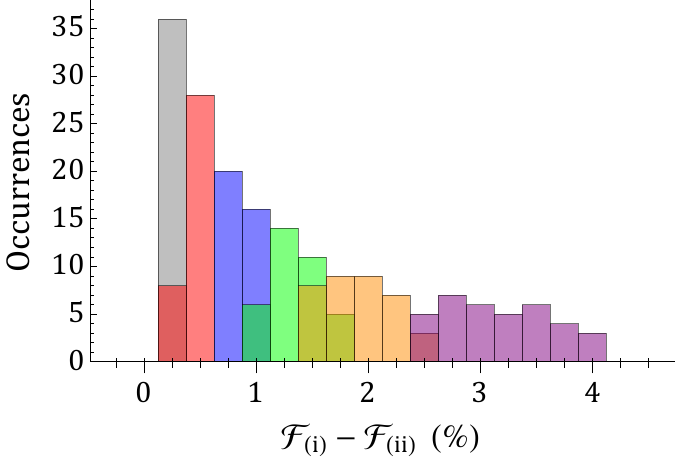}}
    \caption{Contrasting the fidelity performance of the (i) and (ii) variants of the 5P5M and 6P6M$\pm$ circuits. For each corresponding circuit we overlay histograms of the fidelity differences, $\cF_{(\rm i)} - \cF_{(\rm ii)}$, for each $\Omega$ considered in \figref{fig:freq_full_fidelity_grid} (shown in different colors). 
    Each histogram for a fixed circuit and $\Omega$ is comprised of $36$ data points corresponding to the values of $\g_s$ and $\g_{id}$ swept over for each density plot in \figref{fig:freq_full_fidelity_grid}.
    }\label{fig:variant_histograms}
\end{figure}

Beyond the comparison using the fidelity annotation coloring in \figref{fig:freq_full_fidelity_grid}, in \figref{fig:variant_histograms} we give histograms contrasting the performance of these variants for the various considered $\Omega$ values.
For both the 5P5M and 6P6M circuits we see that circuit (i) performs significantly better for quite lossy beamsplitters, e.g., performing $\cO(3\%)$ better for $\Omega = 20$, which corresponds to a 50:50 beamsplitter efficiency of $93\%$ and a full-swap efficiency of $90\%$. The difference in performance of course vanishes in the limit of perfect beamsplitters, $\Omega \ra \infty$, however it does so more quickly for the 5P5M variants. The variants considered are not meant to be exhaustive, but rather to illustrate the extent to which implementations are sensitive not just how much loss there is overall, but also where it occurs. 

\subsection{When beamsplitter loss lumps}
Generally, the introduction of lossy beamsplitters, especially in the frequency domain, goes beyond the scope of the lumped-loss model as seen in \figref{fig:4P5M_HBSG_shifting_loss}. However, the 4P6M and 4P8M circuits have the property that beamsplitter loss can faithfully be shifted to the end of the circuit using properties of AD channels, as do the $n$-qubit GHZ generation schemes of Refs.~\citenum{gimeno2016towards} and \citenum{chin2024heralded}, see \figref{fig:hybrid_circuits}. Namely, suppose the beamsplitter efficiency depends on the reflectivity $\sin^2\theta$, as is the case for the driven coupled microring frequency beamsplitters, see \eqref{eq:etaOmegaFull}.

For the 4P8M circuit all eight beamsplitters are 50:50 though some have different mode separations, $j-i \in \{1, 2, 4\}$. We suppose the efficiencies can vary but are the same for a given mode separation, $\eta^{(j-i)}_{1/2}$. Then one can simply include beamsplitter loss by making the replacements 
$\eta_b \ra \eta_b \cdot \eta^{(4)}_{1/2}$
and 
$\eta_h \ra \eta_h \cdot \eta^{(4)}_{1/2} \eta^{(1)}_{1/2} \eta^{(2)}_{1/2}$
in \eqref{eq:lumpedLossGeneral}.
For the 4P6M circuit, which uses beamsplitters with reflectivities of $1/2$ and $2/3$ acting on adjacent modes with respective efficiencies $\eta_{1/2}$ and $\eta_{2/3}$, we can similarly make the replacements
$\eta_b \ra \eta_b \cdot \eta_{1/2} \sqrt{\eta_{2/3}}$ and
$\eta_h \ra \eta_h \cdot \eta_{1/2} \eta_{2/3}$.
The $\sqrt{\eta_{2/3}}$ term simply arises due to the fact that only half of the output modes experience the $\sin^2\theta = 2/3$ beamsplitter loss.
These properties allow us to understand the $\g_{id}=0$ behavior of the 4P6M and 4P8M (including hybrid) schemes as shown in \figref{fig:freq_full_fidelity_grid}. 
Moreover, they ease the analytic calculation of how loss propagates in subsequent operations such as fusions.
	
\end{document}